\documentclass[aps,pre,superscriptaddress,10pt]{revtex4}
\usepackage[paperwidth=210mm,paperheight=297mm,centering,hmargin=2cm,vmargin=2.4cm]{geometry}
\usepackage{graphicx}
\graphicspath{{../Imgs/}}
\usepackage{graphics,subfigure,tikz}

\usepackage{comment}
\usepackage[american]{babel}
\usepackage{amsmath,amsthm,amsfonts,amssymb}
\usepackage{color}
\usepackage[normalem]{ulem}
\usepackage[colorlinks]{hyperref }

 \newcommand{\re}{\text{Re}}

\DeclareMathOperator{\sech}{sech}

\newcommand{\rmi}{\mathrm{i}}

\begin{document}
\newtheorem{corollary}{Corollary}[section]
\newtheorem{remark}{Remark}[section]
\newtheorem{definition}{Definition}[section]
\newtheorem{theorem}{Theorem}[section]
\newtheorem{proposition}{Proposition}[section]
\newtheorem{lemma}{Lemma}[section]
\newtheorem{help1}{Example}[section]
\renewcommand{\theequation}{\arabic{section}.\arabic{equation}}

\newcommand{\tl}{\textlatin}
\newcommand{\no}{\noindent}

%%%%%%%%%%%%%%%%%%%%%%%%%%%%
\newcommand{\bbR}{\mathbb{R}}

\newcommand{\bbN}{\mathbb{N}}
\newcommand{\bbC}{\mathbb{C}}
\newcommand{\bbZ}{\mathbb{Z}}

%%%%%%%%%%%%%%%%%%%%%%%%%%%%%%%%%%%%%%%%%%%%%%%%%%%%%%%%%
\newcommand{\bfR}{\mathbf{R}}
\newcommand{\bfE}{\mathbf{E}}
\newcommand{\bfC}{\mathbf{C}}
%%%%%%%%%%%%%%%%%%%%%%%%%%%%%%%%%%%%%%%%%%%%%%%%%%%%%%%%%%%%%%%%%%%%%%%%%
\newcommand{\mcL}{\mathcal{L}}
\newcommand{\mcU}{\mathcal{U}}
\newcommand{\mcA}{\mathcal{A}}
\newcommand{\mcR}{\mathcal{R}}
\newcommand{\ran}{\mathrm{ran}}
\newcommand{\mcV}{\mathcal{V}}
\newcommand{\mcB}{\mathcal{AB}}   
\newcommand{\dom}{\mathrm{dom}}
\newcommand\fh{{\mathfrak{H}}}
\newcommand\ch{{\mathcal{H}}}
\newcommand\ci{{\mathcal{I}}}
\newcommand\cb{{\mathcal{B}}}
\newcommand\cj{{\mathcal{J}}}
\newcommand\ck{{\mathcal{K}}}
\newcommand\cd{{\mathcal{D}}}
\newcommand\cw{{\mathcal{W}}}
\newcommand\ct{{\mathcal{T}}}
\newcommand\cx{{\mathcal{X}}}
\newcommand\cf{{\mathcal{F}}}

\newcommand\cm{{\mathcal {M}}}
\newcommand\bbc{{\mathbb{C}}}
\newcommand\bfh{{\mathbf{B}(\mathfrak {H}})}

\newcommand\hilbert{{\tl {Hilbert}\;}}
\newcommand\inpr{{\langle\cdot, \cdot\rangle }}

\newcommand{\Th}{\widehat{T}_{\max}} 
\newcommand{\To}{\overline{T}_{\max}}

\title{Collapse dynamics for the discrete nonlinear Schr\"odinger equation with gain and loss}

\author{G. Fotopoulos}
\affiliation{Department of Mathematics, Statistics and Physics, College of Arts and Sciences, Qatar University, P.O. Box 2713, Doha, Qatar}
\author{N.\,I. Karachalios}
\affiliation{Department of Mathematics, University of the Aegean, Karlovassi, 83200 Samos, Greece}
\author{V. Koukouloyannis}
\affiliation{Department of Mathematics, Statistics and Physics, College of Arts and Sciences, Qatar University, P.O. Box 2713, Doha, Qatar}
\author{K. Vetas}
\affiliation{Department of Mathematics, University of the Aegean, Karlovassi, 83200 Samos, Greece}
\begin{abstract}
We discuss  the finite-time collapse, also referred as blow-up, of the solutions of
a discrete nonlinear Schr\"{o}dinger (DNLS) equation incorporating
linear and nonlinear gain and loss. This DNLS system 
appears in many inherently discrete physical contexts  as
a more realistic generalization of the Hamiltonian DNLS lattice.   By using
energy arguments in finite  and infinite dimensional phase spaces (as guided by the boundary conditions imposed), we prove analytical upper and lower bounds for the collapse time, valid for both the defocusing and focusing cases of the model. In addition, the existence of a critical value in the linear loss parameter is underlined, separating finite time-collapse from energy decay. The numerical simulations, performed for a wide class of initial data, not only verified the validity of our bounds, but also revealed that the analytical bounds can be useful in identifying two distinct types of collapse dynamics, namely, extended or localized.  Pending on the discreteness /amplitude regime, the system exhibits either type of collapse and the actual blow-up times approach, and in many cases are in excellent agreement, with the upper or the lower bound respectively. When these times lie between the analytical  bounds, they are associated with a nontrivial mixing of the above major types of collapse dynamics, due to the corroboration of defocusing/focusing effects and energy gain/loss, in the presence of discreteness and nonlinearity. 
\end{abstract}
\maketitle
\section{Introduction}\label{sec1}
%The discrete nonlinear Schr\"odinger (DNLS) equation 
%\begin{equation}\label{eq001}
%	\mathrm{i}\dot{\psi}_n +k(\psi_{n+1}-2\psi_n+\psi_{n-1})+ f(|\psi_n|^2) \psi_n =0,
%\end{equation} 
%is one of the fundamental nonlinear lattice dynamical systems. In Eq.~\eqref{eq001}, $\psi_n(t)$ is the complex dynamical variable of the coupled unit at the $n$-th site of the 1D-lattice, the dot stands for time-differentiation and $k=1/2h^2$ denotes the discretization parameter;
%$h$ is the lattice spacing, so as to control
%the distance from the continuum limit as $h \rightarrow 0$.  The sufficiently smooth function $f:\mathbb{R}\rightarrow\mathbb{R}$ represents a generic nonlinearity. Formally, the continuum limit of the DNLS Eq.~(\ref{eq001}) corresponds to its continuous counterpart, the 1D-nonlinear Schr\"odinger (NLS) partial differential equation (PDE). 

The discrete nonlinear Schr\"odinger (DNLS) equation is linked to the mathematical description of numerous  physical processes, ranging from the propagation of self-trapped modes in biomolecules and in arrays of nonlinear optical fibers, to the tight-binding description of Bose--Einstein condensates in optical lattices,  \cite{JCE1, kevrekidis2001,KevreDNLS, reviewsD}. Furthermore, the  DNLS equation can be associated to 
%another important lattice, 
the discrete Klein--Gordon (DKG) lattice and it  actually consists a normal form of the DKG in the small energy - small amplitude regime \cite{PalPen}. 
%\begin{eqnarray}
%	\label{eq002}
%	\ddot{U}_n-\varepsilon(U_{n+1}-2U_n+U_{n-1})+W'(U_n)=0.
%\end{eqnarray}
%In Eq.~(\ref{eq002}), the parameter $\varepsilon$ stands for the coupling parameter between adjacent sites, while the function $W(U)$ corresponds to the nonlinear (on-site) external potential. 
The DKG equation appears in numerous distinct physical contexts; these include 
%the study of
crystals and metamaterials, ferroelectric and ferromagnetic domain
walls, Josephson junctions, nonlinear optics, topological excitations in hydrogen-bonded chains or chain of base pairs in DNA, complex
electromechanical devices,
% and
and so on --cf. the monographs \cite{DPbook, Chrisbook}  and the review articles \cite{reviewsA,reviewsB,reviewsC}. 
%The derivation of the DNLS equation \eqref{eq001} results when seeking for the DKG equation (\ref{eq002}) extended  (plane wave) solutions of the form:
%\begin{eqnarray}
%	\label{mi3}
%	U_n(t)=\psi_n(t)e^{-\mathrm{i}\omega t}+ \hbox{c.c},\quad\psi_n\in\mathbb{C},
%\end{eqnarray}
%where c.c. stands for complex conjugation.
%
%For instance, it is shown in \cite{Yuri}, that in the case of Eq.~(\ref{eq002}) subject to a cubic or quartic potential $W(U)$, the envelope $\psi_n(t)$ of the solutions (\ref{mi3}), if  slowly-varied in time (with respect to the main oscillation at frequency $\omega$), then it satisfies the cubic DNLS ($f(\rho)=\rho$).

%applications such as the oscillations of nanomechanical cantilever arrays and the denaturation and related phase transformations of the DNA double strand. It can also be used to describe the phenomenon of freak waves (or rogue waves) in the ocean. For a detailed and exhaustive study see, \cite{kevrekidis2001} and \cite{KevreDNLS} and the references therein.

However, a crucial difference between the  DNLS and the DKG models concerns the issue of global-in-time existence of their solutions. More closely to its continuous counterpart (the Klein--Gordon PDE), the DKG equation may exhibit solutions which may collapse (blow-up) in finite time, depending on the type of the on-site potential and the size of initial data. Generally, the collapse phenomenon features the domination of
%the
nonlinearity when the initial data are sufficiently large; however, the DKG equation provides an example that it may arise
%since
in more subtle cases, where the initial data
may be contained even deeply within the domain of attraction of a stable
state. Such a process indicates the existence of interesting energy
exchange and transfer mechanisms between the interacting units, which may lead to their escape dynamics towards infinity. The simplest example relevant to the above phenomenology is the DKG equation with a repulsive quartic on-site potential \cite{V13}. On the other hand, the %corresponding cubic 
DNLS equation with generic nonlinearities, 
% still validly derived through the aforementioned slow-variation approximation of solutions (\ref{mi3}), 
does not exhibit finite-in time collapse dynamics. While this is still in similarity with the global existence of solutions of 
its continuous analogous,  
the focusing 1D-cubic NLS,
%{\bf VK: mono cubic??})
% the solutions of DNLS models with generic nonlinearities $f$ always exist globally in-time, 
it is in contrast with the 
%corresponding ones of 
behavior of the generic NLS equations with non-cubic nonlinearities, \cite{sulem}. This remarkable difference between the DNLS and its continuous 
%[\textbf{GF: why only continuous}] 
counterpart holds independently of the size of the initial data, the growth described by the nonlinearity and even of the lattice dimension, \cite{karachalios2005}.

Therefore, the emergence of finite time blow-up in DNLS-type lattices could occur in the presence of additional linear/nonlinear effects. In this perspective, we investigate in the present paper, the finite time collapse for the DNLS equation:
\begin{equation}\label{eq01}
	\mathrm{i}\dot{\psi}_n  -sk(\psi_{n+1}-2\psi_n+\psi_{n-1})+ |\psi_n|^2 \psi_n = \mathrm{i}\gamma \psi_n + \mathrm{i}\delta |\psi_n|^2 \psi_n \quad \gamma,\delta\in\bbR,\qquad s=\pm1.
\end{equation} 
Equation (\ref{eq01}) governs the dynamics of an arbitrary number of oscillators, which are placed equidistantly. The parameter $s$ assigns the defocusing/focusing nature of the lattice: the case $s=1$ corresponds to the so-called defocusing case, while the  $s=-1$ case corresponds to its focusing counterpart. Together with the  initial conditions, 
we supplement the lattice~\eqref{eq01}, with either {\it periodic boundary conditions} of with {\it Dirichlet boundary conditions}.
In the case of an infinite lattice, we can consider {\it  the vanishing boundary conditions}. For simplicity, in what follows, the periodic initial-boundary value problem will be called as ($\mathcal{P}$), while the Dirichlet or vanishing initial-boundary value problem, will be called as ($\mathcal{D}$).

%The DNLS equation (\ref{eq01})  governs the dynamics of an arbitrary number of $N+1$ oscillators, which are placed equidistantly at the interval $\Omega=[-L,L]$ with lattice spacing $h=2L/N$
% of length $2L$
%. The case $s=1$ ($s=-1$) corresponds to the so-called defocusing (focusing) case, while $k=1/2h^2$ stands for the discretization parameter. 

%Together with the  initial conditions 
%\begin{equation}\label{eq03}
%	\psi_n(0) = \psi_{n,0},
%\end{equation}
%we supplement the lattice~\eqref{eq01}, with either {\it periodic boundary conditions} 
%\begin{equation}\label{eq02} 
%	\psi_n = \psi_{n+N},
%\end{equation} 
%or {\it Dirichlet boundary conditions}
%\begin{equation}
%\psi_0=\psi_N=0,\label{vanc}
%\end{equation}
%or, we may even consider an infinite lattice, supplemented with {\it  the vanishing boundary conditions}
%\begin{equation}
%\label{vancv} 
%	\lim_{|n|\rightarrow\infty}\psi_n=0.
%\end{equation}

%For simplicity, the periodic initial-boundary value problem \eqref{eq01}-\eqref{eq03}-\eqref{eq02} will be called as ($\mathcal{P}$), while the Dirichlet or vanishing initial-boundary value problem \eqref{eq01}-\eqref{eq03}-\eqref{vanc} [(\ref{vancv})], will be called as ($\mathcal{D}$).

Eq.~(\ref{eq01}) is one of the simplest DNLS models
incorporating linear and nonlinear gain/loss effects.
In particular, the parameter $\gamma$ describes linear loss ($\gamma<0$) [or gain
($\gamma>0$)],
while $\delta$ describes
nonlinear loss ($\delta<0$) [or gain ($\delta>0$)].
The presence of these
effects is physically relevant in the context of nonlinear optics, especially for the evolution of localized modes in optical waveguides, see \cite{Efrem1,Boris1,Boris2,Boris3} (and the references therein).  In this context, $\gamma$ describes a
linear absorption ($\gamma<0$) [or linear amplification ($\gamma>0$)], while $\delta$
stands for nonlinear amplification ($\delta>0$) [or gain saturation ($\delta<0$)].
We also refer to \cite{Nail1, KodHas87, Agra1, Agra2, akbook, Gagnon}, for the role of such effects in various discrete and continuous set-ups; generally,
for physically relevant settings, such terms play a crucial role in the stability of
the localized modes. However, the potential destabilization of such structures leading to a finite time collapse has been proved to be a major characteristic of the dynamical behavior of DNLS models incorporating gain/loss effects, \cite{Boris2, Boris3}.

The presentation of the results has as follows:   After some preliminary results on the well-posedness of the problems ($\mathcal{P}$) and ($\mathcal{D}$) stated in Section~\ref{sec2A}, we consider in Section~\ref{sec2B1}, the initial-boundary value problem ($\mathcal{P}$),  describing the analytical arguments on the finite time collapse for its solutions. The methods are motivated from the results and questions posed on the dynamics of  the continuous counterpart, discussed in Ref.~\cite{anastassi2017}.
Extending the energy arguments of \cite{anastassi2017} from the continuum to the discrete ambient space, and by using a spatially averaged
power-energy functional, we prove an analytical upper bound for the blow-up time. This upper bound is a function of the averaged power of the initial data
and the gain/loss parameters $\gamma$ and $\delta$.
%This is due to the fact that
%the relevant energy inequality and the analytical upper bound of the blow-up time were found to depend
%only on these parameters. Two of the main scenarios, finite collapse or decay to the zero-state
%characterized by the choices of $\gamma$ and $\delta$, are as follows.
%illustrated in Figure~\ref{Final}:
%%%%%%%%%%%%%%%%%%%%%%%%%%%%%%%%%%%%%%%%%%%%%%%%%%%%%%%%%%%%%%%%%%%%%%%%
While regimes of collapse and decay of solutions are defined by $\gamma,\delta>0$ and $\gamma,\delta<0$, interestingly, in the regime
%The 1st $\gamma,\delta>0$ and 3rd $\gamma,\delta<0$ quadrants define the regimes of collapse and decay of solutions.for
%in the 2nd
$\gamma<0,\,\delta>0$, %-quadrant,
we derive 
%when $\delta>0$, 
%[\textbf{GF: why we have $\delta>0$ again?}]
%have recovered
the existence of a critical value $\gamma^*$ on the linear loss,
%$\gamma^*=-M(0)\delta$, where $M(0)$ stands for the initial averaged norm of the initial data.
separating the finite-time collapse (for $\gamma>\gamma^*$) from the decay (for $\gamma<\gamma^*$) of solutions as in the continuous case. 

Next, in Section~\ref{sec2B2}, we consider the initial-boundary value problem ($\mathcal{D}$).
For this problem, the natural phase space  is the  $\ell^2$-space of square summable sequences. By using the inclusion relations $\ell^q\subset\ell^p$ for $1\leq q\leq p\leq\infty$ between the $\ell^p$-spaces, we derive this time, a lower bound for the collapse time of solutions. 
%While the lower bound bears the same functional dependence on the lattice parameters, it is proved to depend on the power of the initial conditions  {\bf (VK: phrasing. Don't understand)}. 
%%Remarkably, the novel estimate is not only fulfilled as a lower bound but in the focusing case is proved to be in excellent agreement with the numerical blow-up times. Such a saturation  was also  observed in the defocusing case when the discretization parameter was increased. On the one hand, the numerical results verify the previous analysis on the role of energy dispersion and discreteness in defining the transient dynamics prior to collapse; the analytical arguments on the phase space of truly localized states effectively describes the consequent localized type of collapse in the presence of the focusing effect, which is enhanced in the discrete regime. On the other hand, 
It is important to remark that the above analytical argument is valid  only for the inherently discrete model and not for its continuous limit; the $L^p(\Omega)$ spaces embed into one another in the reverse order to the embeddings of $\ell^p$ when $\Omega\subset\mathbb{R}^K$, $K\geq 1$ has finite volume, while the $\ell^p$-inclusions are not valid when $\Omega=\mathbb{R}^K$, \cite{AdamsS}. This can be viewed as a functional manifestation of the discreteness effect being responsible for the translational invariance breaking; the latter is a major feature for the differences between the discrete and continuous siblings, \cite{KevreDNLS}. 

%\textcolor{blue}{The proofs of all the above  the analytical results are included for the convenience of the reader, in the complementary Appendix \ref{proofs}.}

Section~\ref{sec3}, is devoted to  numerical investigations for the dynamics of both problems ($\mathcal{P}$) and ($\mathcal{D}$), in order to examine the validity of the analytical results, as well as, to examine the behavior of the systems in various parametric regimes and different types of initial conditions.
% We consider three different examples of initial conditions, namely, spatially extended, vanishing, and decaying to a finite background. 

First, in Section~\ref{sec3A}, we study a set of spatially extended initial conditions in the form of a discrete plane wave which are associated to the problem ($\mathcal{P}$). In this case, we observe an excellent agreement between the analytical upper bound and numerical blow-up times, independently of the various parameters of the lattice and the initial conditions, which is also analytically justified. %This excellent agreement for both the focusing the defocusing case, is completely justified by the fact that the timescales for collapseof the background, and the analytical upper bound for the collapse-time of the averaged power-functional coincide. Actually, the evolution of the power-functional is proved to be that of the extended initial wave-background governed by a cubic ordinary differential equation (ODE). 
In this case, the collapse of the solution is manifested through an increase of the amplitude of the initial state in the whole lattice. This kind of behavior defines the so-called {\it weak} or {\it extended blow-up}. 
%Concerning the critical value $\gamma^*$, it is verified
%numerically that its analytical estimation is very accurate; %and the important feature, that when
%for $\gamma<\gamma^*$ the solution decays and
%when
%for $\gamma>\gamma^*$ the solution blows-up in finite-time.

In Section~\ref{sec3B}, we present the results for a case of vanishing initial conditions. These initial data have the form of a discretized, $\mathrm{\sech}$--profile, reminiscent of a ``discrete bright soliton''. The behavior of the system under these spatially localized initial data is found to be far more complex in comparison with the one emerged from the spatially extended initial conditions.  Since the localized initial data are compliant asymptotically to both problems  ($\mathcal{P}$) and ($\mathcal{D}$), we test the numerical blow-up times against both the analytical upper and lower bound, respectively. In the case of high discreteness and for high amplitudes of the initial conditions the  {\it strong or localized type of collapse} appears. This type is dominated by energy localization by an evolution, prior to collapse, self-similar to the initial conditions. This behavior is mainly exhibited in the focusing case. When localized collapse occurs, the numerical blow-up times converge to an excellent agreement with the  analytical lower bound. By considering a less  discrete lattice or/and higher-amplitude
initial conditions we observe a behavior between the two kinds of collapse, in which there is, in various extends, some energy dispersion prior to collapse. In these intermediate cases, also the blow-up times lie between the two analytical bounds. 

The numerical investigations conclude, in Section~\ref{sec3C} and~\ref{sec3D}, with the study of the dynamics for two additional types of initial data. The first one is a decaying to a finite (nonzero)  background, which has the form of a discretized $\mathrm{tanh}^2$ function and resembles the profile of a density dip of a ``discrete dark soliton'',   while the second one has the form of a rectangular box.  
%Such initial data are compliant with problem ($\mathcal{P}$) 
%Note that we do not use Dirichlet boundary conditions in the numerical calculations in this case of considered initial conditions we still consider the lower bound calculated using them in order to defend our claim that it consists a bound for all kinds of inital conditions independently of the boundary ones. Indeed the results show that the blow-up times lie between the two bounds. 
The numerical findings in the defocusing case of the $\mathrm{tanh}^2$-case yet justify the excellent agreement of the numerical blow-up times with the analytical bound, as in the case of the spatially extended initial conditions. In the corresponding focusing case, as well as in the case of the box-initial conditions intermediate behaviors are observed and discussed. Note that in all the above numerically studied cases a very sharp agreement, of the role of $\gamma^*$ in the collapse or decay of the solutions, is established. %Although analytically the lower bound is not valid for the above class of initial conditions---since they do not satisfy the problem ($\mathcal{D}$)---it is still considered in the numerical simulations, and it is found that it is still fulfilled as a lower bound; the numerical blow-up times lie between the analytical estimates. Furthermore, in the focusing case, we observe that the numerical blow-up times approach closely the lower bound as the system approaches its continuous limit, while when the lattice spacing is increasing, the numerical blow-up times converge towards the upper bound. This remarkable behavior is quite on the contrary to the findings in the case of localized initial conditions, and its potential analysis which has still to be explained, motivates for further future explorations.

Finally, in Section~\ref{sec4}, we summarize and discuss the implications of our results with an eye towards future work. 
%%%%%%%%%%%%%%%%%%%%%%%%%%%%%%%%%%%%%%%%%%%%%%%%%%%%%
\section{Collapse in finite time}\label{sec2}
\setcounter{equation}{0}
\subsection{Local-in-time existence of solutions}\label{sec2A}
\label{presec}
In this section, we recall
some preliminary information on the functional setting of the problem, as well as a local-in-time, existence of solutions result. Equation~\ref{eq01} describes the dynamics of an arbitrary number of $N+1$ oscillators, which are placed equidistantly at the interval $\Omega=[-L,L]$ with lattice spacing $h=2L/N$, while $k=1/2h^2$ stands for the discretization parameter. We remind here that $s=1$  corresponds to the so-called defocusing  case, while $s=-1$ stands for the focusing case.

Together with the  initial conditions 
\begin{equation}\label{eq03}
	\psi_n(0) = \psi_{n,0},
\end{equation}
we supplement the lattice~\eqref{eq01}, with either {\it periodic boundary conditions} 
\begin{equation}\label{eq02} 
	\psi_n = \psi_{n+N},
\end{equation} 
or {\it Dirichlet boundary conditions}
\begin{equation}
\psi_0=\psi_N=0,\label{vanc}
\end{equation}
or, we may even consider an infinite lattice, supplemented with {\it  the vanishing boundary conditions}
\begin{equation}
\label{vancv} 
	\lim_{|n|\rightarrow\infty}\psi_n=0.
\end{equation}

The periodic initial-boundary value problem \eqref{eq01}-\eqref{eq03}-\eqref{eq02} will be called thereafter as ($\mathcal{P}$), while the Dirichlet or vanishing initial-boundary value problem \eqref{eq01}-\eqref{eq03}-\eqref{vanc} [(\ref{vancv})], will be called as ($\mathcal{D}$).

For brevity, we refer only to the focusing case of the problem ($\mathcal{P}$), noting that the corresponding results are still valid for both cases of the problem [($\mathcal{D}$), ($\mathcal{P}$)] and the variants (focusing and defocusing) of the DNLS, and can be proved in an analogous manner.  
 
The problem ($\mathcal{P}$) will be considered in complexifications of real periodic sequences of period $N$,  denoted by 
\begin{eqnarray}
\label{lp}
{\ell}^p_{\mathrm{per}}:=\left\{\psi=(\psi_n)_{n\in\mathbb{Z}}\in\mathbb{C}:\quad\psi_n=\psi_{n+N},\quad
\|\psi\|_{\ell^p_{\mathrm{per}}}:=\left(h\sum_{n=0}^{N-1}|\psi_n|^p\right)^{\frac{1}{p}}<\infty\right\}, \quad 1\leq p<\infty.
\end{eqnarray}
The definition of the norm of ${\ell}^p_{\mathrm{per}}$, follows by the standard 
%trapezoidal rule of the 
numerical approximation of the $L^p(\Omega)$-norm. Note that 
%\begin{eqnarray*}
%||\psi||^p_{L^p(\Omega)}\approx h\left(\frac{|\psi_0|^p+|\psi_N|^p}{2}+\sum_{n=1}^{N-1}|\psi_n|^p\right),	
%\end{eqnarray*}
%for the continuous limit $h\rightarrow 0$, when the periodic boundary conditions (\ref{eq02})  are applied.
 the case $h=O(1)$ corresponds to the discrete regime of our system, while, when $h\rightarrow0$, we approximate the continuous NLS equation. Setting $p=2$ in (\ref{lp}), we get the usual Hilbert space of square-summable (complex) periodic sequences endowed with the real scalar product
\begin{eqnarray*}
\label{lp2}
(\phi,\psi)_{\ell^2_{\mathrm{per}}}=h\,\mathrm{Re}\sum_{n=0}^{N-1}\phi_n\overline{\psi}_n,\quad\phi,\,\psi\in\ell^2_{\mathrm{per}}.
\end{eqnarray*}
We shall also use for convenience, the short-hand notation $\Delta_{d}$ for the
one-dimensional discrete Laplacian
\begin{equation*}\label{eqD}
\Delta_{d}: {\ell}^2_{\mathrm{per}} \to {\ell}^2_{\mathrm{per}}, \quad
\left\{\Delta_{d} \psi\right\}_{n\in\bbZ}=\psi_{n+1}-2\psi_n+\psi_{n-1},
\end{equation*}
defining a $\mathbb{C}$-linear, self-adjoint negative  operator on $\ell^2_{\mathrm{per}}$, as it can be shown in the following lemma. The proof of the lemma as well as the following proofs can be found in Appendix~\ref{proofs}.
\begin{lemma}
\label{DL} For any $\psi\in {\ell}^2_{\mathrm{per}}$, we consider the linear operator  $(B\psi)_{n\in\mathbb{Z}}=\psi_{n+1}-\psi_{n}$. The operator $\Delta_{d}: {\ell}^2_{\mathrm{per}} \to {\ell}^2_{\mathrm{per}}$ satisfies the relations
\begin{eqnarray}
\label{lp6}
(\Delta_{d}\psi,\psi)_{\ell^2_{\mathrm{per}}}&=&-\|B\psi\|_{\ell^2_{\mathrm{per}}}\leq 0,\\
\label{lp7}
(\Delta_{d}\phi,\psi)_{\ell^2_{\mathrm{per}}}&=&-(B\phi,B\psi)_{\ell^2_{\mathrm{per}}}=(\phi,\Delta_{d}\psi)_{\ell^2_{\mathrm{per}}}, \quad \phi,\;\psi\in {\ell}^2_{\mathrm{per}}.
\end{eqnarray}
\end{lemma}
%Note that in the case of the infinite lattice $\bbZ$, the following inequalities for the sequence spaces $\ell^p$ and $\ell^q$
%\begin{equation}
%||\psi||_{q} \leq  ||\psi||_{p},\;\;1\leq p\leq q\leq\infty
%\end{equation}
%and
%\begin{equation}\label{eqF}
%0\leq (-\Delta_{d} \psi_n, \psi_n)_{\ell^2} \leq 
%4 \sum_{n\in\bbZ}|\psi_n|^2
%\end{equation}
%hold, being valid also in the finite-dimensional setup. 
%
%In what follows we restrict to the the finite dimensional subspace $\ell^2_{N+1}$.
Consequently, we have from Lemma \ref{DL}, that the operator
$\mathrm{i}\Delta_{d}:\ell^2_{\mathrm{per}}\rightarrow \ell^2_{\mathrm{per}}$ is $\mathbb{C}$-linear and skew-adjoint and $\rmi\Delta_{d}$
generates a group $(\mathcal{T}(t))_{t\in\mathbb{R}}$, of
isometries on $\ell^2_{\mathrm{per}}$. Clearly, the same properties hold for the operator $k \Delta_{d} $.  Thus, for fixed $T>0$ and $(\psi_{n,0})_{n\in\mathbb{Z}}:=\psi^0\in\ell^2_{\mathrm{per}}$, a function $\psi\in\mathrm{C}^1([0,T],\ell^2_{\mathrm{per}})$ is a solution of the problem $(\mathcal{P})$ %(\ref{eq01})-(\ref{eq02}), 
for $s=-1,$ if and only if
\begin{eqnarray}
	\label{milds}
	\psi(t)=\mathcal{T}(t)\psi^0+\rmi\int_{0}^{t}\mathcal{T}(t-\tau)F(\psi(\tau))d\tau,\quad
	(F(\psi))_{n\in\mathbb{Z}}:=\mathrm{i}\gamma\psi_n+(\mathrm{i}\delta-1)|\psi_n|^2\psi_n.
\end{eqnarray}
The existence of local-in-time solutions is guaranteed by the following theorem, whose proof follows the lines of \cite[Theorem 2.1. p. 94]{karachalios2005}.
\begin{theorem}\label{th01}
Let $\psi^0\in\ell^2_{\mathrm{per}}$ and $\gamma, \delta \in\bbR.$ Then, there exists a function $T_{\mathrm{max}}:\ell^2_{\mathrm{per}}\rightarrow (0,\infty]$ with the following properties:\vspace{.2cm}\\
(a)\  For all $\psi^0\in\ell^2_{\mathrm{per}}$, there exists  $\psi\in\mathrm{C}^1([0,T_{\mathrm{max}}(\psi^0)),\ell^2_{\mathrm{per}})$  such that
for all $0<T<T_{\mathrm{max}}(\psi^0)$, $\psi$ is the unique solution of the problem (\ref{eq01})-(\ref{eq03}) in $\mathrm{C}^1([0,T],\ell^2_{\mathrm{per}})$ (well-posedeness).\vspace{.2cm}\\
(b)\  For all $t\in [0,T_{\mathrm{max}}(\psi^0))$,
\begin{eqnarray*}
\label{maxT}
T_{\mathrm{max}}(\psi^0)-t\geq \frac{1}{2(L(R)+1)}:=T_R,\quad R=2\|u(t)\|_{\ell^2_{\mathrm{per}}},
\end{eqnarray*}
where $L(R)$ is the Lipschitz constant for the map $F:\ell^2_{\mathrm{per}}\rightarrow\ell^2_{\mathrm{per}}$. Moreover, the following alternative holds: (i) $T_{\mathrm{max}}(\psi^0)=\infty$, or (ii) $T_{\mathrm{max}}(\psi^0)<\infty$ and $\lim_{t\uparrow T_{\mathrm{max}}(\psi^0)}\|\psi(t)\|_{\ell^2_{\mathrm{per}}}=\infty$ (maximality).\vspace{.2cm}\\
(c)\ $T_{\mathrm{max}}:\ell^2_{\mathrm{per}}\rightarrow (0,\infty]$ is lower semicontinuous.
In addition, if $\{\psi_{n0}\}_{n\in\mathbb{N}}$ is a sequence in
$\ell^2_{\mathrm{per}}$ such that $\psi_{n0}\rightarrow \psi^0$ and if $T<T_{\mathrm{max}}(\psi^0)$,
then $S(t)\psi_{n0}\rightarrow S(t)\psi^0$ in $\mathrm{C}^1([0,T],\ell^2_{\mathrm{per}})$, where $S(t)\psi^0=\psi(t)$, $t\in [0,T_{\mathrm{max}}(\psi^0))$, denotes the solution of Eq.~(\ref{milds})
(continuous dependence on initial data).
\end{theorem}
\subsection{Estimates on the finite collapse time}
\subsubsection{Upper bound for the collapse time for periodic boundary conditions}\label{sec2B1}
In this paragraph,  we focus on the scenario (b)(ii) of Theorem~\ref{th01} for  the problem ($\mathcal{P}$), that is, in describing the parametric regimes in which $T_{\mathrm{max}}(\psi^0)<\infty$ and $\lim_{t\uparrow T_{\mathrm{max}}(\psi^0)}\|\psi(t)\|_{\ell^2_{\mathrm{per}}}=\infty$. Furthermore, we will also prove an upper estimate for the collapse $T_{\mathrm{max}}(\psi^0)$. Instead of applying energy arguments on the standard power functional (the $\ell^2_{\mathrm{per}}$-norm $\|\psi(t)\|_{\ell^2_{\mathrm{per}}}$), it is more convenient to apply such arguments on the functional  
\begin{equation}\label{eq06}
M(t) = \frac{e^{-2\gamma t}}{N} \sum_{n=0}^{N-1}|\psi_n|^2.
\end{equation}
The consideration of the functional (\ref{eq06}) is motivated by the energy balance law satisfied by the solutions of the problem (\ref{eq01})-(\ref{eq03}):
\begin{equation}
\frac{d}{dt}\sum_{n=0}^{N-1}|\psi_n|^2 = 2\gamma\sum_{n=0}^{N-1}|\psi_n|^2  +2\delta \sum_{n=0}^{N-1}|\psi_n|^4 .
\label{cl}
\end{equation}
We may observe the similarity of the balance law (\ref{cl}) to the corresponding balance laws satisfied by the solutions of the continuous NLS counterparts incorporating gain and loss \cite{anastassi2017,SD}. Indeed,  Eq.~(\ref{cl}) can be derived by repeating a similar algebra to the continuous case, this time in the discrete set-up: Eq.~(\ref{eq01}) is multiplied by $\overline{\psi}_n$, then, by summing on the account of Lemma \ref{DL} and keeping the imaginary parts of the resulting equation we derive Eq.~(\ref{cl}).  It is evident that the $\exp(-2\gamma t)$-term is the integrating factor for the linear (with respect to $\sum_{n=0}^{N-1}|\psi_n|^2$) part of Eq.~(\ref{cl}). Note that, for $\gamma=\delta=0$, Eq.~(\ref{cl}) is nothing but the conservation of the $\ell^2_{\mathrm{per}}$-norm in
the standard conservative DNLS model. Thus, the functional (\ref{eq06}) is the spatially averaged power  $\frac{1}{N} \sum_{n=0}^{N-1}|\psi_n|^2$ (a rather natural choice due to the boundary conditions considered) multiplied by the above integrating factor. It also satisfies a differential equality given in the following lemma.
\begin{lemma}
\label{DifEq}
For  $\psi^0\in\ell^2_{\mathrm{per}}$, we consider the unique solution $\psi\in\mathrm{C}^1([0,T_{\mathrm{max}}(\psi^0)),\ell^2_{\mathrm{per}})$ of the problem ($\mathcal{P}$). Then $M(t)$ for this solution, satisfies the differential equality
\begin{equation}\label{diff.form.M}
\frac{d M(t)}{dt} = \frac{2e^{-2\gamma t}}{N}\left(\delta \sum_{n=0}^{N-1}|\psi_n|^4\right).
\end{equation}
\end{lemma}

%Therefore we see that $J_1=J_2=J_3=0$ and we have for eq~(\ref{eq07}) that the second term is 
%\begin{equation}\label{eq09}
%\frac{2e^{-2\gamma t}}{2L+1} Re\sum_{n=0}^{K}\dot{\psi}_n\overline{\psi}_n = 
%\frac{2e^{-2\gamma t}}{2L+1}\left(\gamma \sum_{n=0}^{K}|\psi_n|^2\right) + 
%\frac{2e^{-2\gamma t}}{2L+1}\left(\delta \sum_{n=0}^{K}|\psi_n|^4\right).
%\end{equation} 
%Using the above differential form, we will prove a finite blow-up time estimate for the solution of the problem \eqref{eq01}--\eqref{eq03}, which is stated in the following theorem.
%This differential form of the function $M$, will play a crucial role in the proof of the next important Theorem, where we conclude that the solution diverges in finite-time under appropriate assumptions on its initial value at time $t=0$.
With Lemma \ref{DifEq} in hand, we can prove the parametrically conditioned collapse of solutions in finite time, and an upper bound for their collapse time. The result is stated in the following theorem.
\begin{theorem}\label{The2a}
For $\psi^{0} \in{\ell^2_{\mathrm{per}}}$, consider the unique solution $\psi\in\mathrm{C}^1([0,T_{\mathrm{max}}(\psi^0)),\ell^2_{\mathrm{per}})$ of the problem ($\mathcal{P}$). Assume that the parameter $\delta>0$ and that the initial condition $\psi^0$ is such that $M(0)>0.$ Then, the finite collapse time $T_{\max}$ satisfies the following upper bounds:
\begin{align}
\label{eqTh1}
&(i) \quad T_{\max}\leq \frac{1}{2\gamma} \ln \left[1+\frac{\gamma}{\delta M(0)}\right]:= \widehat{T}_{\max}[\gamma,\delta,M(0)]\\[2ex] 
\label{eqTh2}
&\mbox{for} \quad \gamma\neq 0 \quad  \mbox{and} \quad \gamma > \gamma^*:=-\delta M(0),\\[2ex]
\label{eqTh3}
&(ii)\quad T_{\max}\leq \frac{1}{2\delta M(0)}:=\widehat{T}^{\,0}_{\max}[\delta, M(0)], \quad \mbox{for} \quad \gamma = 0.
\end{align}
\end{theorem}

\paragraph{Definition of a critical value on the linear gain/loss parameter, separating global existence from collapse.}
From the definition of the analytical upper bound of the blow-up time \eqref{eqTh1} 
\begin{eqnarray*}
\label{eqUB}
\widehat{T}_{\mathrm{max}}[\gamma,\delta,M(0)]=\frac{1}{2\gamma }\ln \left[ 1+\frac{\gamma }{\delta M\left( 0 \right)} \right],
\end{eqnarray*}	
we  may define a critical value of the linear gain/loss parameter as
\begin{eqnarray*}
\label{CRIT}
\gamma^{*}=-\delta M(0).
\end{eqnarray*}
We observe that
\begin{eqnarray}
\label{eqUB2}	
\lim_{\gamma\rightarrow\gamma^*}\widehat{T}_{\mathrm{max}}[\gamma,\delta,M(0)]=+\infty,
\end{eqnarray}
while $\widehat{T}_{\mathrm{max}}[\gamma,\delta,M(0)]$ is finite if
\begin{eqnarray*}
\label{eqUB2a}
\gamma>\gamma^*,
\end{eqnarray*}
according to condition \eqref{eqTh1}. Then, Eq.~\eqref{eqUB2} suggests that when $\delta>0$, the critical %point
value $\gamma^*$ may act as a critical
point separating the two dynamical behaviors:  blow-up in finite-time for $\gamma>\gamma^*$ and global existence for $\gamma<\gamma^*$.  We shall investigate this  scenario numerically in the next section.

We also remark that the analytical upper bound for the blow-up time \eqref{eqTh3} in the $\gamma=0$ case,
\begin{eqnarray*}
\label{eqUB0}
\widehat{T}^{\,0}_{\mathrm{max}}[\delta,M(0)]=\frac{1}{2\delta M\left( 0 \right)},
\end{eqnarray*}
is actually the limit of the analytical upper bound \eqref{eqTh1} for $\gamma>0$ as $\gamma\rightarrow 0$, i.e.,
\begin{eqnarray*}
\label{eqUB01}	
\lim_{\gamma\rightarrow 0}\widehat{T}_{\mathrm{max}}[\gamma,\delta,M(0)]=\widehat{T}^{\,0}_{\mathrm{max}}[\delta,M(0)].
\end{eqnarray*}
% % % % % % % % % % % % % % % % % % % % % % % % % %
\begin{remark} 
We may observe the identicality of the functional form of the analytical upper bounds for the problem ($\mathcal{P}$), to the corresponding bounds derived for the continuous NLS counterparts incorporating gain and loss, when the latter are supplemented with periodic boundary conditions, \cite{anastassi2017,SD}. This is due to the similarity of the balance law (\ref{cl}), and the analogy of the  Cauchy--Schwarz inequality \eqref{DCSI} with its continuous counterpart for integrals on finite intervals. However, it should be recalled that inequality  \eqref{DCSI} is only valid for the finite-dimensional cases of the system,  as the one defined by the periodic boundary conditions of the problem  ($\mathcal{P}$). 
\end{remark}
\subsubsection{Lower bound for the collapse time for Dirichlet or vanishing boundary conditions}\label{sec2B2}
In this paragraph, we derive a lower bound for the finite collapse time, which is valid for problem ($\mathcal{D}$). In both cases of boundary conditions, Dirichlet (\ref{vanc}), or vanishing (\ref{vancv}) the problem  ($\mathcal{D}$) may be considered in the standard infinite dimensional sequence space
\begin{eqnarray}
\label{lpinf}
{\ell}^p:=\left\{\psi=(\psi_n)_{n\in\mathbb{Z}}\in\mathbb{C}:\quad
\|\psi\|_{\ell^p}:=\left(\sum_{n\in\mathbb{Z}}|\psi_n|^p\right)^{\frac{1}{p}}<\infty\right\}.
\end{eqnarray}
Let us recall that the spaces $\ell^p$, satisfy the following elementary embedding relation  \cite[p. 35]{AdamsS}
\begin{eqnarray}
\label{lp1}
\ell^q\subset\ell^p,\quad \|u\|_{\ell^p}\leq \|u\|_{\ell^q}, \quad 1\leq q\leq p\leq\infty.
\end{eqnarray}	
Note that the inclusions (\ref{lp1}) are still valid in the case of a finite lattice supplemented with Dirichlet boundary conditions:  denoting the finite dimensional subspaces of $\ell^p$ associated to the boundary conditions (\ref{vanc}) by $\ell^p_0$, we just note that each element of $\ell^p_0$ can be extended to an element $\psi=(\psi_n)_{n\in\mathbb{Z}}\in \ell^p$, by setting $\psi_n=0$ for $n\geq N$ and $n\leq 0$.  Thus, for simplicity, the proof will be presented only for the case of Dirichlet boundary conditions. We will consider again the corresponding functional $M(t)$ \eqref{eq06}, however, the standard norm (power) 
\begin{equation}P(t)=\sum_{n=1}^{N-1}|\psi_n(t)|^2,
\label{p}\end{equation} 
will also come into play. 
\begin{theorem}\label{The2}
	For $\psi^{0} \in\ell^2_0$, consider the unique solution $\psi\in\mathrm{C}^1([0,T_{\mathrm{max}}(\psi^0)),\ell^2_0)$ of the problem ($\mathcal{D}$). Assume that the parameter $\delta>0$ and that the initial condition $\psi^0$ is such that $P(0)>0.$ Then, the finite collapse time $T_{\max}$ satisfies the following lower bounds
	\begin{align}
	\label{eqAlt1}
	&(i) \quad T_{\max}\ge \frac{1}{2\gamma} \ln \left[1+\frac{\gamma}{\delta P(0)}\right]:= \overline{T}_{\max}[\gamma,\delta,P(0)]\\[2ex] 
	\label{eqAlt2}
	&\mbox{for} \quad \gamma\neq 0 \quad  \mbox{and} \quad \gamma > \gamma^*:=-\delta P(0),\\[2ex]
	\label{eqAlt3}
	&(ii)\quad T_{\max}\ge \frac{1}{2\delta P(0)}:=\overline{T}^{\, 0}_{\max}[\delta, P(0)], \quad \mbox{for} \quad \gamma = 0.
	\end{align}
\end{theorem}
%%%%%%%%%%%%%%%%%%%%%%%%%%%%%%%%%%%%%%%%%%%%%%
\begin{remark} 
We may observe that  in the case of the problem ($\mathcal{D}$), the application of the reverse inequality (\ref{lp1}) has lead   to the reverse differential inequality (modulo the constant $N$) of  \eqref{help1R}. This explains why both the upper and the lower bounds share the same functional form. 
\end{remark}
%%%%%%%%%%%%%%%%%%%%%%%%%%%%%%%%%%%%%%%%%%%%%%

\section{Numerical results}\label{sec3}
%In this section, we perform a numerical investigation in order to study the collapse/decay dynamics for three different types of initial conditions: (a) spatially extended initial conditions, (b) $\sech$-profiled initial conditions, resembling a ``discrete bright soliton''  and (c) $\tanh^2$-profiled initial conditions, resembling a ``discrete dark soliton''. In all the above cases, we compare the results for both the focusing  ($s=-1$)  and the defocusing ($s=1$) DNLS equation (\ref{eq01}).  The numerical integration is based on a fourth-order Runge-Kutta scheme,  incorporating an embedded error estimator that makes it possible to efficiently determine appropriate step sizes, \cite{sulem,Cho}. We denote as time of collapse the time where the numerical scheme detects a singularity in the dynamics. The numerical singularity is detected when the time step is becoming of order $<10^{-12}$.

In this section, we perform a numerical investigation in order to study the collapse dynamics of the solutions of \eqref{eq01} for four different types of initial conditions: (a) spatially extended, plane-wave-profiled, initial conditions, (b) $\sech$-profiled initial conditions, resembling a ``discrete bright soliton'', (c) $\tanh^2$-profiled initial conditions, resembling a ``discrete dark soliton''  and (d) initial conditions of the form of a rectangular box. In addition, we discuss briefly the decay properties of the examined system which are suggested by our analytic results. In all the above cases, we compare the results for both the focusing  ($s=-1$)  and the defocusing ($s=1$) DNLS equation (\ref{eq01}). The numerical integration, of the system under investigation, is based on a fourth-order Runge-Kutta scheme, with variable step size $\Delta t\propto\min\left[\Delta t_0,1/\max |\psi_n|^2\right]$, where $\Delta t_0$ is an appropriate initial integration step (see e.g.~\cite{Cho}).Then, the blow-up time is considered the one in which the value of $\Delta t$ becomes smaller, or equivalently, the value of $\max|\psi_n|^2$ exceeds a proper threshold. 

Concerning the implementation of the Dirichlet boundary conditions for case (b) of the initial data, let us also recall that these conditions are strictly satisfied only asymptotically, as $L\rightarrow\infty$; %and that 
However, for a finite $L$, the induced error has
negligible effects in the observed dynamics, as it is  of order $\exp(-L)$, and the smallest value for $L$ used herein is $L=50$.

%We denote as time of collapse the time where the numerical scheme detects a singularity in the dynamics. 

%%%%%%%%%%%%%%%%%%%%%%%%%%%%%%%%%%%%%%%%%%%%%%%%
\subsection{Spatially extended (plane-wave profiled) initial conditions.}
\label{sec3A}
\setcounter{equation}{0}
The first example of the numerical study considers spatially extended initial data of the form:
\begin{equation}\label{plane_wave}
\psi_n(0) = Ae^ {-\rmi\frac{\pi K x_n}{L}}, 
\end{equation}
where  $A > 0$ is the amplitude, $\widetilde{K}:=\pi K/L, K\in \mathbb{N}$ is the wavenumber and the discrete spatial coordinate  $x_n$ is given by
\begin{equation}\label{spatial_cor}
	x_n=-L+nh, \quad n= 0,1,2,\ldots,N.
\end{equation}

We may show that the analytical upper estimate (\ref{eqTh1})-(\ref{eqTh2}) on the collapse time $\widehat{T}_{\mathrm{max}}$ is sharp for the initial data (\ref{plane_wave}), as it is also in its continuous counterparts \cite{anastassi2017, SD}. In fact, it can be verified that the evolution of either discrete or continuous plane-waves is governed by the same ODE dynamics, as we may effectively transfer the arguments of \cite{anastassi2017, SD}, from the continuous, to the discrete set-up: substituting in \eqref{eq01} the ansatz $\psi_n = W(t)e^{\rmi\widetilde{K}x_n},$ we find that $W$ satisfies the equation
\begin{equation}\label{eqW}
\rmi\dot{W} - skCW +|W|^2 W = \rmi\gamma W +\rmi\delta |W|^2 W,
\end{equation}
where $C = -4\sin^2(\widetilde{K}/2)$. %$C=2(\cos(K)-1)$. 
Introducing now the phase factor $W(t)=e^{\rmi\phi t}w(t)$ with $\phi=-skC$, which absorbs the second term in \eqref{eqW}, we derive the equation 
\begin{equation*}
\rmi\dot{w}+|w|^2 w = \rmi\gamma w + \rmi\delta |w|^2 w.
\end{equation*}
%%%%%%%%%%%%%%%%%5
\begin{figure}[tbh!]
	\centering 
	\begin{tabular}{ccc}
	(a)& &(b)\\
	\includegraphics[scale=0.4]{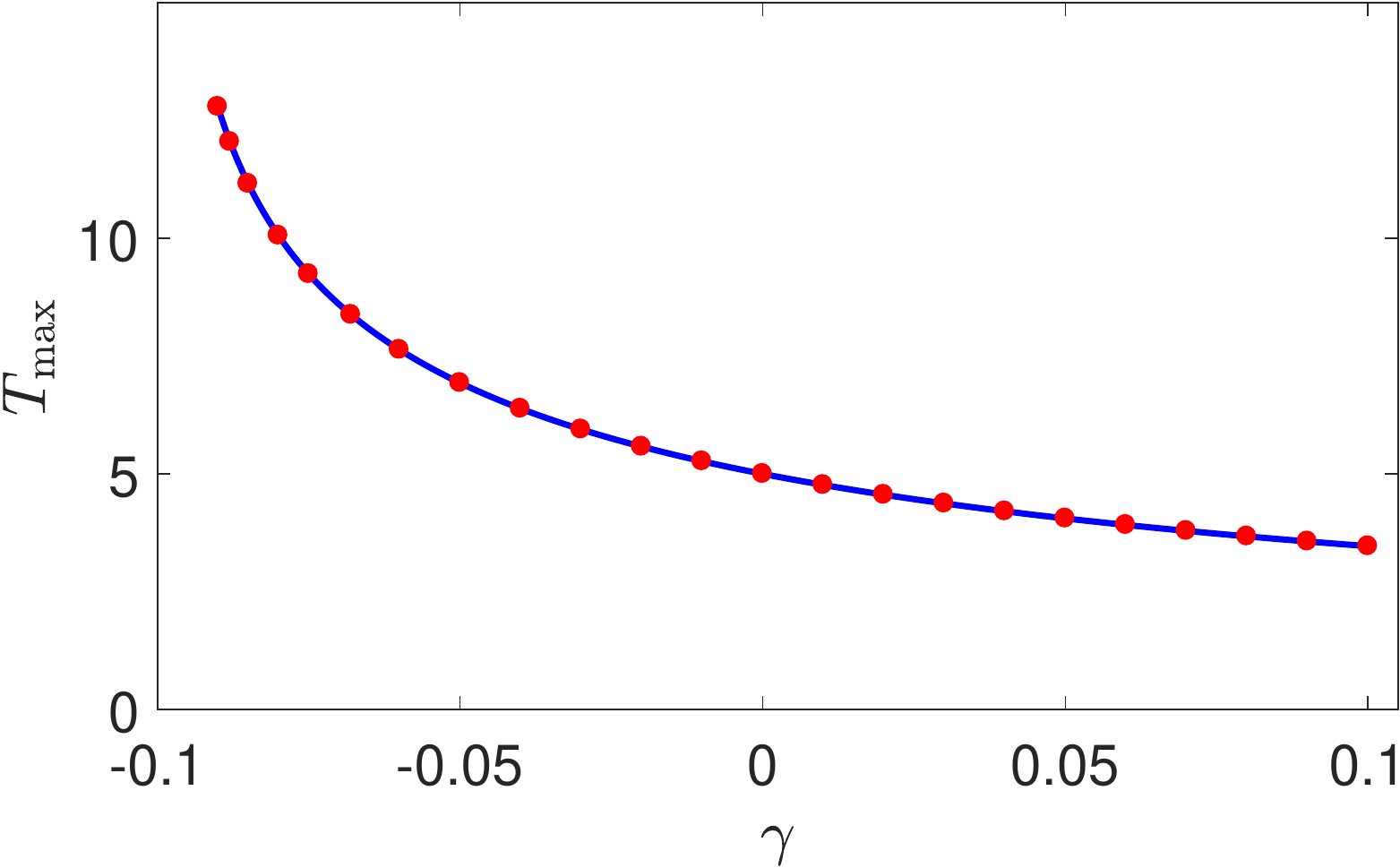}
	& &
	\includegraphics[scale=0.4]{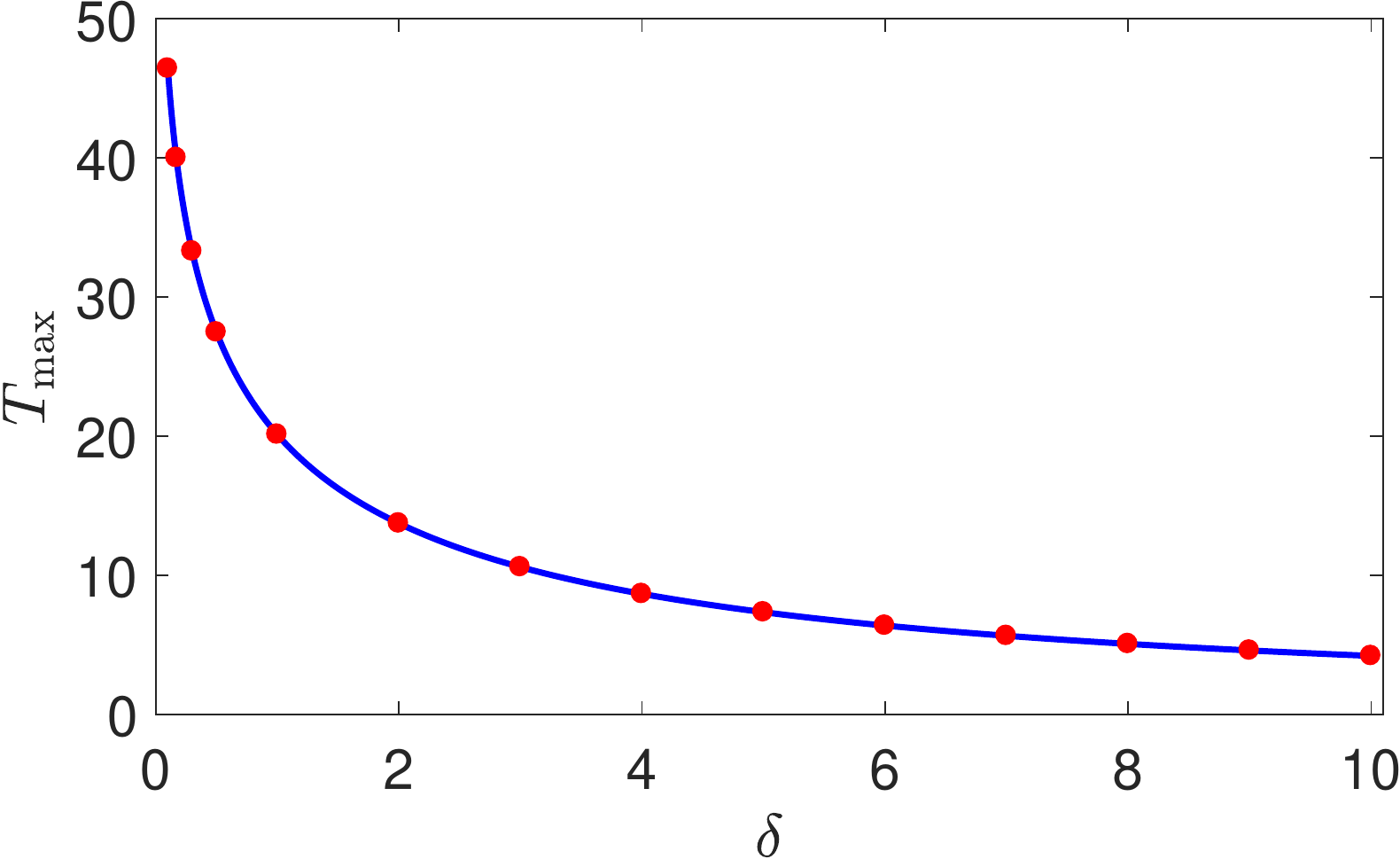}
	\\[5pt]
	(c)& &(d)\\
	\includegraphics[scale=0.4]{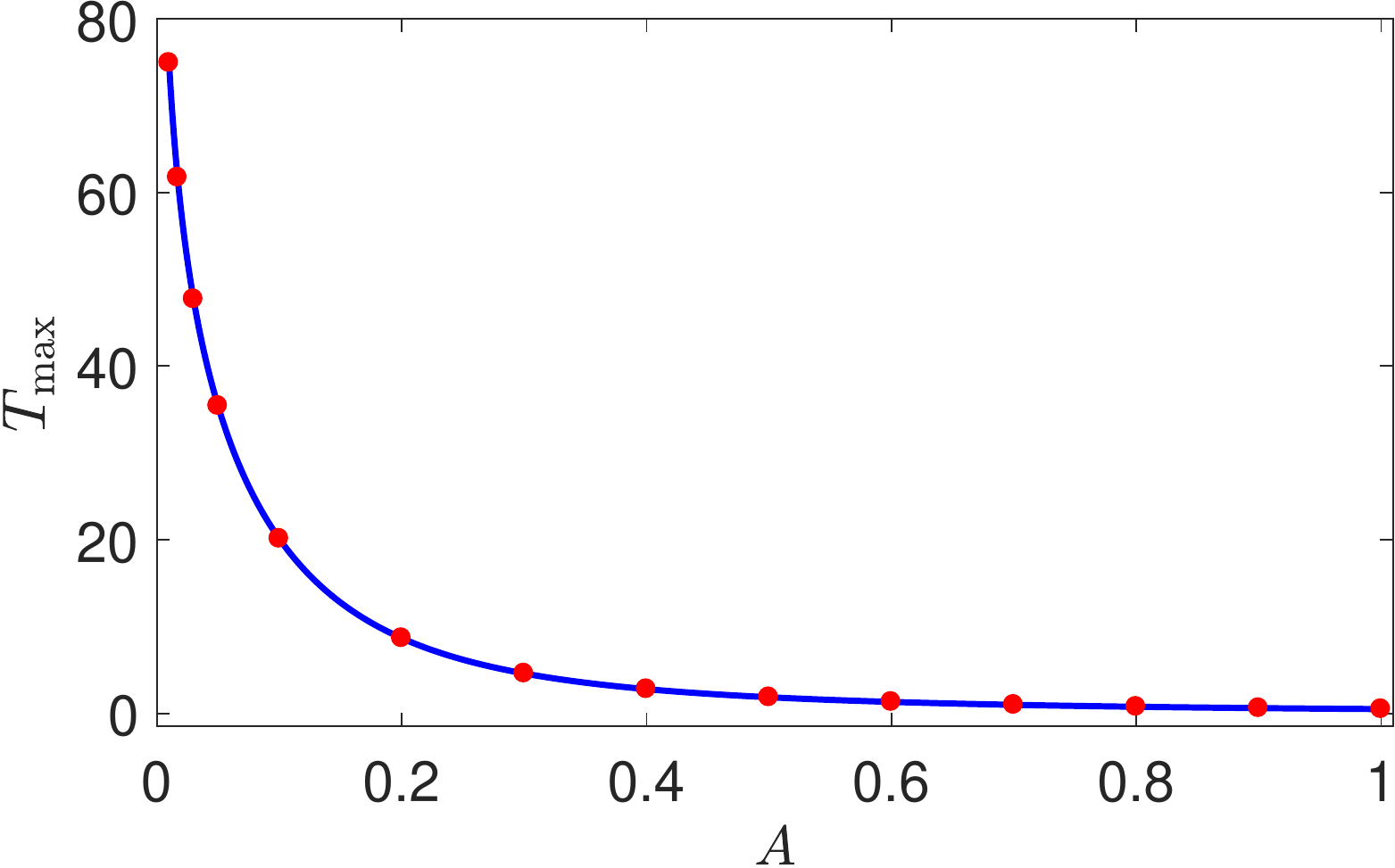}
	& &
	\includegraphics[scale=0.4]{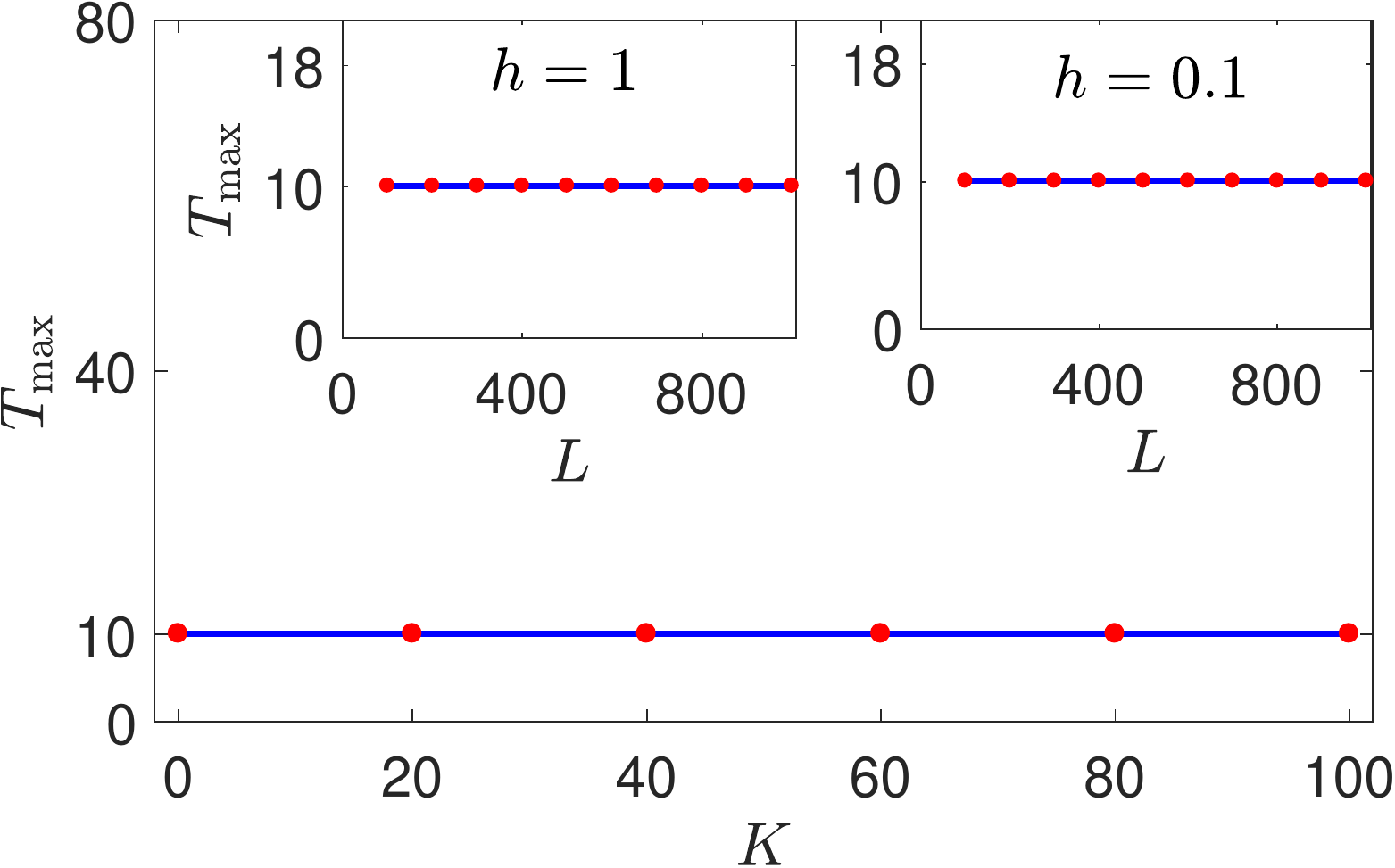}  
	\end{tabular}
	\caption{Comparison of the analytical upper bound $\widehat{T}_\text{max}$, which is depicted by a continuous (blue) curve against the numerically calculated blow-up times, shown as (red) dots, in the case of the spatially extended initial condition \eqref{plane_wave}, and the defocusing case $s=1$.  Panel (a): Comparison with respect to $\gamma\in[-0.09,0.1]$. Parameters: $\delta=0.1, A=1, L=100, N=100$ and $K=10$.  Panel (b): Comparison with respect to $\delta\in[0.1,10]$. Here, $\gamma=0.04, A=0.1$ and the rest parameters are fixed as before. Panel (c): Comparison with respect to $A\in[0.01,1]$ for $\gamma=0.04, \delta=1$ and the rest parameters fixed as in panel (a). Panel (d): The main panel,  shows the independence of $K$. The insets show the independence of the half-length $L$ for $K=10$; in the left inset $h=1$ and in the right one, $h=0.1$.  Other parameters are: $\gamma=-0.08, \delta=0.1, A=1$ and $L=100$.} 
	\label{Fig1}
\end{figure}
%%%%%%%%%%%%%%%%%%%%%%%%%%5
Finally, by substituting the polar decomposition $w(t) = f(t)e^{\rmi\theta(t)}$, we get the Bernoulli-type  ODE
\begin{equation}\label{eqbern}
\dot{f}=\gamma f+\delta f^3
\end{equation}
and $\dot{\theta}=f^2$ and $f:\mathbb{R}\rightarrow\mathbb{R}$. The ODE \eqref{eqbern} supplemented with the initial condition $f(0)=f_0$, has the unique solution 
\begin{equation}\label{eqf}
f^2(t) = \frac{\gamma f_{0}^{2} e^{2\gamma t}}{\gamma +\delta f_{0}^{2} - \delta f_{0}^{2} e^{2\gamma t}}.
\end{equation}
The solution \eqref{eqf} blows-up in finite time
\begin{equation}
\label{exact}
\widetilde{T}_{\max}=\frac{1}{2\gamma}\ln\left[1+\frac{\gamma}{\delta f^2_{0}}\right],
\end{equation}
under the same conditions on the parameters $\gamma$ and $\delta$ as stated in Theorem \ref{The2a}. For the solution  \eqref{eqf}, the critical value of $\gamma$ is  $\gamma^*=-\delta f_0^2$.  On the other hand, we observe that for the initial condition \eqref{plane_wave} it is
\begin{equation*}
M(0)=\frac{1}{N}\sum_{n=0}^{N-1}A^2 \left|e^{\frac{-\rmi K\pi n}{L}}\right|^2 = A^2,
\end{equation*}
while, according to \eqref{eqTh1}, the upper bound of the blow-up time is
\begin{equation}\label{pwupbound}
\widehat{T}_{\max}[\gamma, \delta, A]=\frac{1}{2\gamma} \ln\left[1+\frac{\gamma}{\delta A^2} \right],
\end{equation}
with critical value $\gamma^* = -\delta A^2$. Thus, the exact blow-up time $\widetilde{T}_{\max}$ (\ref{exact}) and the analytical upper estimate $\widehat{T}_{\max}$ (\ref{pwupbound}) coincide. Also, according to \eqref{eqTh3}, when $\gamma=0$, it is
\begin{equation*}\label{upbound0}
\widehat{T}^{\,0}_{\max}[\delta,A]=\frac{1}{2\delta A^2}. 
\end{equation*}
\begin{remark}\emph{
In the case of the spatially extended initial data \eqref{plane_wave}, the collapse is spatially homogeneous and is manifested by an increase of the amplitude of the discrete plane-wave $f(t)$. In general, the {\it extended or weak collapse} scenario is defined when the blow-up occurs through a simultaneous increase of the amplitude of all (or a vast majority of) the oscillators of the lattice. The collapse dynamics of the initial data \eqref{plane_wave} discussed above is an ideal manifestation of this scenario.}
\end{remark}
The sharpness of the analytical upper bound $\widehat{T}_\text{max}$ (\ref{pwupbound}) is numerically verified. In Figure~\ref{Fig1}, we compare it against the numerically calculated blow-up time, both considered as functions of various parameters. The illustrated results refer to the defocusing case $s=1$, for $N=100$. 
%%We experimented with larger numbers of oscillators as well, e.g, $N=201, 301, 1001$ with no deviation in the results.    
{}%Furthermore, we remark that in the focusing case analogous results were obtained, hence no figures were supplemented in this case.
In panel (a), we compare the analytical upper bound against the numerical blow-up times as functions of $\gamma$. The rest of parameters are fixed as $\delta=0.1, A=1, L=100$  and $K=10$. We observe that $\widehat{T}_\text{max}$ is a very accurate estimate for the numerical collapse time. This accuracy was also observed in the case $\gamma = 0$, for which the analytical upper bound coincide with the numerical blow-up time at the value  $\widehat{T}^0_{\max}[0.1,1] = T_{\mathrm{num}}=5$.

This remarkable agreement of $\widehat{T}_\text{max}$ with the numerical collapse times is also observed when the comparison between them is performed with respect to the rest of the parameters. Panel (b), portrays  the comparison with respect to $\delta$, for fixed linear gain strength $\gamma=0.04, A = 0.1$, and the rest of parameters fixed as above. Indeed, panel (c), depicts the comparison with respect to the amplitude $A$ of the initial condition (\ref{plane_wave}); the rest of parameters are fixed as $\gamma=0.04, \delta=1, L=100$ and $K=10$.
%%%%%%%%%%%%%%%%%%%%%%%%%%%%%%%%%%%%%5
%%%%%%%%%%%%%%%%%%%%%%%%%%%%%%%%%%%%%%
\begin{figure}[tbh!]
	\centering 
\includegraphics[scale=0.43]{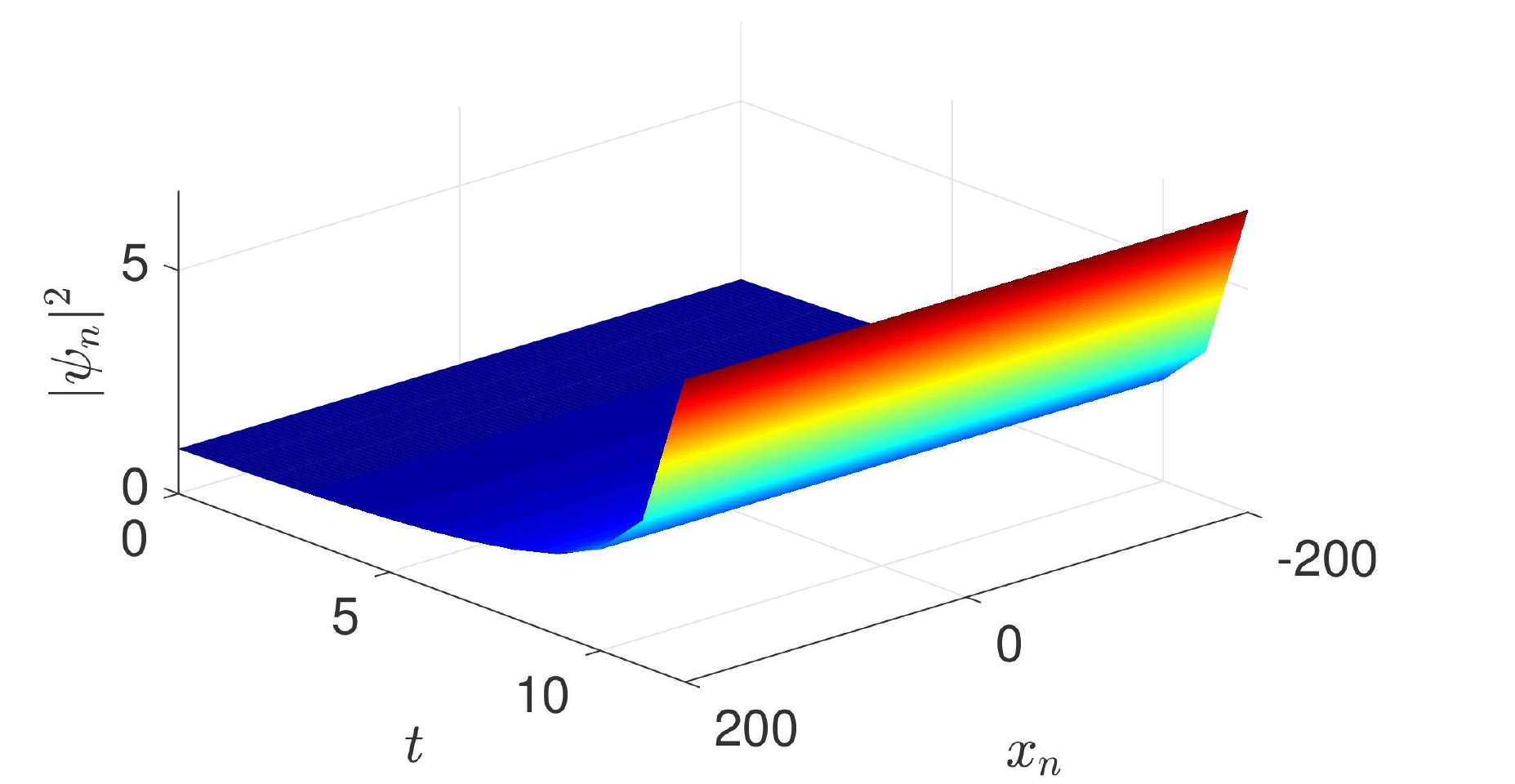}
\quad
\includegraphics[scale=0.43]{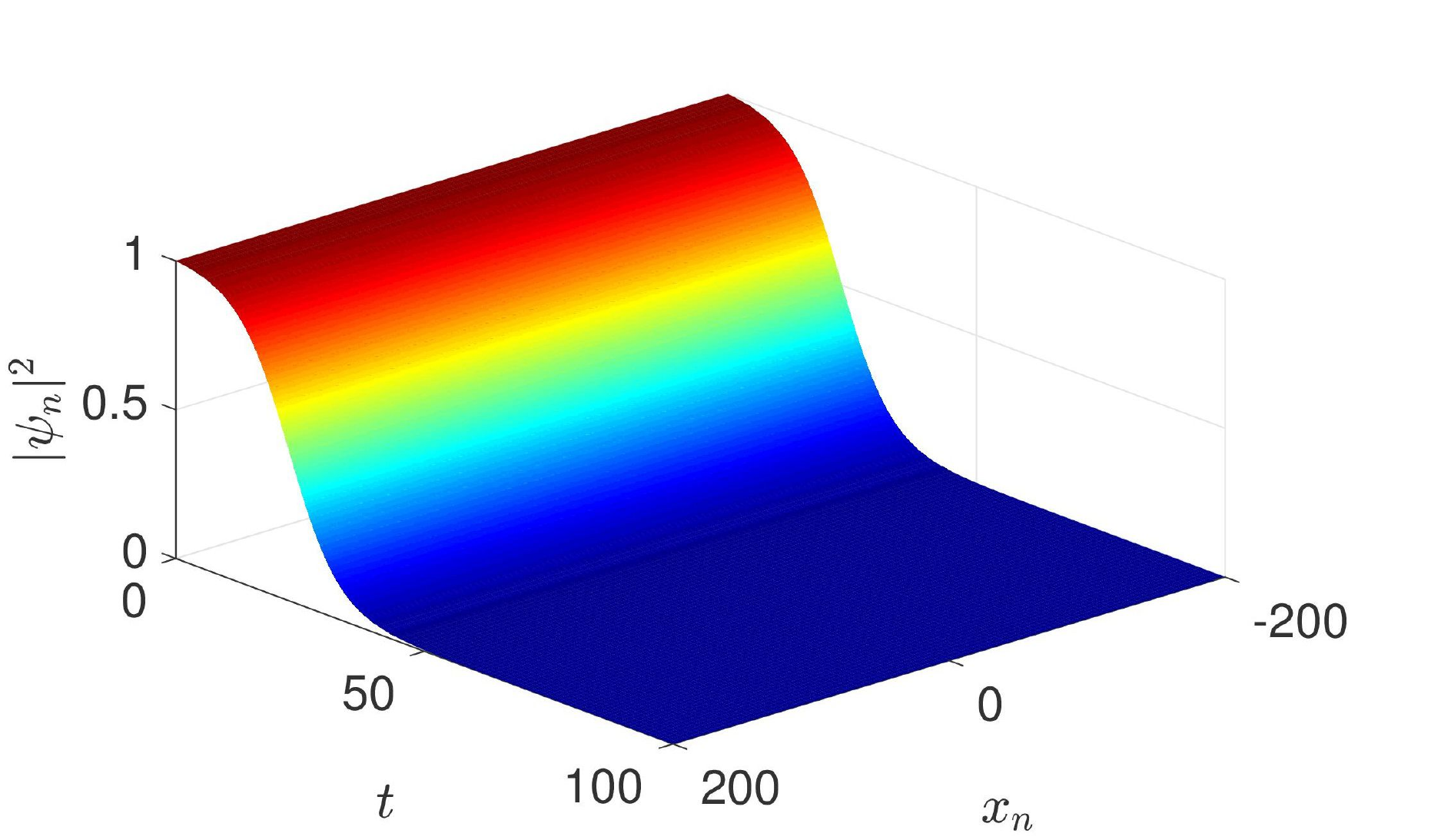}
%\vspace{0cm}
	\caption{Left panel: Collapse of the density $|\psi_n|^2$ for the spatially extended initial condition \eqref{plane_wave}, for $\delta=0.1, A=1, L=200, K=10, N=100$ and $\gamma=-0.09$. Right panel: Decay of the density $|\psi_n|^2$ for the spatially extended initial condition \eqref{plane_wave} for the same parameters as before, but for $\gamma=-0.11$. Both panels consider the defocusing case $s=1$.}
	\label{Fig2}
\end{figure}
%%%%%%%%%%%%%%%%%%%%%%%%%%%%%%%%%%%%%%%%%%%%%%%
From \eqref{pwupbound}, we notice that the analytical upper bound  is independent of the wavenumber $K$ and the half-length $L$. This interesting feature is verified  for the numerical blow-up times as well, and it is shown in panel (d) of Figure~\ref{Fig1}.
The loss/gain parameters are  $\gamma=-0.08, \delta=0.1$, for a half-length $L=100$, $K=10$ and initial amplitude $A=1$.  The insets depict the numerical results for two choices of the spacing parameter, $h = 1$ (left inset) and $h = 0.1$ (right inset), recalling that the number of oscillators is calculated as $N = 2L/h$. 

Next, we note that the considered linear loss strength $\gamma=-0.08>\gamma^*=-\delta A^2=-0.1$, finely satisfies the parametric condition for the finiteness of the collapse time. 
The  parametric separatrix $\gamma^*=-\delta A^2$ was numerically verified to be also sharp, in  distinguishing between finite time collapse (for $\gamma>\gamma^*$) and global existence dynamics (for $\gamma<\gamma^*$). Figure~\ref{Fig2} illustrates the evolution of the density $|\psi_n|^2$ for the above initial conditions, for two distinct values of linear gain/loss strength $\gamma$, and the defocusing variant of the model ($s=1$). The left panel portrays the evolution of the density for $\gamma = -0.09 > \gamma^*$, satisfying the parametric condition for finite time collapse at $T_{\mathrm{num}}=12$ according to the ODE-type of the collapse dynamics governed by (\ref{eqbern}), and consists a representative example of the extended blow-up scenario.  The right panel depicts the evolution of the density for $\gamma = -0.11 < \gamma^*$, where global existence in time is expected. We observe that global existence is actually associated with the decay of the density.  The portrayed decay dynamics are in accordance with the decay of the solution (\ref{eqf}), which occurs when $\gamma<-\delta f_0^2<0$. 

We conclude the presentation of the numerical results for the spatially extended initial data, by noting that exactly the same results (not shown here) were verified for the focusing $s=-1$  case of the model. 
\begin{remark}\emph{
Generically,  discrete plane wave solutions are prone to modulational instability (MI) effects. Here, we are interested in the instability leading to finite time blow-up, and according to the analytical results of Theorem \ref{The2a} (and as illustrated numerically), this is driven by the gain/loss effects, in the parametric regime $\gamma>\gamma^*$ and $\delta>0$, and not by any potential MI effects. On the other hand, the second parametric regime we are interested herein, namely $\delta>0$, $\gamma<\gamma^*$, is associated with global existence and energy decay, and yet MI effects are irrelevant to this type of dynamics. MI effects can be potentially relevant in the regime $\gamma>0$, $\delta<0$ which may be associated with global existence but not necessarily decay of solutions (see \cite{anastassi2017}  for the continuous counterpart).  However, the study of this regime and effects induced by MI, are beyond the scope of the present work.}
\end{remark}
%%%%%%%%%%%%%%%%%%%%%%%%%%%%%%%%%%%%%%%%%%%%%%%
\subsection{Vanishing (\texorpdfstring{$\sech$}--profiled) initial conditions.}\label{sec3B}
%\paragraph{Numerical study on the analytical upper bound for the periodic boundary conditions.}
The second example of our numerical investigation  concerns $\sech$-profiled initial data of the form
\begin{equation}\label{sech}
\psi_n(0) = A \sech x_n,
\end{equation}
where $A>0$ is the amplitude and $x_n$ the discrete spatial coordinate \eqref{spatial_cor}. This initial condition resembles the profile of a ``discrete bright-soliton'' as an initial state. 

The numerical study considers both problems ($\mathcal{P}$) and ($\mathcal{D}$). Thus, the numerical blow-up times which are acquired by using both periodic and Dirichlet boundary conditions will be compared with the analytical upper bounds $\widehat{T}_\text{max}$ (\ref{eqTh1})-(\ref{eqTh3}) and lower bounds $\overline{T}_\text{max}$ (\ref{eqAlt1})-(\ref{eqAlt3}).

\begin{figure}[tbh!]
	\centering 
	\begin{tabular}{cc}
	\large(a)&\large(b)\\
	\includegraphics[scale=0.5]{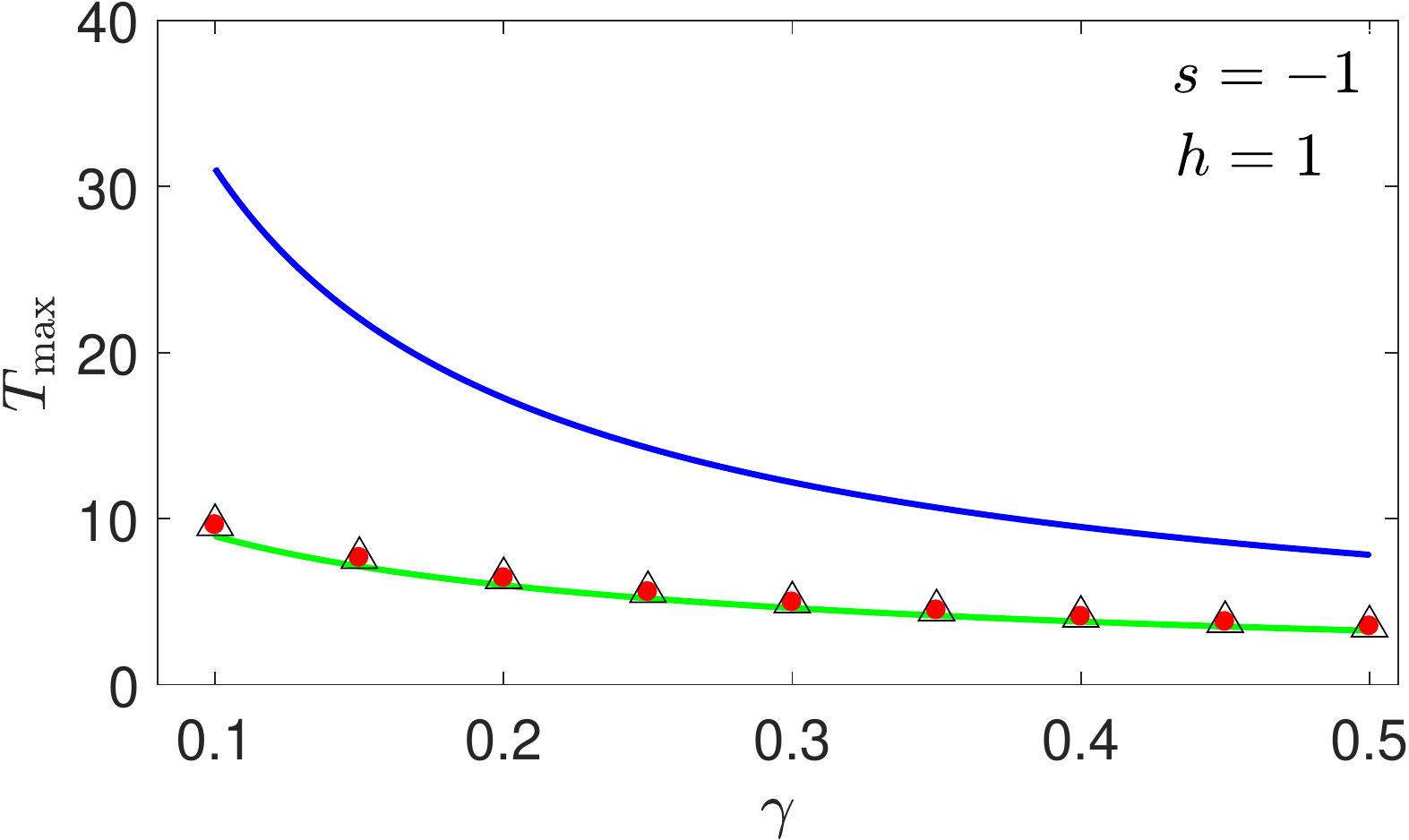}&
	\includegraphics[scale=0.5]{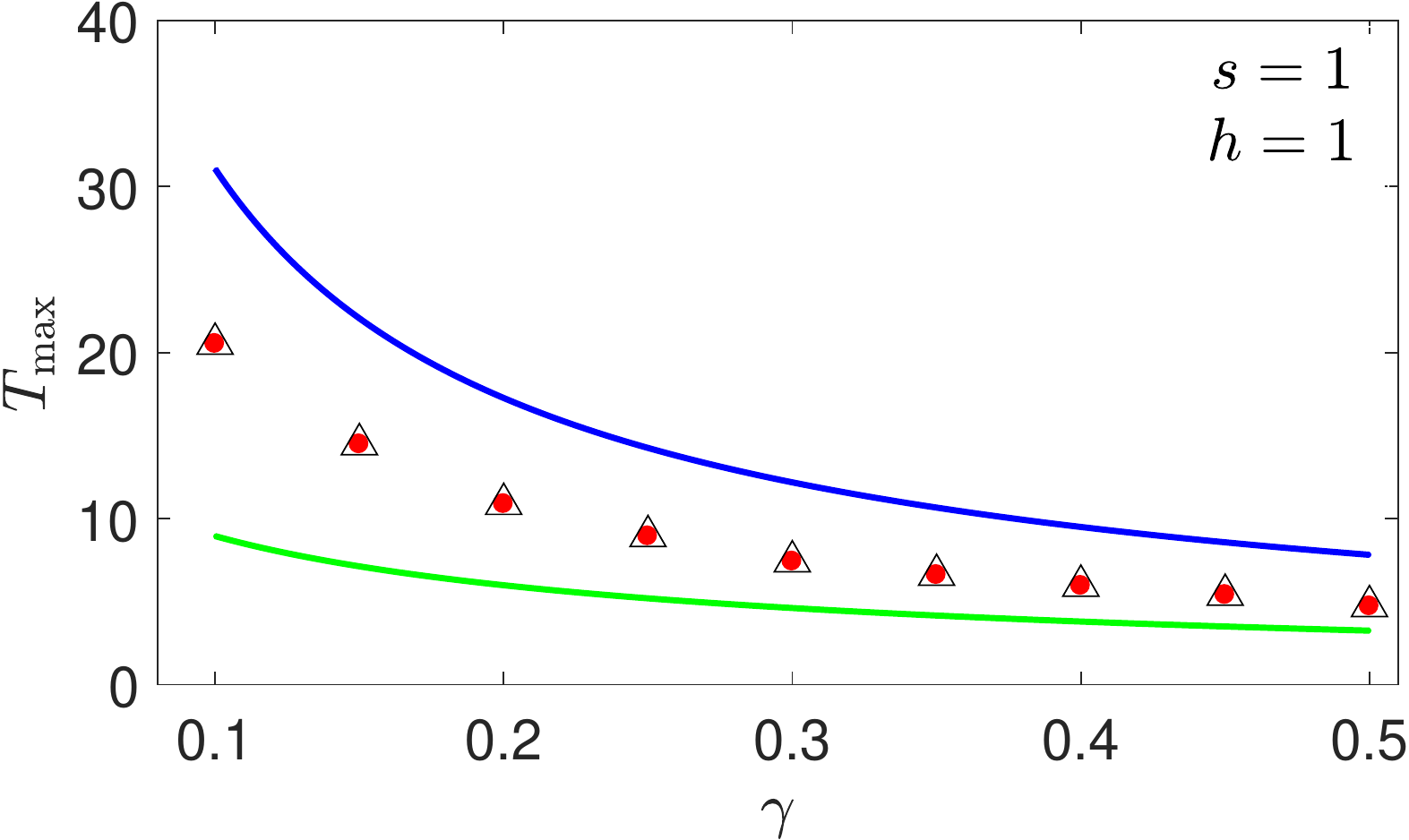}
	\end{tabular}	
	\caption{Numerical blow-up times [(red) dots for the problem ($\mathcal{P}$)-triangles for the problem ($\mathcal{D})$], for the $\sech$-profiled initial condition \eqref{sech},  against the analytical upper  bound $\Th$ [upper (blue) curve] and lower bound $\To$ [lower (green) curve], as functions of $\gamma\in[0.1,0.5]$.  Panel (a) for the focusing case $s=-1$ and panel (b) for the defocusing case $s=1$. Other parameters: $\delta=0.01, A=1, L=50$ and $h=1$ ($N=100$).}
	\label{Fig3}
\end{figure}

In Figure \ref{Fig3}, the results of the comparison with respect to the parameter $\gamma \in [0.1, 0.5]$  are depicted. Panel (a) shows the results for the focusing case $s=-1$, and panel (b) shows the results for the defocusing case $s=1$. The other parameters are $\delta = 0.01, A = 1, L = 50$ and $h=1$ ($N = 100$). We observe that for both choices for $s$, the numerical collapse times of the numerical solutions of problems ($\mathcal{P}$) [(red) dots] and ($\mathcal{D}$) [triangles] practically coincide.  This coincidence justifies that, in this parameter region, the boundary conditions do not affect the dynamics in accordance to the introductory remarks of Section~\ref{sec3}. Thus, although the two analytical bounds have been extracted for different boundary conditions, in practice, for the numerical investigation they can be used together as a lower and an upper bound for both problems ($\mathcal{P}$) and ($\mathcal{D}$). %differs form ints analytical upper bound systematically by about $30$\% in the defocusing case and $50$\% in the focusing. 
Figure~\ref{Fig3}, justifies  that all the numerically obtained blow-up times lie between the two analytical estimates, validating the analytical considerations of Section~\ref{sec2}. Although this is not shown here, the same numerical justification is observed when the comparison is made with respect to the parameter $\delta$.

\begin{figure}[tbh!]
	\centering 
\includegraphics[scale=0.32]{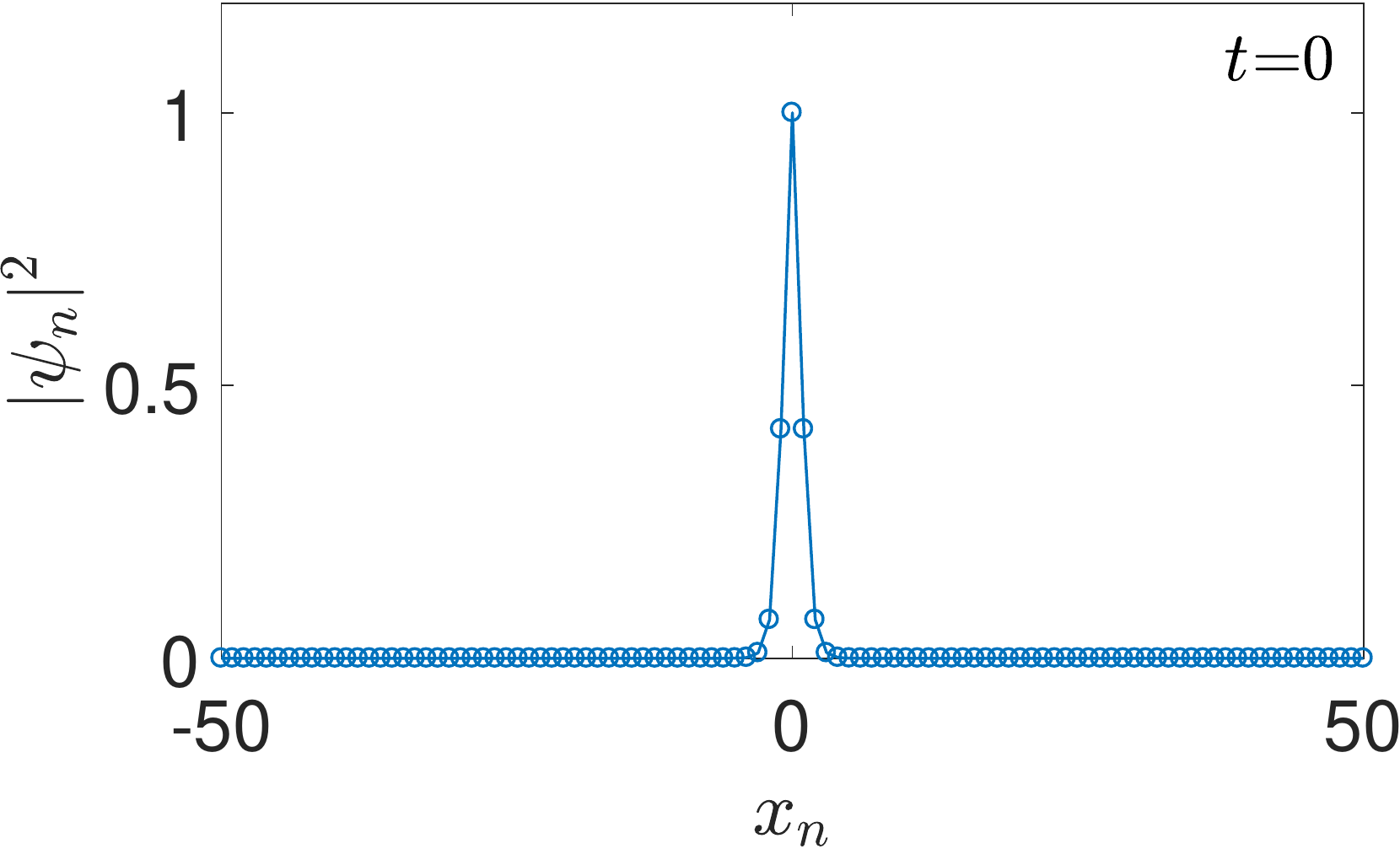}
\hspace{0.2cm}
\includegraphics[scale=0.32]{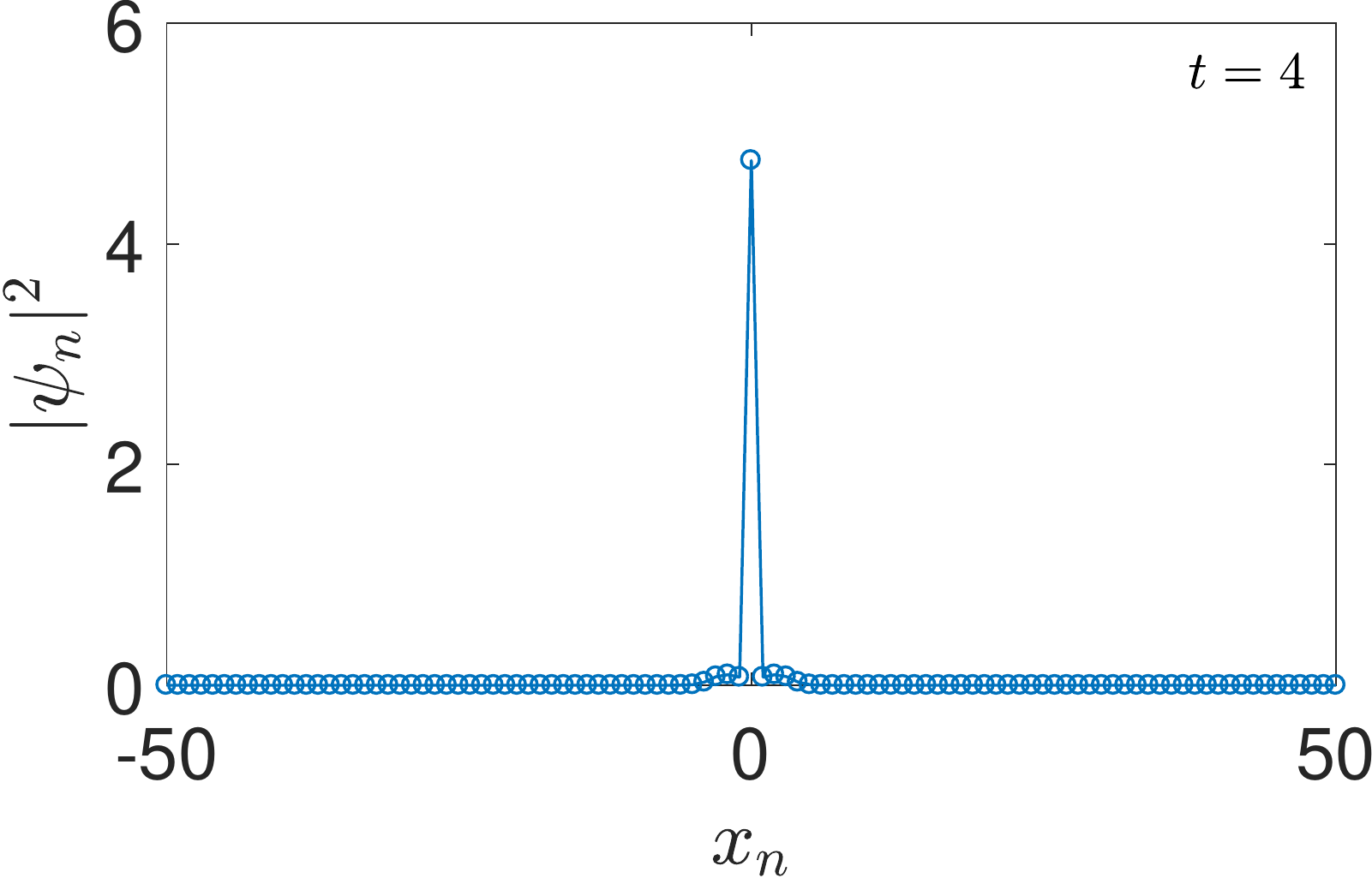}
\hspace{0.2cm}
\includegraphics[scale=0.32]{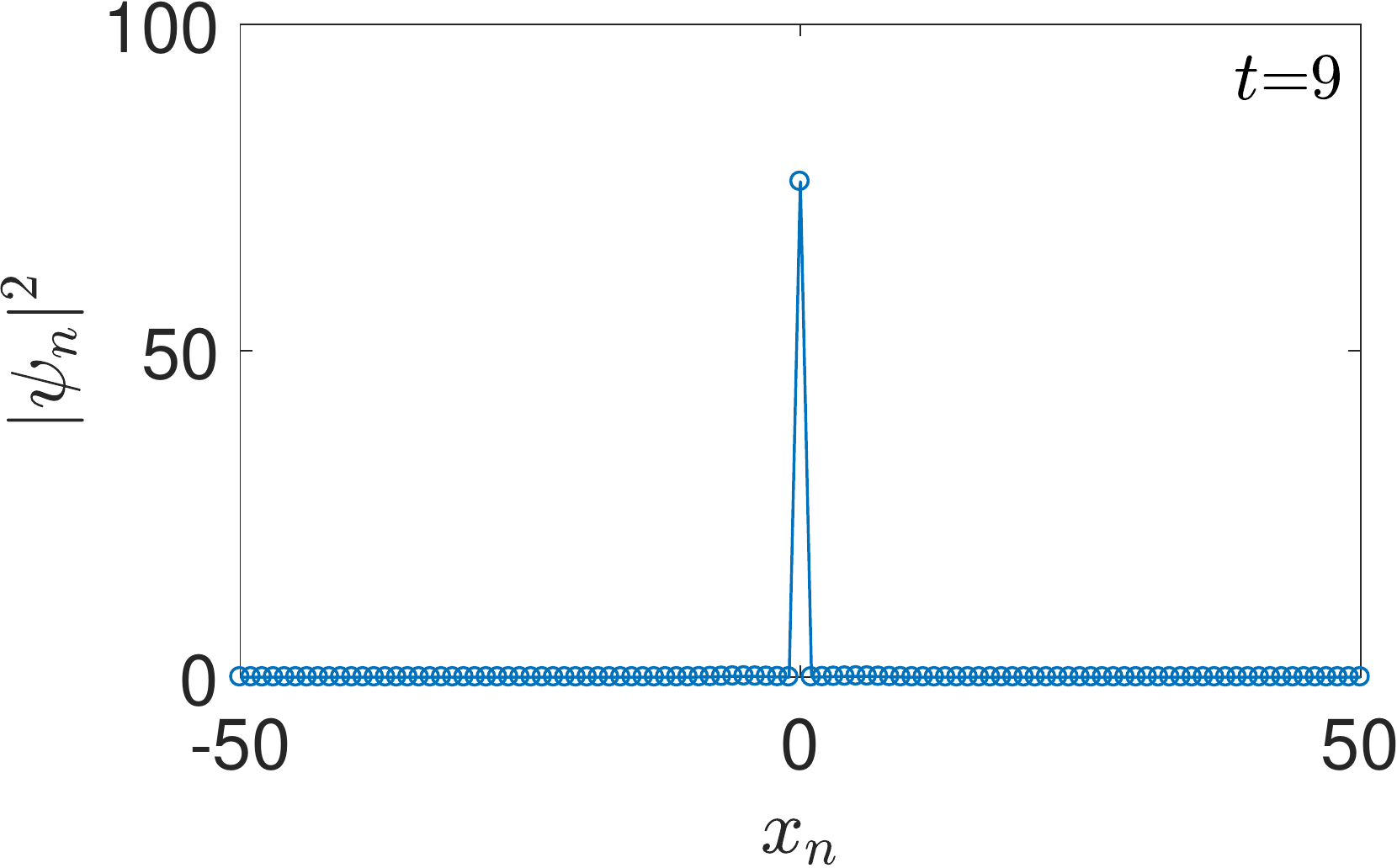}

	\caption{Snapshots of the time-evolution of the density $|\psi_n|^2$ for the focusing case $s=-1$, emerged from the initial condition \eqref{sech} with amplitude $A=1$. Parameters: $\gamma=0.1$,  $\delta=0.01, L=50, N=100$ ($h=1$). }
	\label{blowfoc}
\end{figure}

Next, we analyze further the results portrayed in panel (a) of Figure \ref{Fig3}, where the focusing case $s=-1$ is considered. We observe  that the numerical blow-up times lie far from the upper bound $\widehat{T}_\text{max}$ \eqref{eqTh1}, and are in excellent agreement with the analytical lower bound $\overline{T}_\text{max}$ \eqref{eqAlt1}. This  behavior of the numerical blow-up times suggests a dynamics different from the extended type of collapse discussed in Section~\ref{sec3A}. Indeed, this different type of collapse dynamics is shown in Figure~\ref{blowfoc}, which depicts snapshots of the time-evolution of the density $|\psi_n|^2$, for one of the solutions shown also in Figure~\ref{Fig3}(a) and more specifically the one with $\gamma=0.1$.  We observe the absence of energy dispersion due to the focusing effect, enhanced also by discreteness. As a result, the evolution of the initial condition is self-similar  and the lattice  blows-up directly from the central cite, in which the energy is localized. Consequently, the numerical blow-up times are much smaller than the ones observed for the extended initial conditions. Effectively, they are captured by the analytical lower-bound derived for the localized initial data and solutions of problem  ($\mathcal{D}$). 

\begin{remark} \emph{In the case of localized initial data (\ref{sech}) and the aforementioned parametric region, the transient dynamics prior collapse is self-similar and manifested by an increase of the amplitude of the central site. The {\it localized or strong collapse} scenario is defined when the blow-up occurs by just one site without any (or with very small) prior energy dispersion. The collapse dynamics of the initial data \eqref{sech} discussed above is an ideal manifestation of localized blow-up.}  
\end{remark}

%Since we consider localized initial data, the solution will be also spatially localized by continuity, at least for a finite interval $[0,\tau]$ prior to collapse. In this case, for the height of the wave-background we have that $f^2(t)=0=\min_{n\in [0,N-1]}|\psi_n (t)|^2$ which is the minimum of the density of the solution. Moreover, we have the inequality
%\begin{align}
%\label{colcom2}
%e^{-2\gamma t}f^2(t)&< M(t) = \frac{e^{-2\gamma t}}{N} \sum_{n=0}^{N-1}|\psi_n|^2\nonumber\\
%&< \frac{e^{-2\gamma t}}{N}\max_{n\in [0,N-1]}|\psi_n (t)|^2\;N\nonumber \\
%&= e^{-2\gamma t}\max_{n\in [0,N-1]}|\psi_n (t)|^2
%\end{align}
%satisfied by the amplitude of the density $\max_{n\in [0,N-1]}|\psi_n (t)|^2$ of the localized solution, for all $\;\;t\in[0,\tau]$. Thus, in the case of spatially localized data, we expect that the collapse will be predominately manifested as the increase of the amplitude, which grows faster than the averaged norm $M(t)$, due to (\ref{colcom2}). Consequently, the amplitude may be
%expected to blow-up at an earlier time than $M(t)$, which causes the difference between the numerical blow-up times and the analytical upper bound. {\bf VK: exw diafora pragmata pou den katalabainw sto parapanw}.
%Some more light, at the behavior of the system in the defocusing case, in this parameter region, can be shed by examining the transient evolution of the initial condition (\ref{sech}) prior to collapse.  
%Such an examination may also explain the higher discrepancy observed in the focusing case ($s=-1$).  

\begin{figure}[tbh!]
	\centering 
	\begin{tabular}{ccc}
	\qquad(a)&\qquad\ (b)&\qquad\quad(c)\\
\includegraphics[scale=0.3]{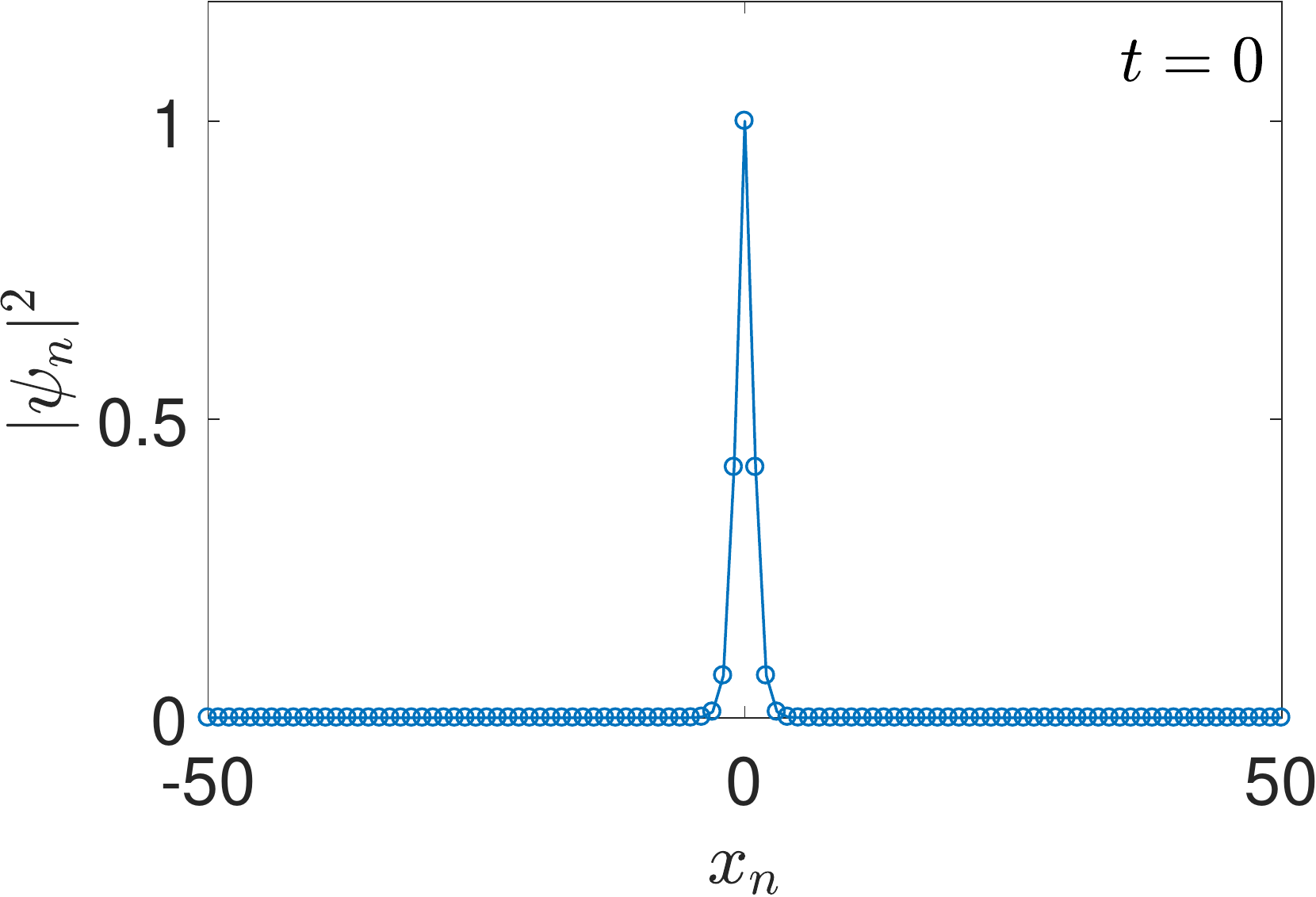}&
\quad
%\includegraphics[scale=0.3]{{fig5b}.pdf}
%\quad 
%\includegraphics[scale=0.3]{{fig5c}.pdf}
%\\[4ex]
\includegraphics[scale=0.3]{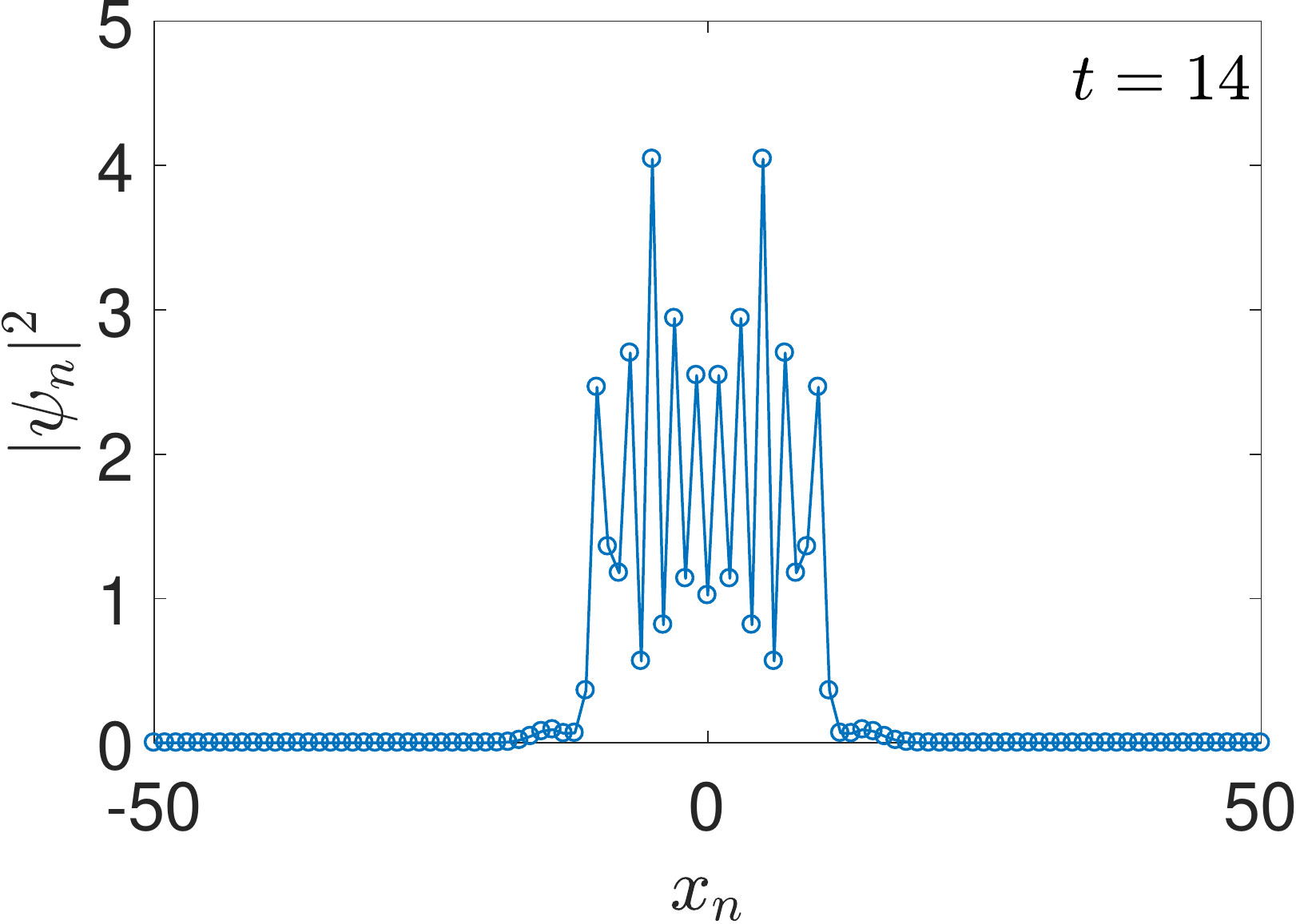}&
%\quad 
%\includegraphics[scale=0.3]{{fig5e}.pdf}
\quad 
\includegraphics[scale=0.3]{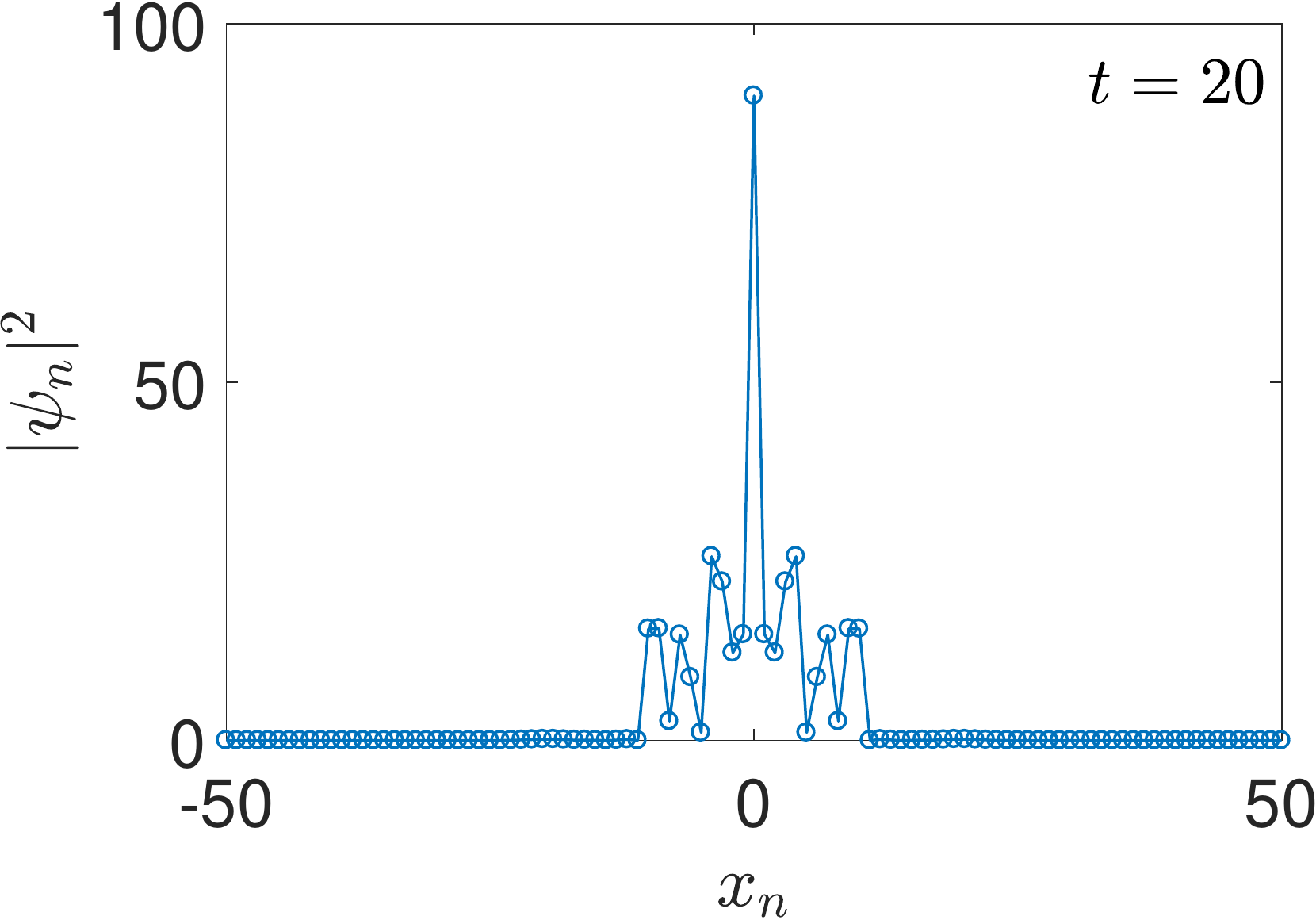}
	\end{tabular}
	\caption{Snapshots of the time-evolution of the density $|\psi_n|^2$ for the defocusing case $s=1$, emerged from the initial condition \eqref{sech} with amplitude $A=1$. Parameters: $\gamma=0.1$,  $\delta=0.01, L=50, N=100$ ($h=1$). }
	\label{Fig4}
\end{figure}

We now move on to examine panel (b) of Figure~\ref{Fig3}, where the defocusing case $s=1$ is considered. We observe that the numerical blow-up times demonstrate a similar qualitative functional trend to the analytical bounds, but quantitatively, they lie intermediately between them. Yet, this behavior of the numerical blow-up times suggests for dynamics different from the extended as well as from the localized type of collapse, that we may now explore. Figure \ref{Fig4} shows snapshots of the time-evolution of the density $|\psi_n|^2$ for the initial condition (\ref{sech}) in the case $s=1$ and $\gamma=0.1$, while the rest of parameters are the same as in Figures~\ref{Fig3}~and~\ref{blowfoc}. The model being defocusing, has the tendency to disperse  the initial localized energy along the lattice. 
%Consequently, the  energy is transferred  to the adjacent neighbors at the early stage of the evolution,
%and as a result, the initial amplitude is reduced, as it is observed in the middle snapshot of the first row for $t=4$. At $t=9$
Progressively, the effects of linear and nonlinear gain become more prominent and they prevent the energy dispersion. The excited units form  a ``plateau"-like state,  resembling in this non-vanishing part of the lattice, the extended state of \eqref{plane_wave}. The length of the chain covered by the excited units gets stabilized and the gained energy is redistributed among them, leading to their oscillating behavior of increasing amplitude (Fig.\ref{Fig4}b). 
%These dynamics are depicted by the snapshots of the second row, describing the time-evolution of the density for $t\in[14,20]$. 
Furthermore, Fig.\ref{Fig4}c 
%the snapshots at times $t=17$ and $t=20$ 
(close to the collapse time), reveals that the redistribution of energy among the excited units, has as a result the concentration of energy to the central unit, and the formation of a critical localized structure prior collapse, that is, a collective energy exchange phenomenon. 
%The impact of this mechanism is evident in the snapshot at time $t=20$, where the amplitude of the central unit is approaching the value $|\psi(x_{50})|^2\sim 100$, as this unit has gained energy from his neighbors, which have at this time much lesser amplitude.

Summarizing, at the first steps of the evolution the dynamics is reminiscent of the {\it extended blow-up} scenario, since the lattice tends to form a delocalized extended state. Next, energy gain domination prevents the spatial extension of this state to the whole lattice, and its evolution prior collapse is reminiscent of the {\it localized blow-up} scenario. This type of transient dynamics between the extended and the localized type of collapse is characterized accordingly, by numerical blow-up times lying between the analytical upper bound $\widehat{T}_\text{max}$ and lower bound $\overline{T}_\text{max}$.
%Summarizing,  in defocusing case $s=1$ the blow-up times of the solutions in the parameter region of Figure~\ref{Fig3}(a) lie between the two analytical bounds $\widehat{T}_\text{max}$ and $\overline{T}_\text{max}$ because, at the first steps of the evolution, our system tend to form a delocalized extended state resembling the {\it extended blow-up} scenario, but finally this state does not populate the whole lattice and the blow-up occurs from one site in a manner close to the {\it localized blow-up} scenario.

\begin{remark}\emph{
\label{rem1} (Classification of the collapse type dynamics). The analytical estimates $\widehat{T}_\text{max}$ and $\overline{T}_\text{max}$ can be useful in classifying the actual type of collapse dynamics for the DNLS (\ref{eq01}).  
\begin{itemize}
	\item In the extended collapse scenario, the numerical blow-up times are in excellent agreement or lie close to the analytical upper bound $\widehat{T}_\text{max}$. The nature of this behavior predisposes, in general, for larger time intervals of existence.
	\item In the localized collapse scenario, the numerical blow-up times are in excellent agreement or lie close to the analytical lower bound $\overline{T}_\text{max}$. This behavior is generically associated to smaller time intervals of existence.
	\item When transient dynamics prior collapse combines those of the extended and of the localized scenario, the numerical blow-up times  lie in between the analytical estimates. Such a behavior is associated to intermediate survival times.
\end{itemize}
%In the extended collapse scenario scenario, the numerical blow-up times are in excellent agreement with the analytical upper bound $\widehat{T}_\text{max}$. blow-up times lie close (or even coincide) with the analytical upper bound $\widehat{T}_\text{max}$. On the other hand when the system follows the localized blow-up scenario, the corresponding times lie close to (or even coincide with) the lower bound $\overline{T}_\text{max}$. When our system's solutions present a behavior that mixes characteristic from these two behaviors, the corresponding blow-up times will lie between the two analytic bounds. 
}
\end{remark}

\begin{remark}\emph{
\label{rem1b} (Defocusing/focusing  mechanisms, resonances with the linear spectrum and associated type of collapse). When vanishing initial conditions are considered, the  extended or the localized type of collapse is associated to the defocusing or to the focusing type of the lattice, respectively. More precisely: 
\begin{itemize}
\item
The  defocusing ($s=1$) DNLS lattice (\ref{eq01}) favors the extended blow-up scenario as, in this case, the system tends to delocalize the initial state (at least at the early stages of the evolution prior collapse). This behavior becomes more evident in the continuous/small amplitude regime. The fundamental dispersion mechanism for the considered initial condition is the resonance of the solution with the linear spectrum of the system. When the continuous limit is approached the linear spectrum becomes more extended. Also, when the amplitude of the solution decreases, the corresponding solution frequencies approach those of the linear spectrum. A combination of the above effects increase the possibility for a potential resonance, enhancing the dispersion of the initial state. As a result, the emerged transient collapse dynamics is reminiscent of the extended blow-up scenario. 
\item
The  focusing ($s=-1$) DNLS lattice (\ref{eq01}) favors the localized blow-up scenario. This behavior becomes  more evident in the discrete/large amplitude regime. When the system is in the  purely discrete regime or is approaching the anticontinuous limit, the linear spectrum shrinks. Also, when the amplitude of the initial condition becomes larger, the frequencies of the solution move away from the  linear spectrum. A combination of these effects enforces the localization of the solution instead of its dispersion. As a result, the  emerged transient collapse dynamics is reminiscent of the localized blow-up scenario. 
\item The discretization regime, which is dictated by the value of $h$, may drastically affect the collapse dynamics exhibited by a lattice, which is parametrically defined (by the choice for the parameter $s=1$)  as defocusing.  It may even reverse its type of collapse. For instance, a defocusing lattice may exhibit a localized type of collapse for large values of the discretization parameter. This phenomenon does not occur, in such extend, in the focusing defined ($s=-1$) lattices.
\end{itemize}
}
\end{remark}

%\begin{remark}
%\label{rem2}
%Although we don't have a proof for this, by examining the agreement of the numerical results with the analytical bounds in Figures~\ref{Fig1}~and~\ref{Fig3}(a) we claim that $\widehat{T}_\text{max}$ and $\overline{T}_\text{max}$ are the best (or at least close to the best) upper and lower bound estimates we could acquire. 
%\end{remark}
To elucidate further the claims of the above remarks, we proceed to a numerical investigation examining the behavior of the system as the discretization parameter $h$ is varied.
%%%%%%%%%%%%%%%%%%%%%%%%%%%%%%%%%%%%%%%%
\begin{figure}[tbh!]
	\centering 
	\includegraphics[scale=0.48]{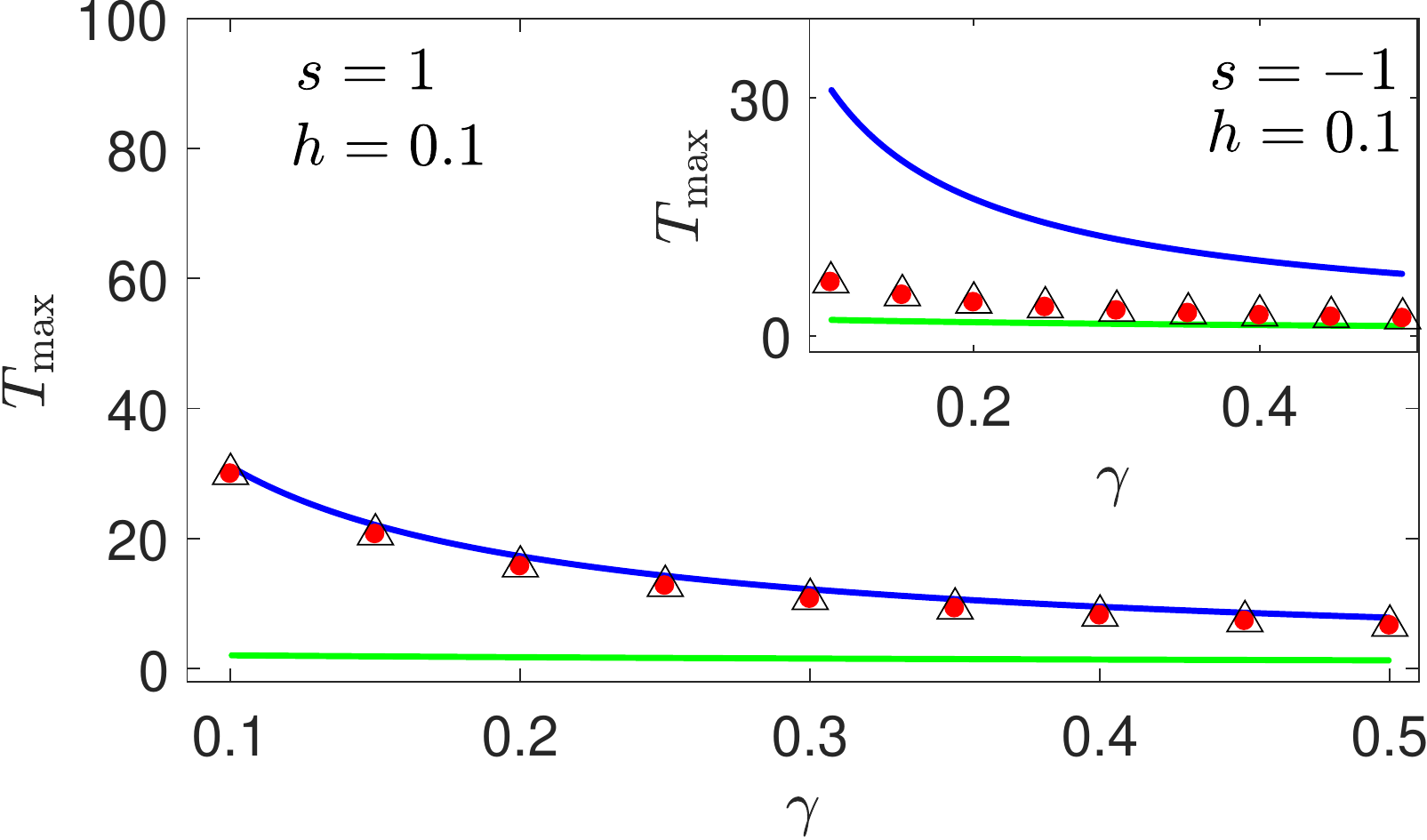}
	\quad
	\includegraphics[scale=0.48]{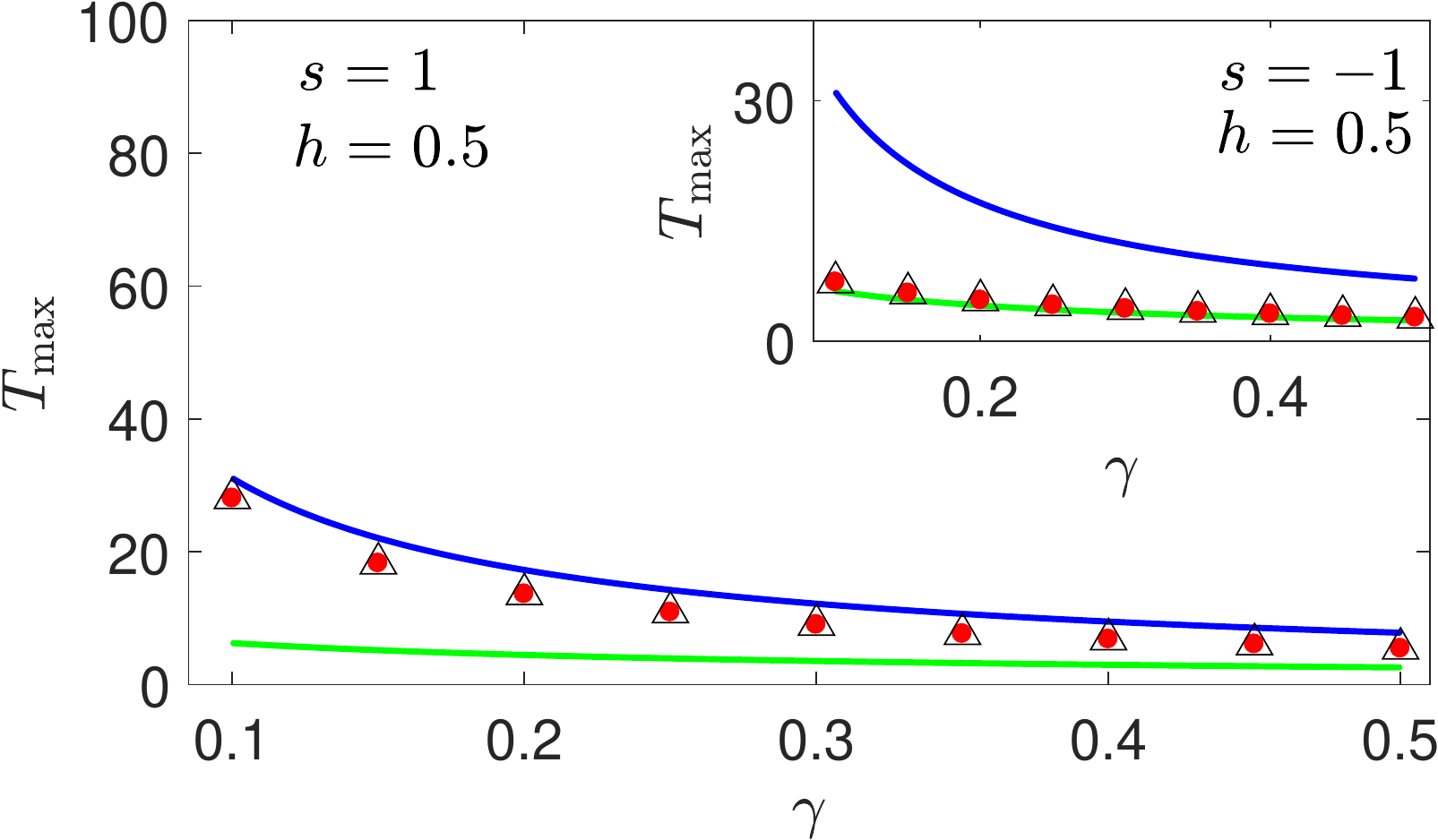}
	\\ [4ex] 
	\includegraphics[scale=0.48]{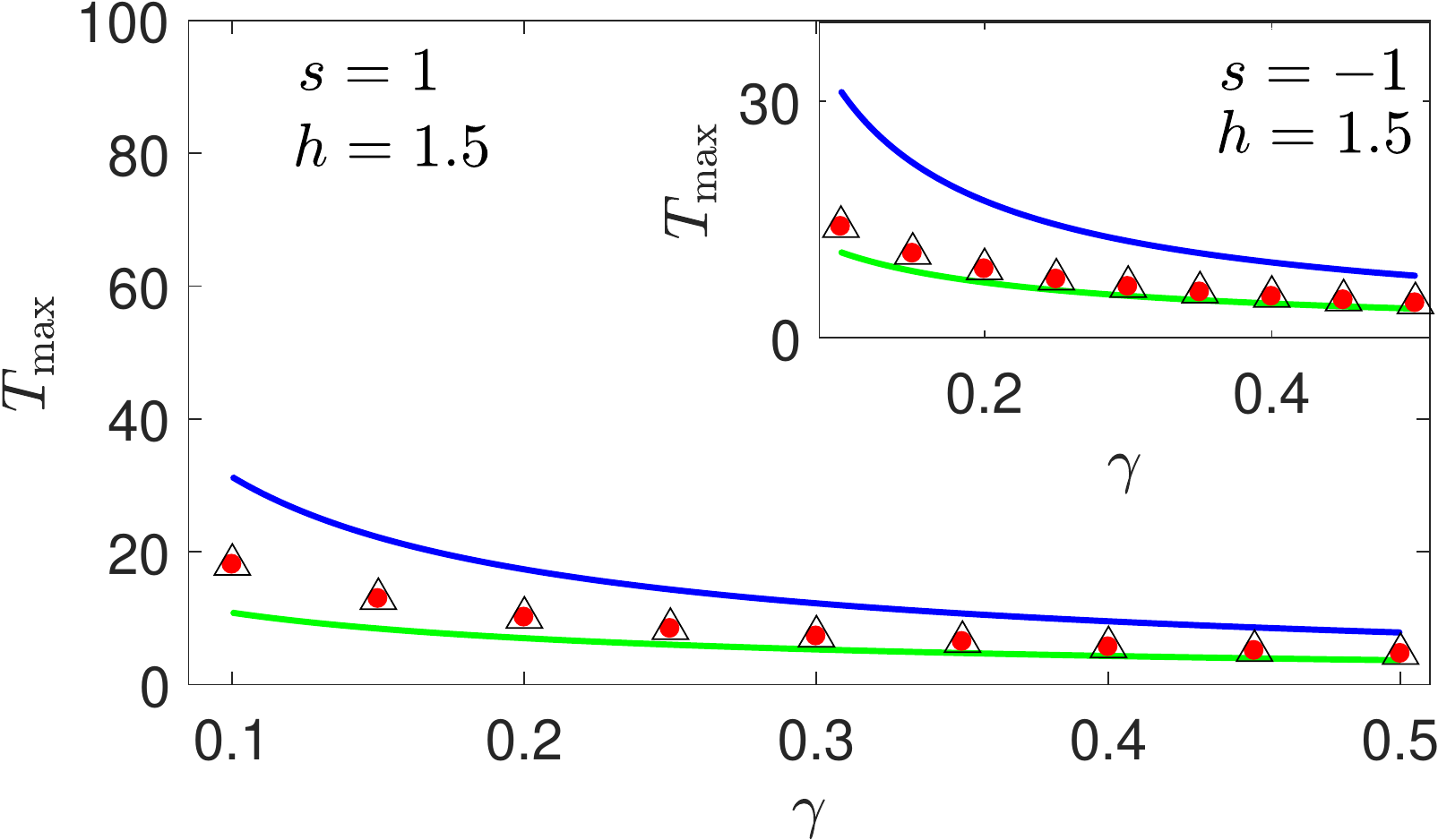}
	\quad
	\includegraphics[scale=0.48]{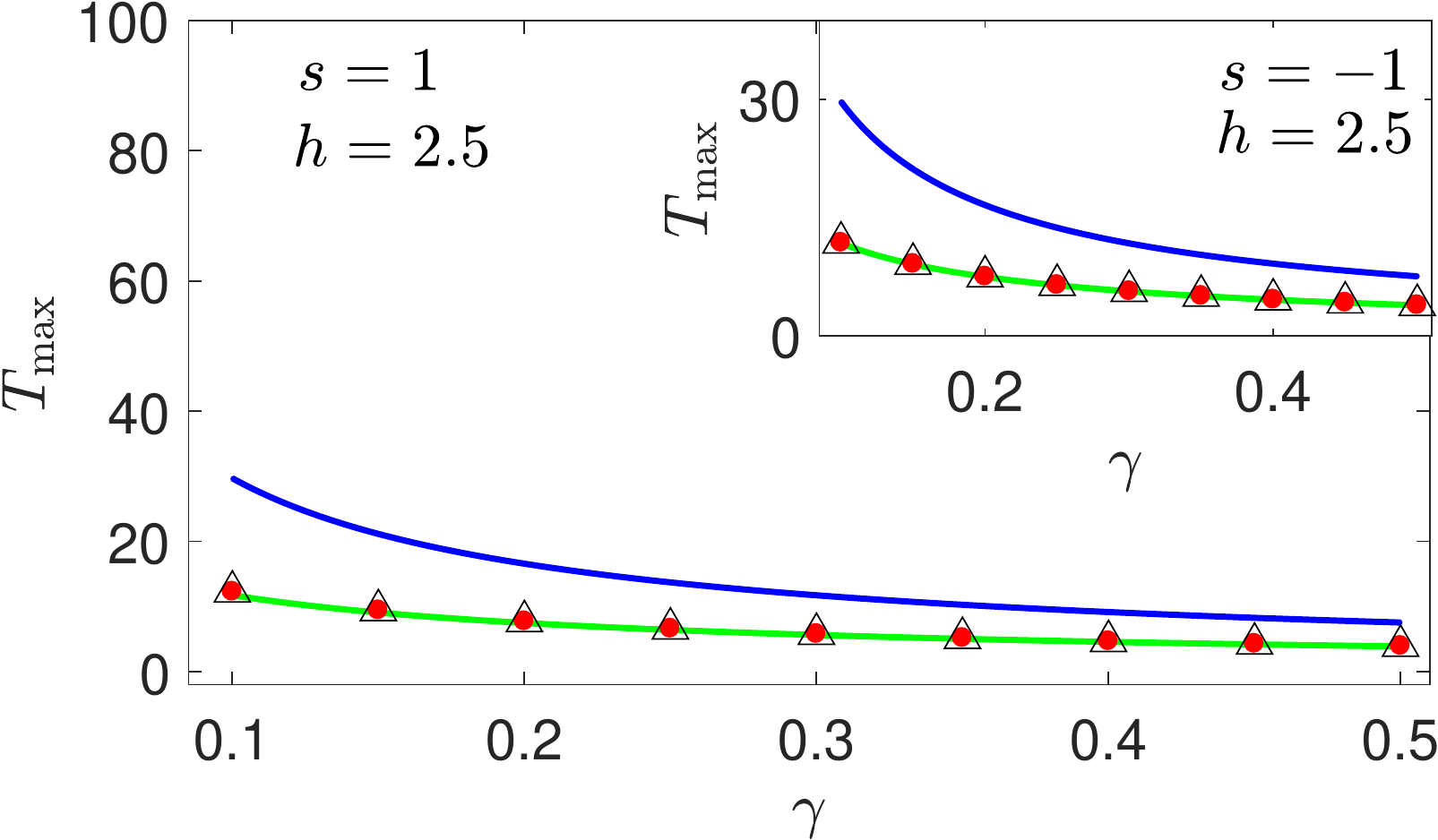}
	\caption{Numerical blow-up times [(red) dots for the problem ($\mathcal{P}$)-triangles for the problem ($\mathcal{D})$], for the $\sech$-profiled initial condition \eqref{sech},  against the analytical upper  bound [upper (blue) curve] and lower bound [lower (green) curve], as functions of $\gamma\in[0.1,0.5]$.  Main panels for the defocusing case $s=1$, and insets for the focusing one $s=-1$. In all panels and insets, $L=50$, $\delta=0.01$ and $A = 1$.   Top left panel: $N=1000$ and $h=0.1$.  Top right panel: $N=200$ and $h=0.5$. Bottom left panel: $N=67$ and $h=1.5$. Bottom right panel: $N=40$ and $h=2.5$.  } 
	\label{Fig7B}
\end{figure}
Figure~\ref{Fig7B} extends the study of Figure~\ref{Fig3}, for increasing values of the lattice spacing  $h=0.1, 0.5, 1.5~\text{and}~2.5$. The main panels depict the results for the defocusing case $s=1$, and the  insets portray the corresponding results for the focusing  case $s=-1$. In the defocusing case, when $h=0.1$, we observe an excellent agreement with the  upper bound $\widehat{T}_{\max}$, which  is in full accordance with the claims of the Remarks~\ref{rem1}-\ref{rem1b}; since the system  is defocusing and falls within the continuous regime,  it favors the extend type of collapse. This choice of behavior of the system is verified in Figure~\ref{Fig4B}, illustrating snapshots of the time evolution of the density, for $s=1$, $h=0.1$ and $\gamma=0.1$.  Approaching the continuous limit, the energy dispersion along the lattice, due to the defocusing effect, is further enhanced. The initial condition is delocalized and a much more extended structure 
%``plateau"  
is formed (if compared with the one observed in the snapshots %at time $t=9$ 
of Figure \ref{Fig4}). This delocalized state evolves reminiscently  of a ``wave background", as shown initially in Figure~\ref{Fig4B}b and which extends to the whole lattice
%the snapshots at $t=19$ and $t=29$ 
just before the collapse of the solution (Fig.~\ref{Fig4B}c).  Such an evolution constitutes another example of the extended blow-up scenario of Remark~\ref{rem1}. Returning to  Figure~\ref{Fig7B}, as $h$  is increasing, and the system moves into more discrete regimes, the blow-up times depart from the $\widehat{T}_{\max}$  and converge towards the estimate $\overline{T}_{\max}$.  When $h=2.5$, the numerical values practically coincide with the analytical lower bound $\overline{T}_{\max}$. The above reversal of the extended collapse dynamics to localized, does not occur in the focusing model $s=-1$, as shown in the insets of   Figure~\ref{Fig7B}.  The focusing nature of the lattice is present even for small values of $h$ but it is enhanced for large values of $h$, and the numerical blow-up times converge to the lower bound $\overline{T}_{\max}$, while they almost coincide for $h=2.5$. 
%%%%%%%%%%%%%%%%%%%%%%%%%%%%%%%%%%%%%%%%
%%%%%%%%%%%%%%%%%%%%%%%%%%%%%%%%%%%%%%%%%%%%55
\begin{figure}[tbh!]
	\centering 
	\begin{tabular}{ccc}
	\qquad(a)&\qquad\ (b)&\qquad\ (c)\\	
\includegraphics[scale=0.3]{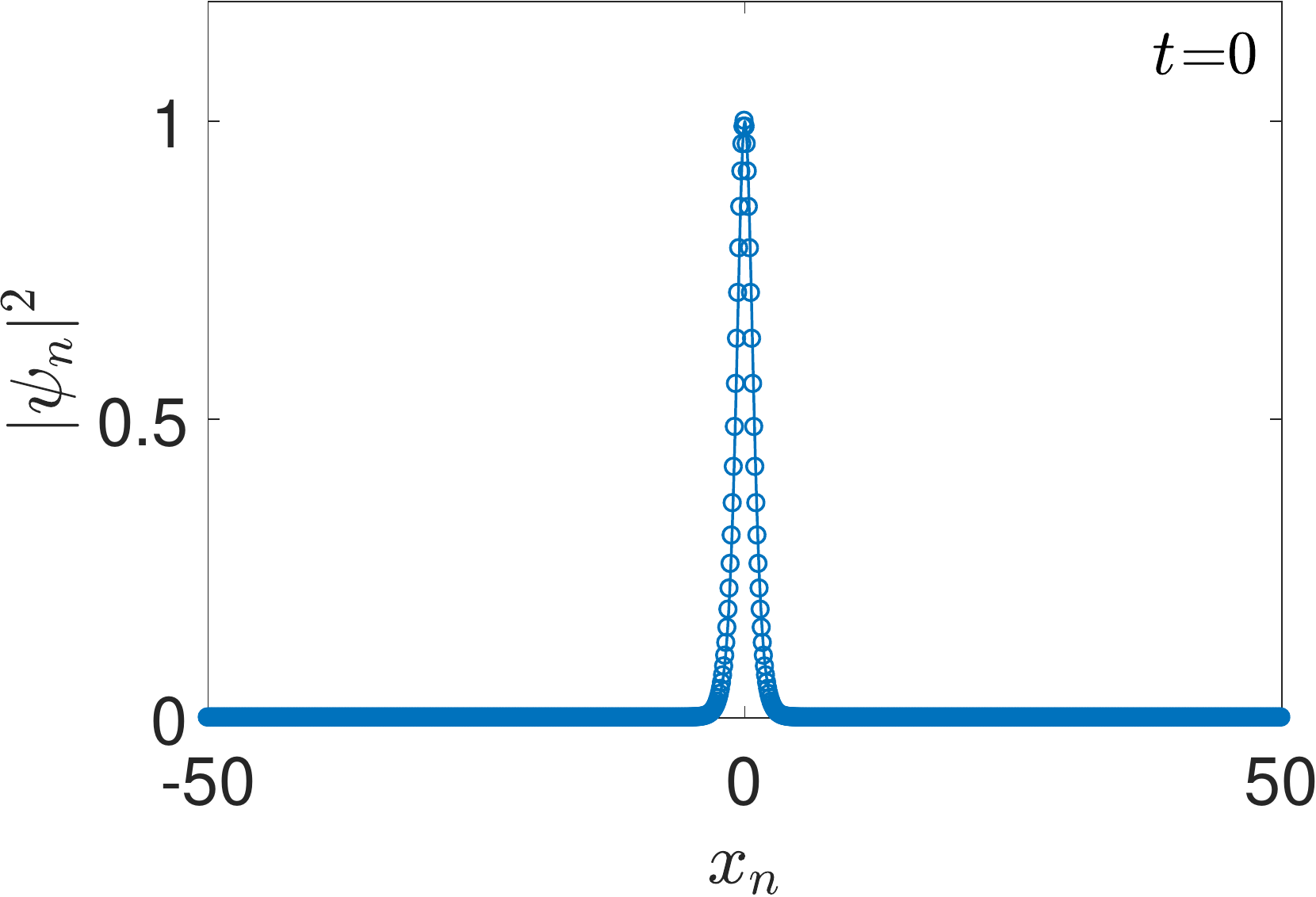}&
\quad
%\includegraphics[scale=0.3]{fig7b.pdf}
%\quad 
%\includegraphics[scale=0.3]{fig7c.pdf}
%\\[4ex]
\includegraphics[scale=0.3]{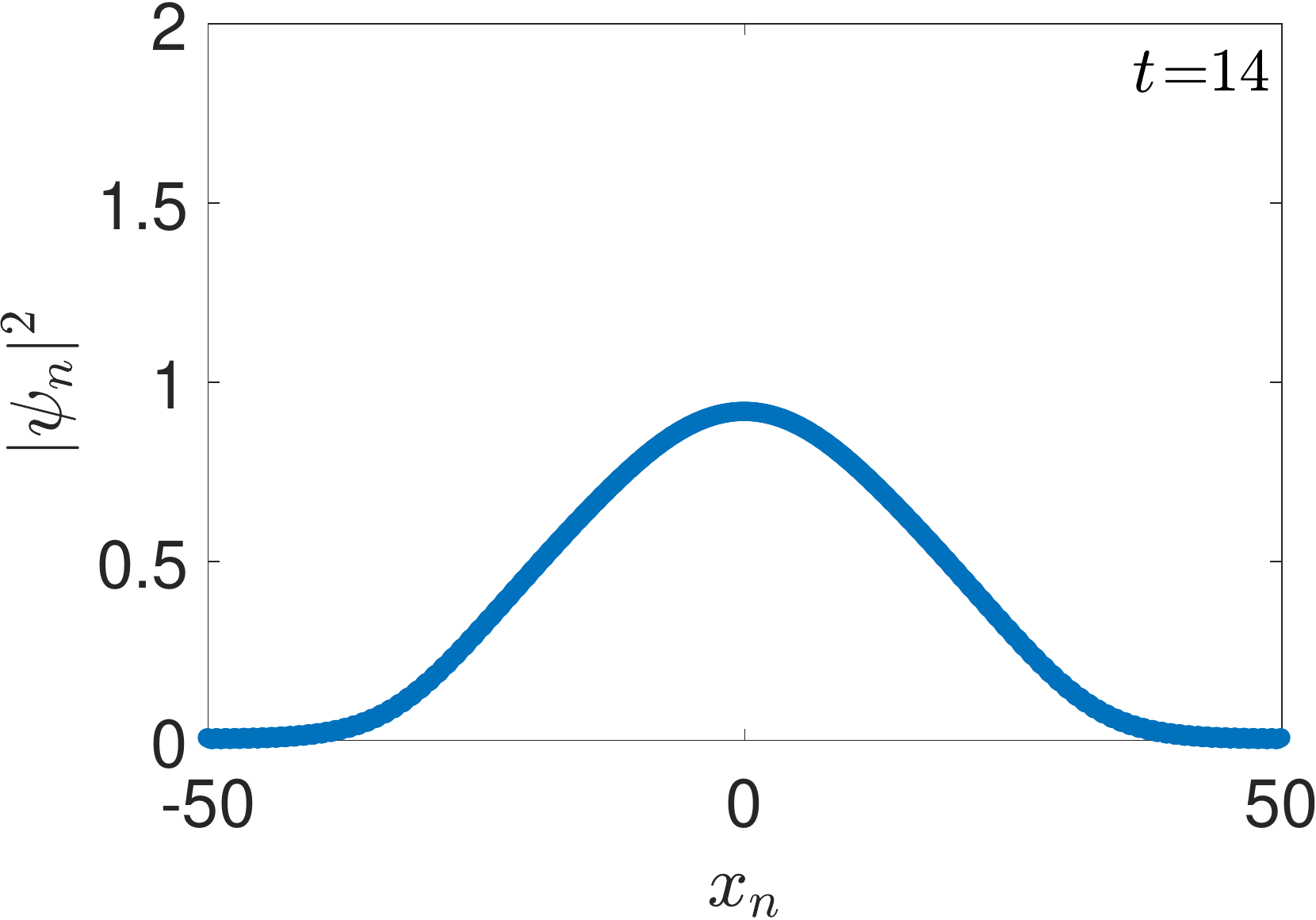}&
\quad 
%\includegraphics[scale=0.3]{fig7e.pdf}
%\quad 
\includegraphics[scale=0.3]{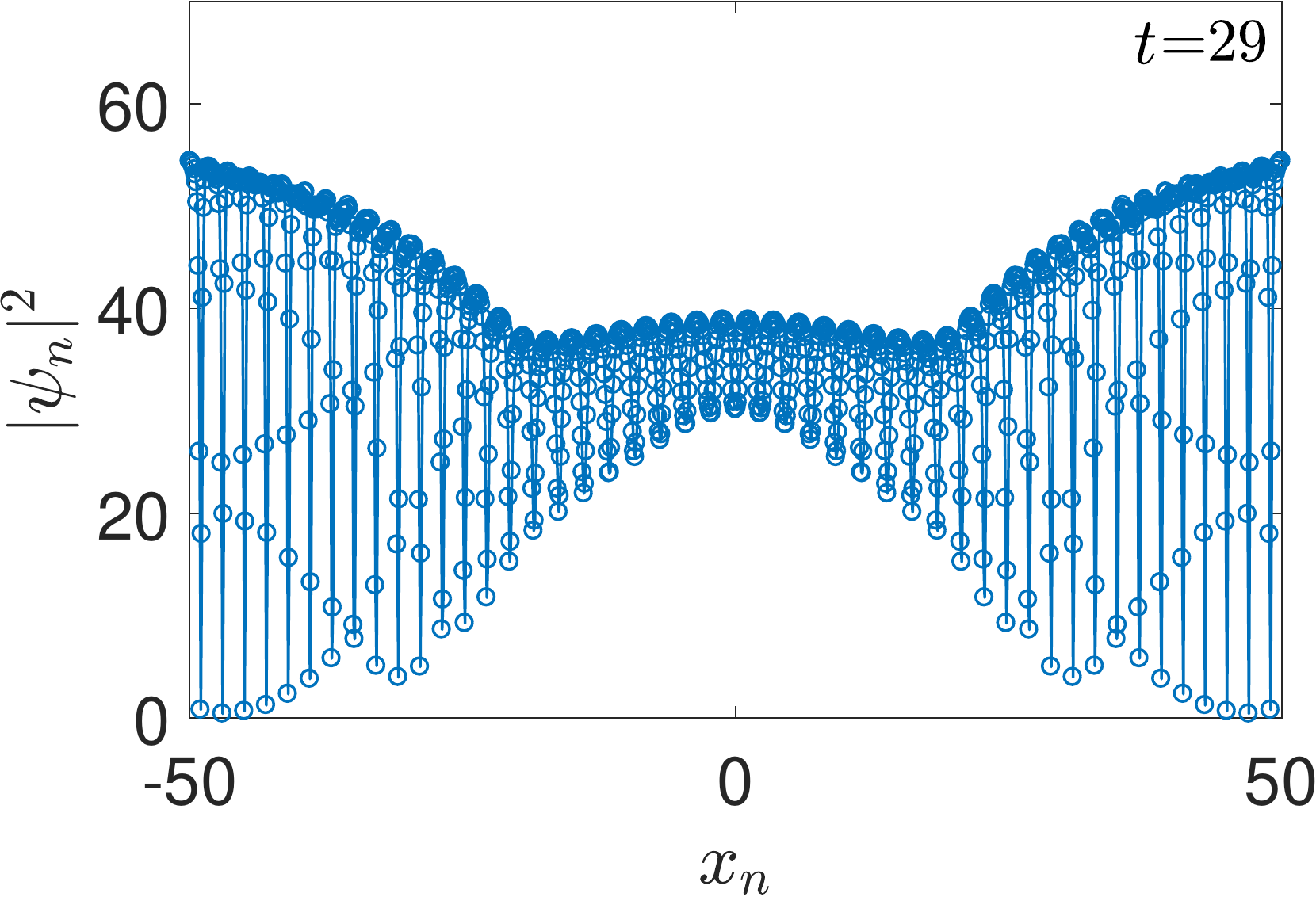}
	\end{tabular}
	\caption{Snapshots of the time-evolution of the density $|\psi_n|^2$ for the defocusing case $s=1$, emerged from the initial condition \eqref{sech} with amplitude $A=1$. Parameters: $\gamma=0.1$,  $\delta=0.01, L=50,~\text{and}~N=1000$ ($h=0.1$).}
	\label{Fig4B}
\end{figure}
%%%%%%%%%%%%%%%%%%%%%%%%%%%%%%%%%%%%%%%%%%%%%%%%%%%%%%%%%%%%%%%%%%%%55555
%%%%%%%%%%%%%%%%%%%%%%%%%%%%%%%%%%%%%%%%%%%%%%%%%%%%%%%%%%%%%%%%%%%%%%%%%%

\begin{figure}[tbh!]
	\centering 
	\begin{tabular}{ccc}
	\large(a)&\large(b)&\large(c)\\[5pt]
	\includegraphics[scale=0.33]{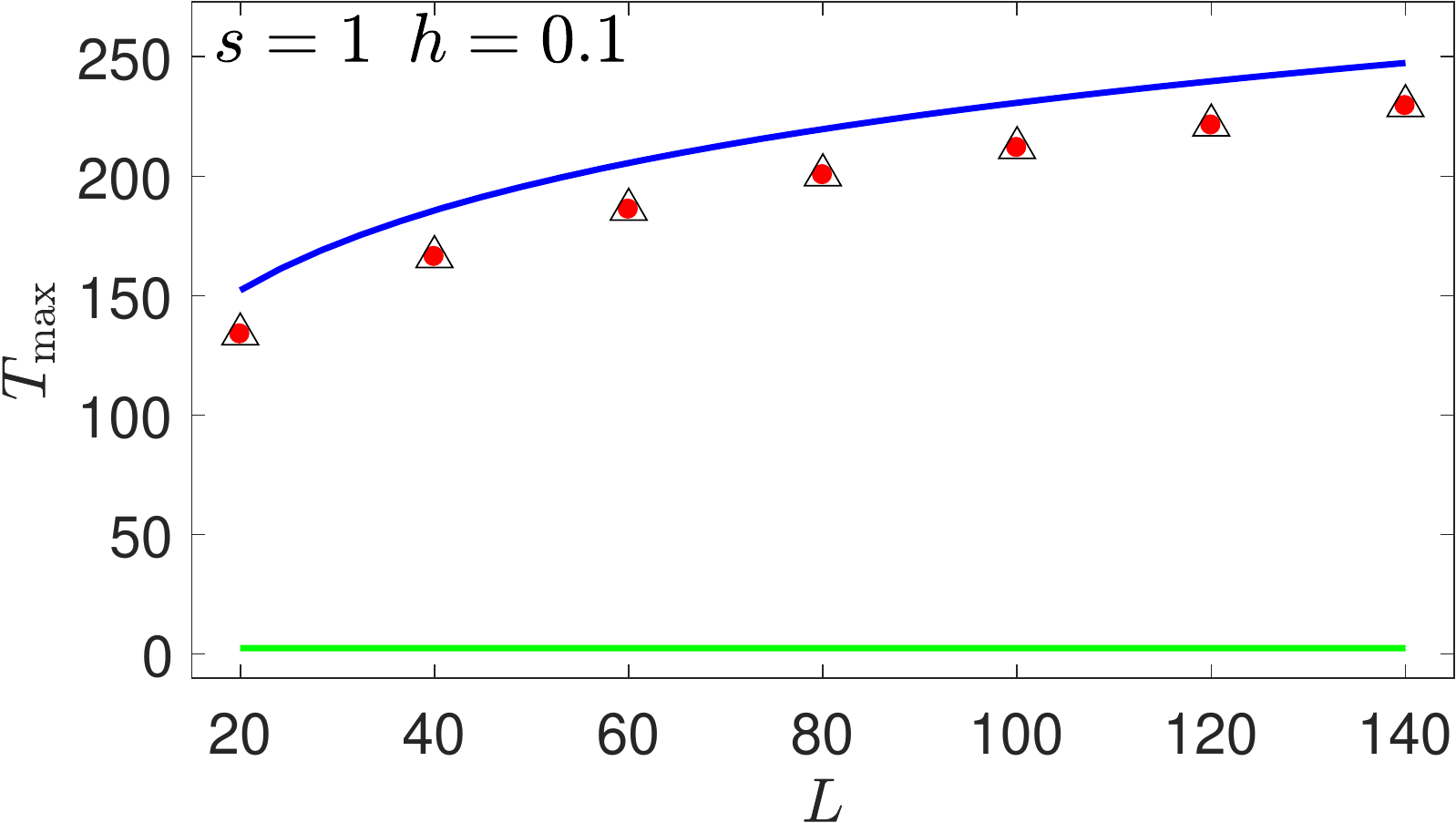}&
	\includegraphics[scale=0.33]{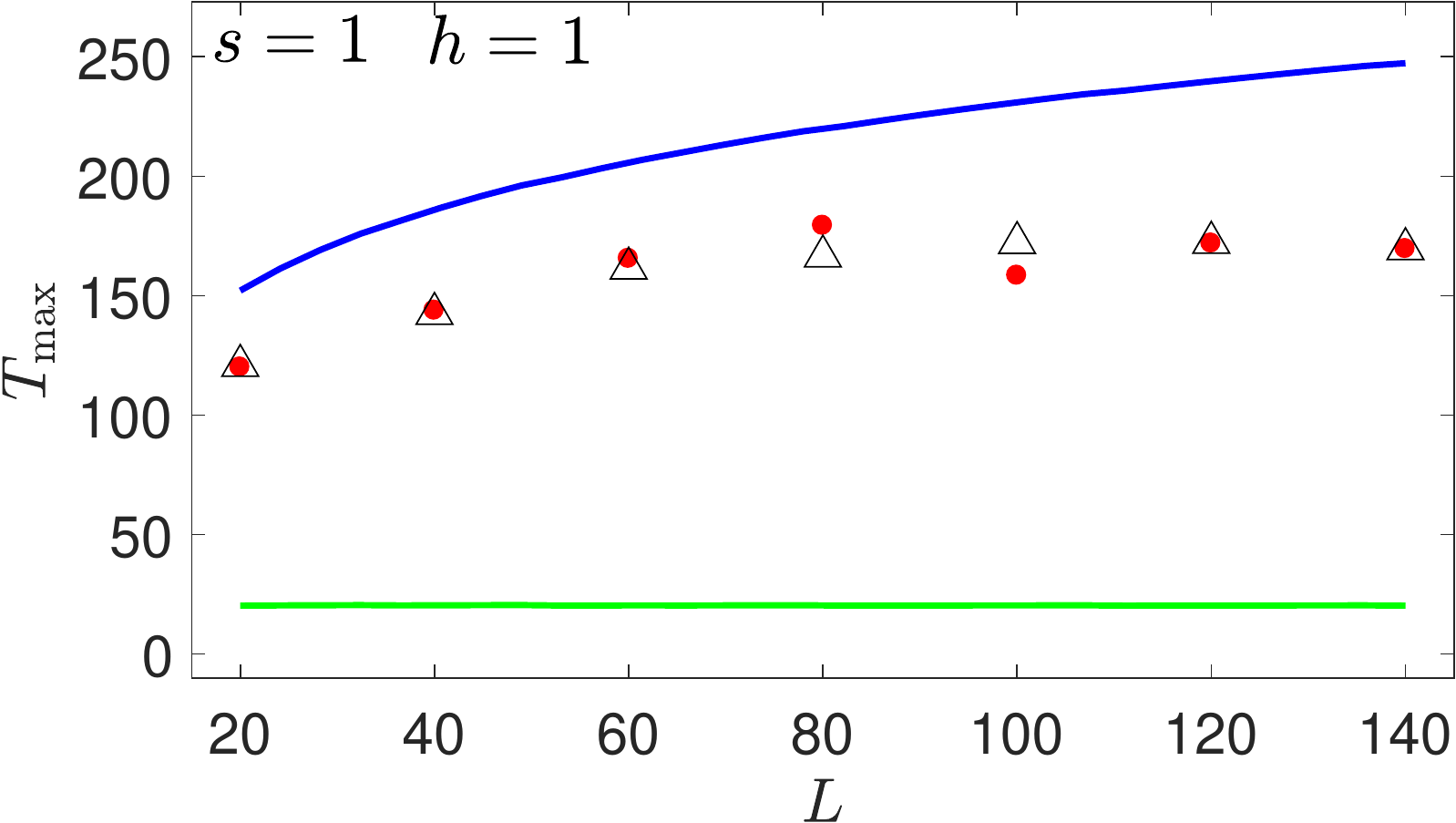}&
	\includegraphics[scale=0.33]{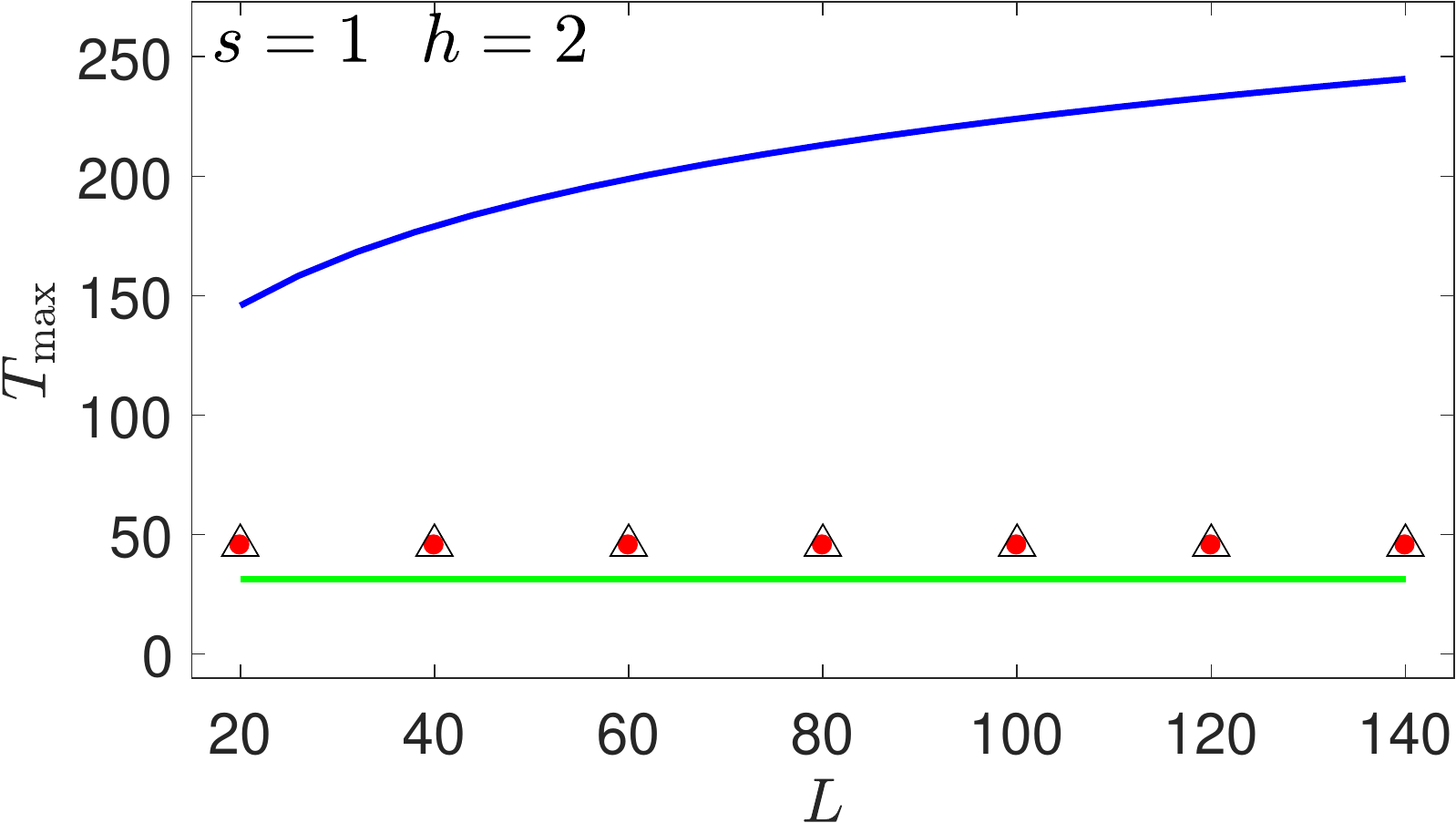}\\[10pt]
	\large(d)&\large(e)&\large(f)\\[5pt]
	\includegraphics[scale=0.33]{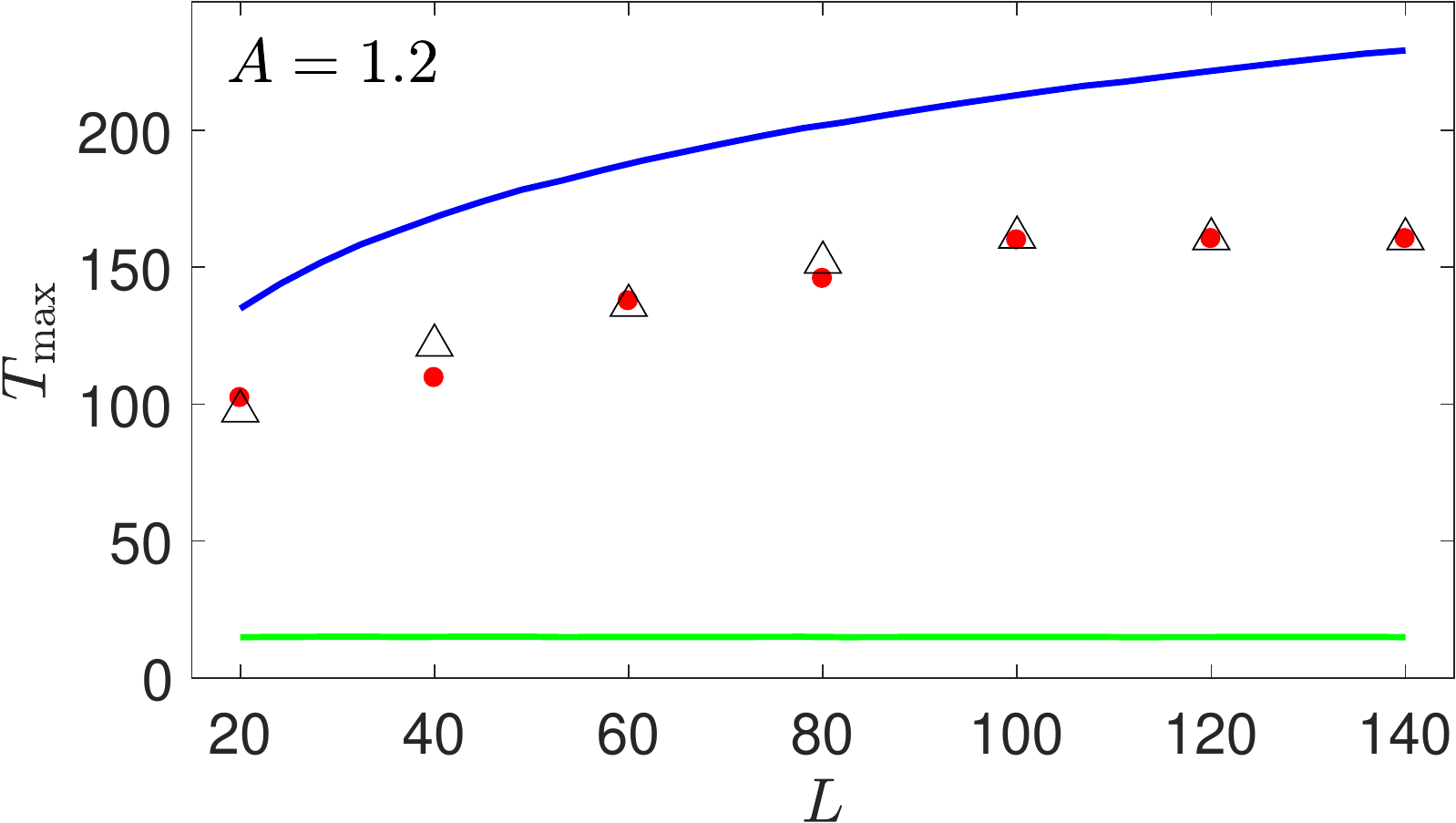}&
 	\includegraphics[scale=0.33]{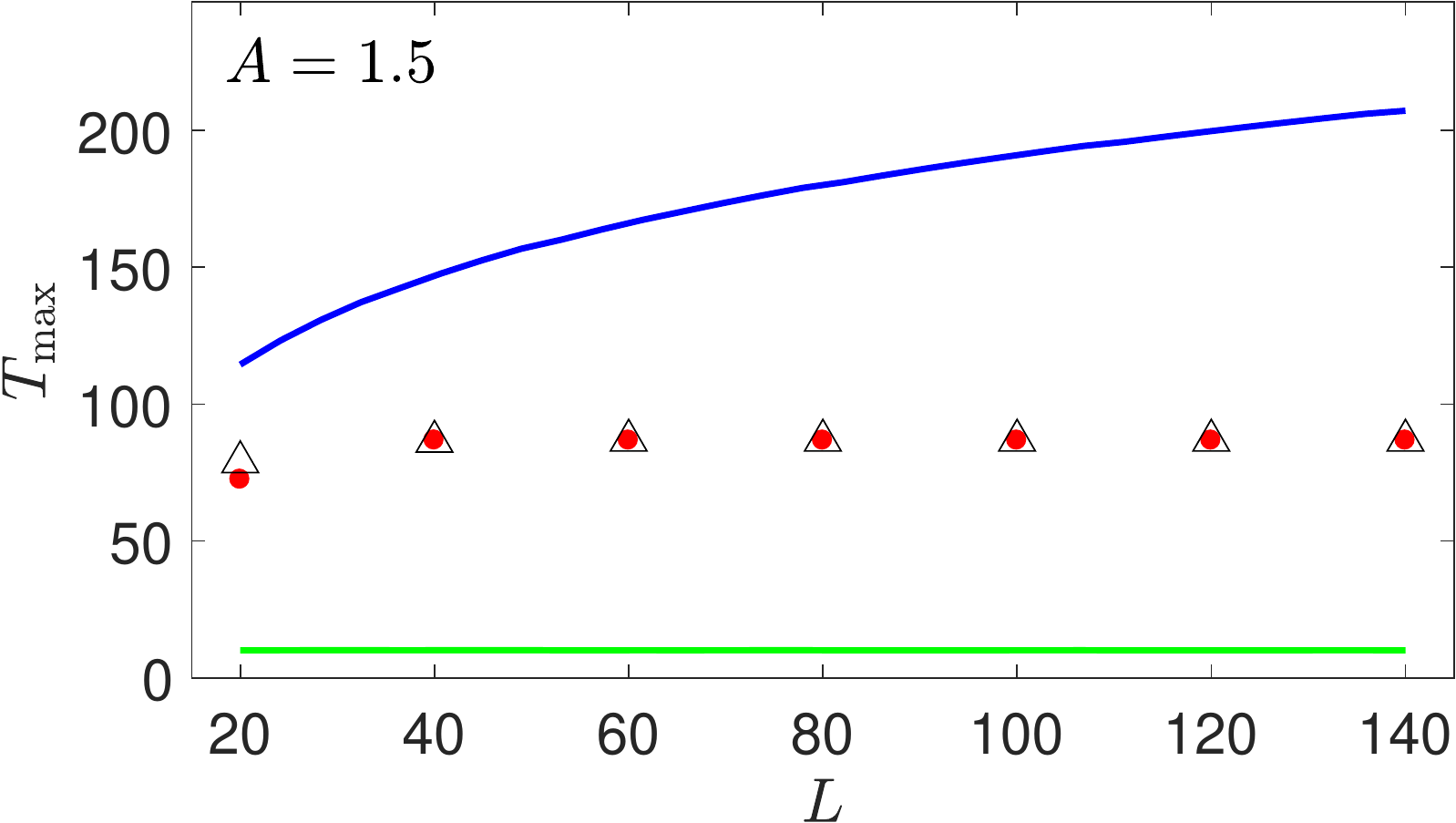}&
	\includegraphics[scale=0.33]{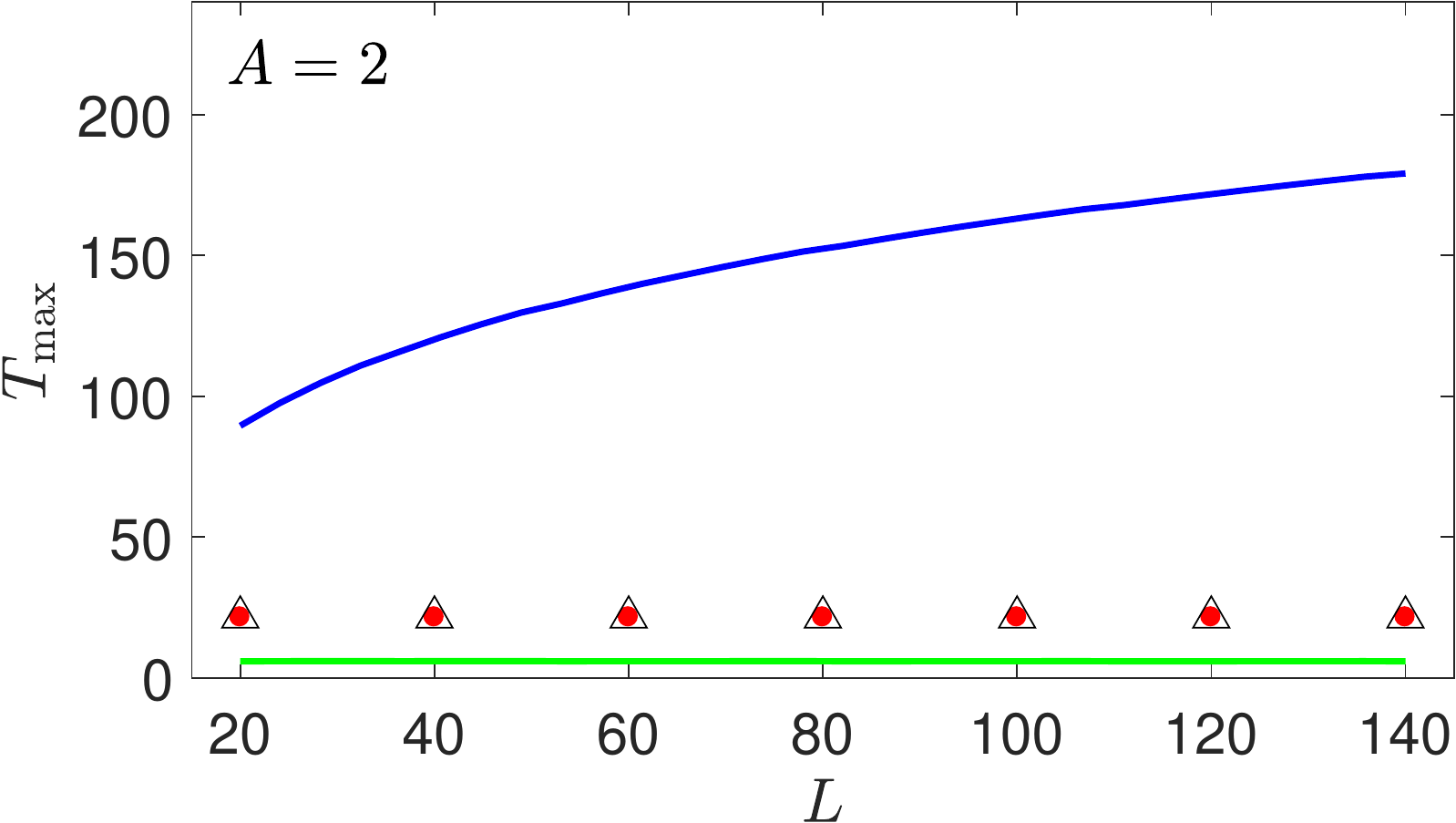}
	\end{tabular}
	\caption{Defocusing case $s=1$: Numerical blow-up times [(red) dots for the problem ($\mathcal{P}$)-triangles for the problem ($\mathcal{D})$], for the $\sech$-profiled initial condition \eqref{sech},  against the analytical upper  bound [upper (blue) curve] and lower bound [lower (green) curve], as functions of $L\in[20, 140]$.  Top (bottom) row, each panel depicts the results of the comparison for increasing values of  the lattice spacing  $h$ (amplitude $A$ of the initial condition) and fixed amplitude $A=1$ (lattice spacing  $h=1$). %Top (bottom) row for the defocusing case $s=1$ (focusing case $s=-1$) .  Parameter $h$: (a) [(d)] $h=0.1$, (b) [(e)] $h=1$ and (c) [(f)] $h=2$. 
	Other parameters: $\gamma = 0.01, \delta = 0.01$.% and amplitude of the initial condition $A = 1$ ($h=1$).
	}
	\label{Fig3B}
\end{figure}

Next, we compare the analytical bounds $\Th$ and $\To$ against the numerical blow-up times as functions of the half length  $L \in [20,140]$, for small values of the gain parameters $\gamma = 0.01, \delta = 0.01$, and amplitude of the initial condition $A=1$. The results of this study are shown in the top row of Figure \ref{Fig3B}, where the results for the defocusing ($s =1$) case, for three values of the lattice spacing ($h = 0.1, 1$~and~$2$), are shown. Once again, we observe the influence of the discreteness in reversing the nature of the collapse dynamics from extended to localized, as the discretization parameter is increased from $h = 0.1$ to $h = 2$. Notably, as shown in panel (b), the reversal is identified by the existence of a critical value of the half-length $L_{\mathrm{cr}}\simeq 90$ with the following property: For  $L<L_{\mathrm{cr}}$, the numerical blow-up times preserve the trend of the logarithmic functional dependence of the upper bound  $\Th$, while for $L\geq L_{\mathrm{cr}}$, the blow-up times approach a constant. The horizontal trend to the lower bound $\To$, suggests that the dispersion along a large portion of the lattice, is prevented when its length exceeds the critical value $L_{\mathrm{cr}}$, and is limited to a finite sub-interval of the lattice. Intuitively, we could think of a large-length defocusing lattice to exhibit collapse dynamics as being ``geometrically similar'' to a focusing one, for large length-scales (recall Figure \ref{Fig4}). It is also interesting to observe in panel (b), a discrepancy between the numerical blow-up times acquired by numerically solving the problems ($\mathcal{P})$  and ($\mathcal{D})$, for values of $L$ close to $L_{\mathrm{cr}}$. The discrepancy suggests for finer interactions with the boundaries in the vicinity of $L_{\mathrm{cr}}$. These interactions are indicating the above transformation of dynamics, and are becoming negligible far from $L_{\mathrm{cr}}$, where its type is transitioned to localized. Although the focusing case $s=-1$ is not illustrated here, let us note that the focusing effect of the system is further enhanced for increasing values of the parameter $h$, and the numerical blow-up times tend to an excellent agreement with the lower bound $\To$. Therefore, both cases justify the claims of the Remark~\ref{rem1b}.

%%%%%%%%%%%%%%%%%%%%%%%%%%%%%%%%%%%%%%%%%%%%%%%%%%%%%%%%%%%%%%
\begin{comment}
Not only the results of the top row of Figure~\ref{Fig3B}, but also those of the bottom row for the focusing  case $s=-1$, justify the claims of the Remark~\ref{rem1b}. The focusing effect is even enhanced for increasing values of the parameter $h$, and the numerical blow-up times tend to an excellent agreement with the lower bound $\To$.
\end{comment}
%%%%%%%%%%%%%%%%%%%%%%%%%%%%%%%%%%%%%%%%%%%%%%%%%%%%%%%%%%%%%%%
\begin{comment}
\begin{figure}[tbh!]
	\centering
	\includegraphics[scale=0.33]{fig9a.pdf}
	\quad
	\includegraphics[scale=0.33]{fig9b.pdf}
	\quad 
	\includegraphics[scale=0.33]{fig9c.pdf}
	\\[4ex]
	\includegraphics[scale=0.33]{fig9d.pdf}
	\quad 
	\includegraphics[scale=0.33]{fig9e.pdf}
	\quad 
	\includegraphics[scale=0.33]{fig9f.pdf}
	\caption{Defocusing case $s=1$: Numerical blow-up times [(red) dots for the problem ($\mathcal{P}$)-triangles for the problem ($\mathcal{D})$], for the $\sech$-profiled initial condition \eqref{sech},  against the analytical upper  bound [upper (blue) curve] and lower bound [lower (green) curve], as functions of $L\in[20, 140]$. Each panel depicts the results of the comparison for increasing values of  the amplitude $A$ of the initial condition. Other parameters: $\gamma=0.01$, $\delta=0.01$,~\text{and}~$h=1$.}
	\label{Fig5}
\end{figure}
\end{comment}
Alternatively, the reversal of the collapse type dynamics, in the defocusing lattice $s = 1$, can be illustrated by examining the behavior of the blow-up times when the amplitude of the initial condition is varied. In the bottom row of Figure~\ref{Fig3B},  each panel (d), (e), and (f) depicts, for increasing values of the amplitude $A =1.2,1.5$ and $2$, respectively of the initial condition \eqref{sech}, the results of the comparison of the numerical blow-up times against the analytical bounds $\widehat{T}_\text{max}$ and $\overline{T}_\text{max}$, as functions of $L$.  The lattice spacing is fixed to $h = 1$, while the other parameters are fixed to $\gamma = \delta = 0.01$. For increasing values of the amplitude, the nonlinearity effects dominate, and enhance the localization of the initial excitation against its dispersion, as described in Remark \ref{rem1b}. For increasing values of $A$, the extended type of collapse persists for small values of $L$. This is manifested by the decreasing value of the critical length $L_{\mathrm{cr}}$ on which the transition, from extended to localized blow-up, occurs. To sustain an extended type of collapse, the length of the lattice should be reduced so that the defocusing-dispersive effects can compensate the nonlinearity-localization ones due to the increase of the amplitude. Still, the existence of the critical length $L_{\mathrm{cr}}$ is located by the observed discrepancy in its vicinity, between the numerical blow-up times acquired by the numerical solutions of the problems ($\mathcal{P}$) and ($\mathcal{D}$). Also, there exists a critical value of the amplitude (herein $A >1.5$), above which, the localized type of collapse dominates. In this high-amplitude regime, the numerical blow-up times remain constant, independently of the length of the lattice. 

Finally, although not shown here, we note  that the accuracy of the separating value $\gamma^*$, in separating the dynamical behavior between finite time collapse (for $\gamma>\gamma^*$) and global existence dynamics (for $\gamma<\gamma^*$), is also verified numerically for this choice of initial conditions, as well as for all the other choices of initial conditions that follow. 
%%%%%%%%%%%%%%%%%%%%%%%%%%%%%%%%%%%%%%%%%%%%%%%%%%%%%%%%%%555
\subsection{Decaying to a finite background (\texorpdfstring{$\tanh^2$}--profiled) initial conditions.}\label{sec3C}
%%%%%%%%%%%%%%%%%%%%%%%%
The next example of initial conditions whose collapse dynamics will be numerically investigated has the form:
\begin{equation}\label{tanh}
\psi_n(0) = A\tanh^2x_n,
\end{equation}
where  $A > 0$ is the amplitude and $x_n$ the discrete spatial coordinate \eqref{spatial_cor}. The discrete function (\ref{tanh}) has a profile reminiscent of the density dip of a ``discrete dark soliton'' over a finite background, which is compliant with the periodic boundary conditions (\ref{eq02}).

\begin{figure}[tbh!]
	\centering 
	\includegraphics[scale=0.34]{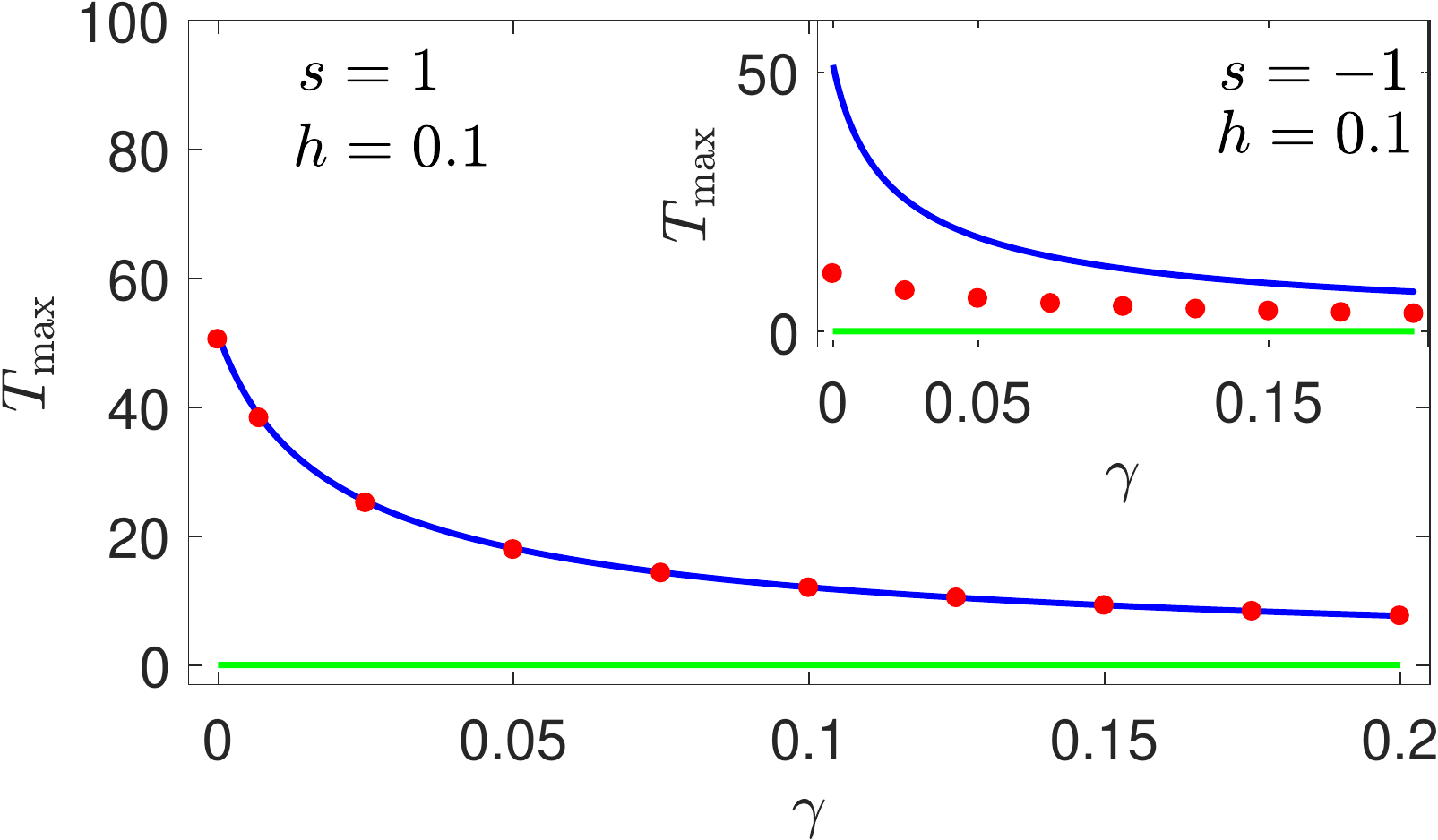}
	\includegraphics[scale=0.34]{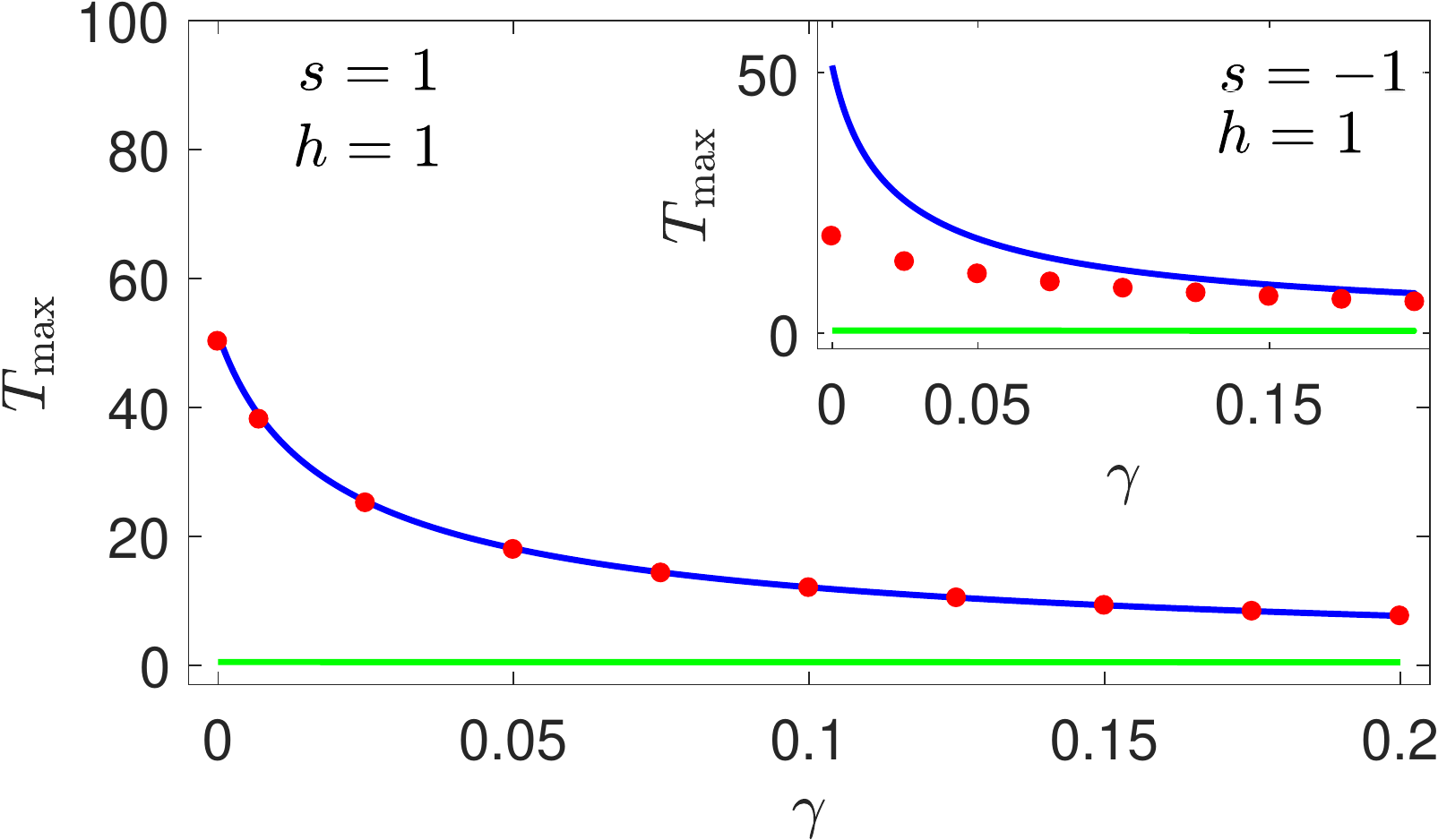}
	\includegraphics[scale=0.34]{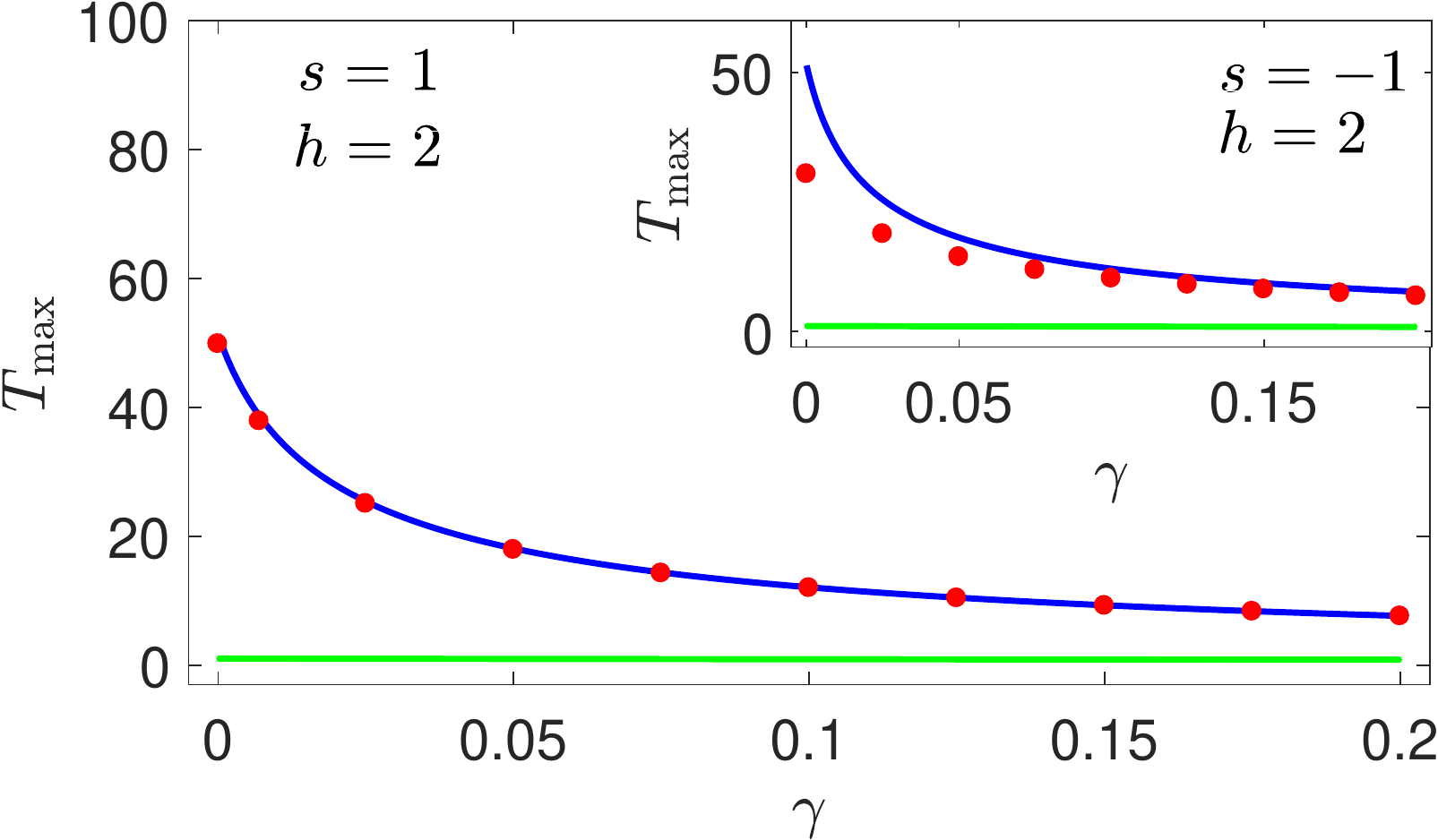}
	\caption{
Numerical blow-up times [(red) dots for the problem ($\mathcal{P}$)], for the $\tanh^2$-profiled initial condition \eqref{sech},  against the analytical upper  bound [upper (blue) curve] and lower bound [lower (green) curve], as functions of $\gamma\in(0,0.2]$. Main panels (insets) for the defocusing case $s=1$ (focusing case $s=-1$) depict the results of the comparison for three values of the spacing parameter $h=0.1, 1~\text{and}~2$. Other parameters: $\delta=0.01, A=1, ~\text{and}~L=50$.}
	\label{Fig8}
\end{figure}

Figure~\ref{Fig8} shows the results of the comparison of the numerical blow-up times against the analytical upper bound $\widehat{T}_\text{max}$ (\ref{eqTh1}) and the lower bound $\overline{T}_\text{max}$ \eqref{eqAlt1}, as functions of $\gamma \in (0, 0.2]$, for three values of the lattice spacing parameter $h=0.1, 1~\text{and}~2$. The other gain and lattice parameters  are: $\delta = 0.01,  L = 50,$ and the amplitude of the initial condition is $A=1$. The main panels depict the results for the defocusing case $s=1$, and the  insets for the focusing case $s=-1$.

\begin{remark}
\label{rem3}
\emph{Although the lower bound $\overline{T}_{\mathrm{max}}$ is not analytically valid for the initial-boundary value problem ($\mathcal{P}$) with which the initial condition (\ref{tanh}) is compliant, we still consider it in the study of Figure~\ref{Fig8}, in order to examine whether it is numerically valid as a lower bound, even for the case of periodic boundary conditions. The aim is to examine if it can provide, in the sense of Remarks \ref{rem1} and \ref{rem1b}, insight for a potential transition between the distinct types of collapse dynamics, even in the case of the initial condition (\ref{tanh}).}
\end{remark}

\begin{figure}[tbh!]
	\centering 
	\includegraphics[scale=0.3]{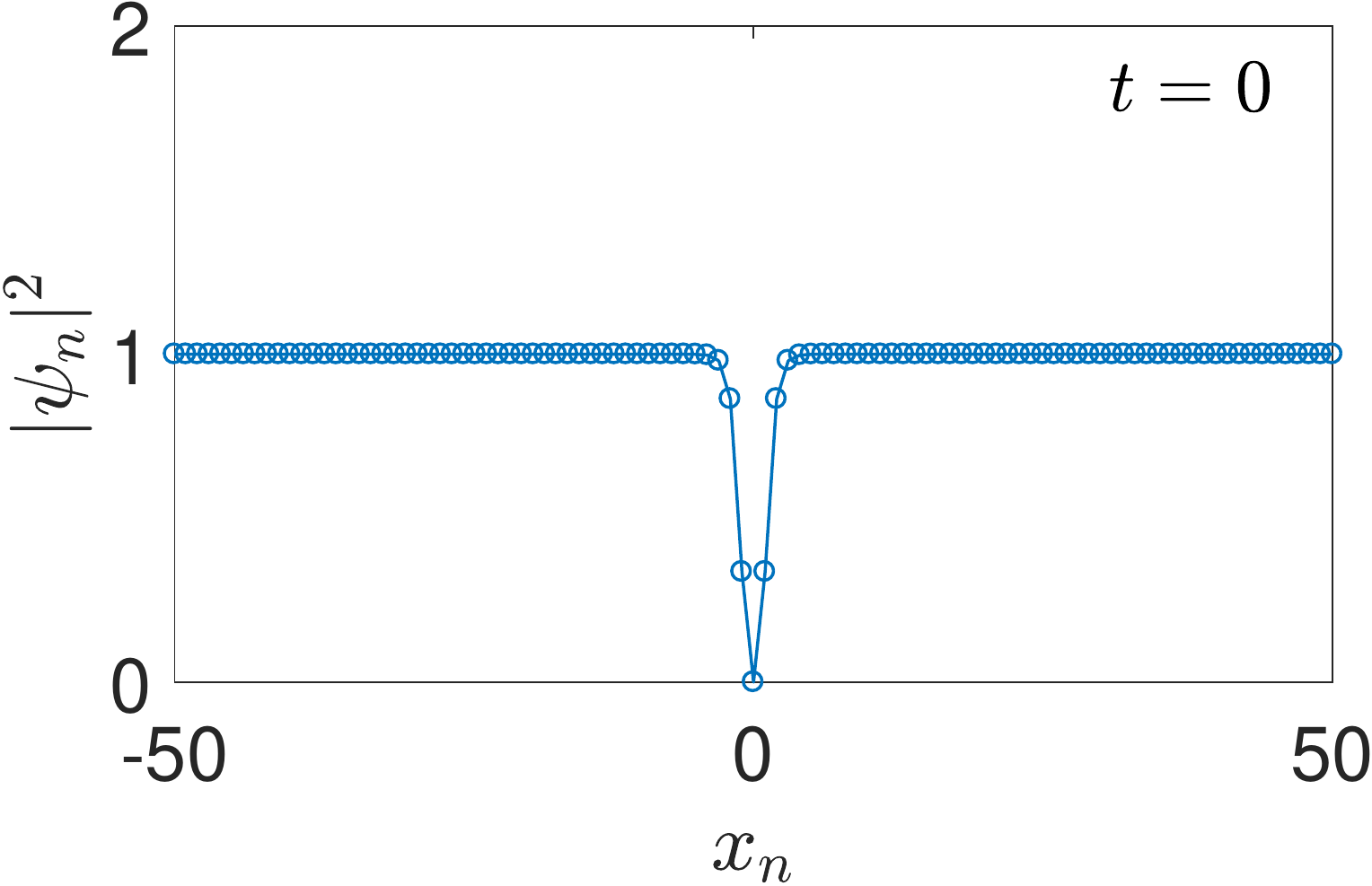}
	\quad
%	\includegraphics[scale=0.3]{fig13b.pdf}
%	\quad 
%	\includegraphics[scale=0.3]{fig13c.pdf}
%	\\ [4ex]
	\includegraphics[scale=0.3]{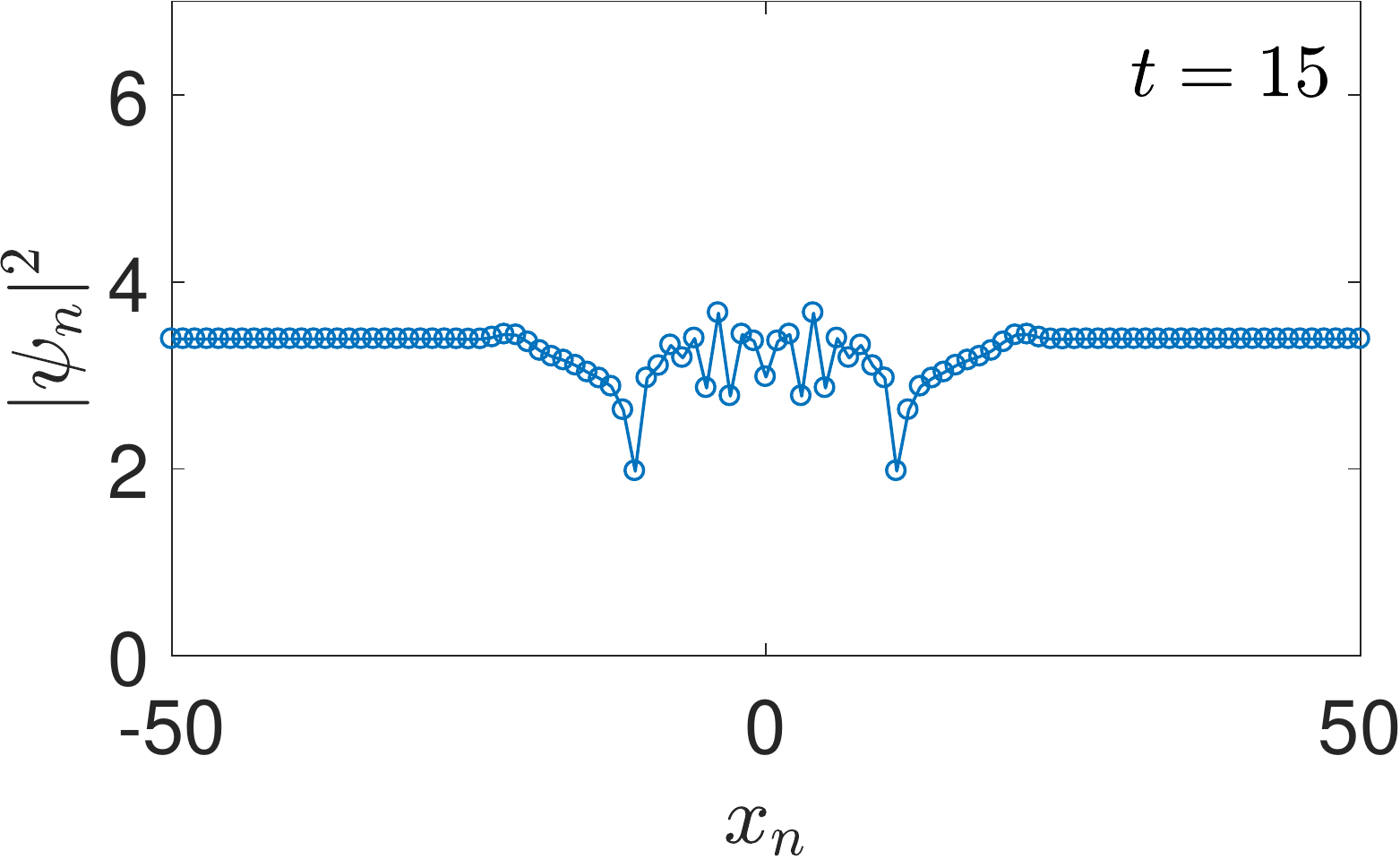}
%	\quad 
%	\includegraphics[scale=0.3]{fig13e.pdf}
	\quad 
	\includegraphics[scale=0.3]{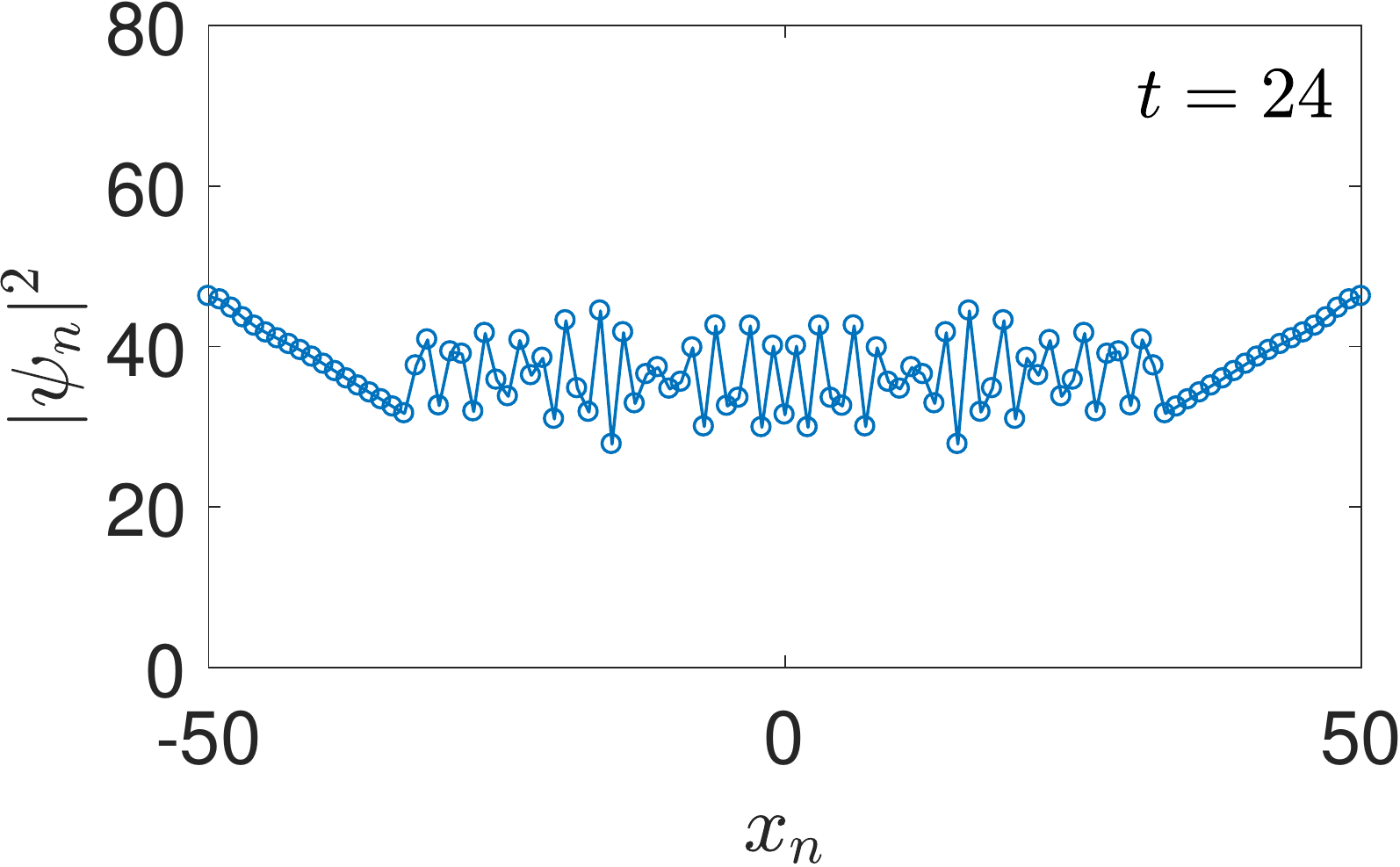}
	\caption {Snapshots of the time-evolution of the density for the initial condition (\ref{tanh})  of amplitude $A=1$, in the defocusing case $s=1$. Other parameters: $\gamma=0.025$, $\delta=0.01$, $h=1$, and $L=50$.}
	\label{Fig8B}
\end{figure}

In the defocusing case $s=1$, the upper bound $\widehat{T}_\text{max}$ is proved to be a sharp estimate for the numerical blow-up times for all values of the parameter $h$. This excellent agreement can be supported by examining the dynamical behavior of the solutions. 
%the expectation that the dip of the density profile has a little effect to the averaged norm $M(t)$, and the main contribution  comes from the  wave background. More precisely, the dynamics of the ODE (\ref{eqbern}) yet effectively predicts the evolution of $M(t)$ in the case of the initial data (\ref{tanh}); this is another characteristic example of the extended blow-up type discussed in the previous sections.
%%%%%%%%%%%%%%%%%%%%%%%%%%%%%%%%%%%%%%%%%%%%
%Indeed, this is the case observed 
In Figure~\ref{Fig8B}, the time-evolution of the density of such a solution, when $\gamma=0.025$ and $h=1$ is depicted. We observe that the increase of the height of the wave background is simultaneously accompanied by the spreading of the perturbation caused by the initial central dip along the lattice, almost in its full extend. Since this behavior is a very good example of the extended blow-up scenario, the compliance of the blow-up times with $\Th$ is expected according to Remarks \ref{rem1}-\ref{rem1b}. 

%\begin{figure}[tbh!]
%	\centering 
%	\includegraphics[scale=0.43]{fig14a.pdf}
%	\includegraphics[scale=0.43]{fig14b.pdf}
%	%
%	\caption{Spatiotemporal evolution of the density for the defocusing case $s=1$, when $\delta = 0.01, L = 50, h=1$, $ N = 100$, and the amplitude of the initial condition (\ref{tanh}) is $A=1$. Left panel: Evolution of the density towards collapse, when $\gamma=0.025>\gamma^*\simeq-0.01$. Right panel: Decay of the density when $\gamma=-0.03<\gamma^*\simeq-0.01$.} 
%	\label{Fig8A}
%\end{figure}

%As we can see the numerical blow-up time is quite close to the analytical upper bound. The critical value of $\gamma$ was calculated numerically and has the value $\gamma^*_{num} \approx -0.02$. However, for the focusing DNLS things are different. As we can see in the inset, the analytical upper bound is preserved but the solution blows-up much earlier, specially for small values of $\gamma$. 
%
%Next, the top right panel of Figure \ref{Fig8} portrays the evolution of the density $|\psi_n|^2$ for the defocusing DNLS, in the regime of finite-time collapse. Here, the parameters are the same as those in the top left panel, but for $\gamma = 0.025$. The numerical collapse time is $Tnum = 16$ and the analytical one is $Tmax = 16.52$. 
%
%The bottom panels illustrate the snapshots of the evolution of the density $|\psi_n|^2$ for the set of parameters  used in the top right panel of Figure \ref{Fig8}. We observe that the the initial density dip separates in two traveling pulses prior to the collapse time.

We turn now in  the insets of Figure~\ref{Fig8} devoted to the focusing case $s=-1$, revealing an intriguing effect. For $h=0.1$, the numerical blow-up times lie close to the lower bound $\overline{T}_\text{max}$% although the initial condition \eqref{tanh} is not satisfying problem ($\mathcal{D}$)
.  As the lattice spacing $h$ is increasing, the actual blow-up times, depart from the lower bound $\overline{T}_\text{max}$, and  progressively,  approach  the upper bound $\widehat{T}_\text{max}$.

\begin{figure}[tbh!]
	\centering 
\includegraphics[scale=0.3]{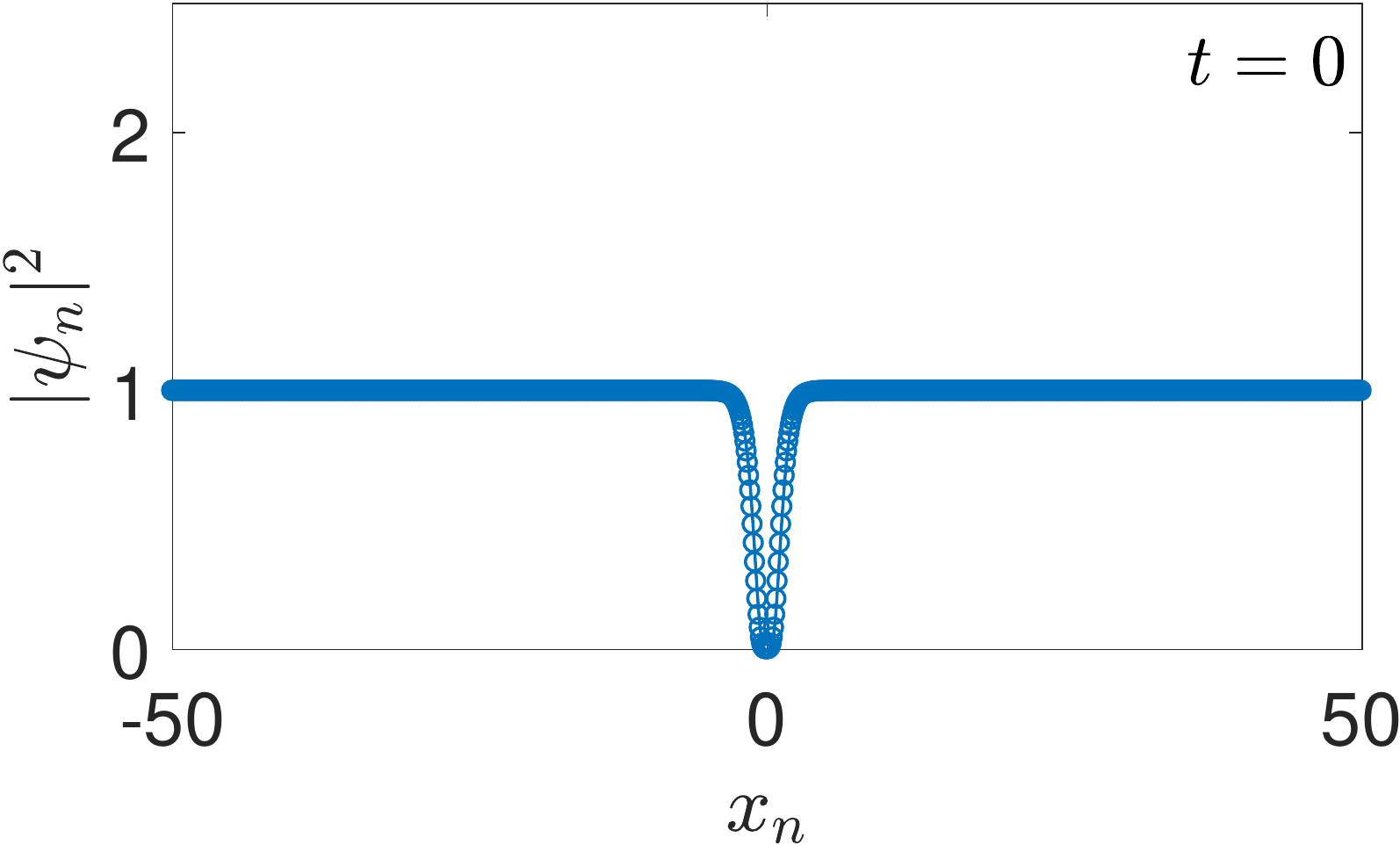}
\quad
\includegraphics[scale=0.3]{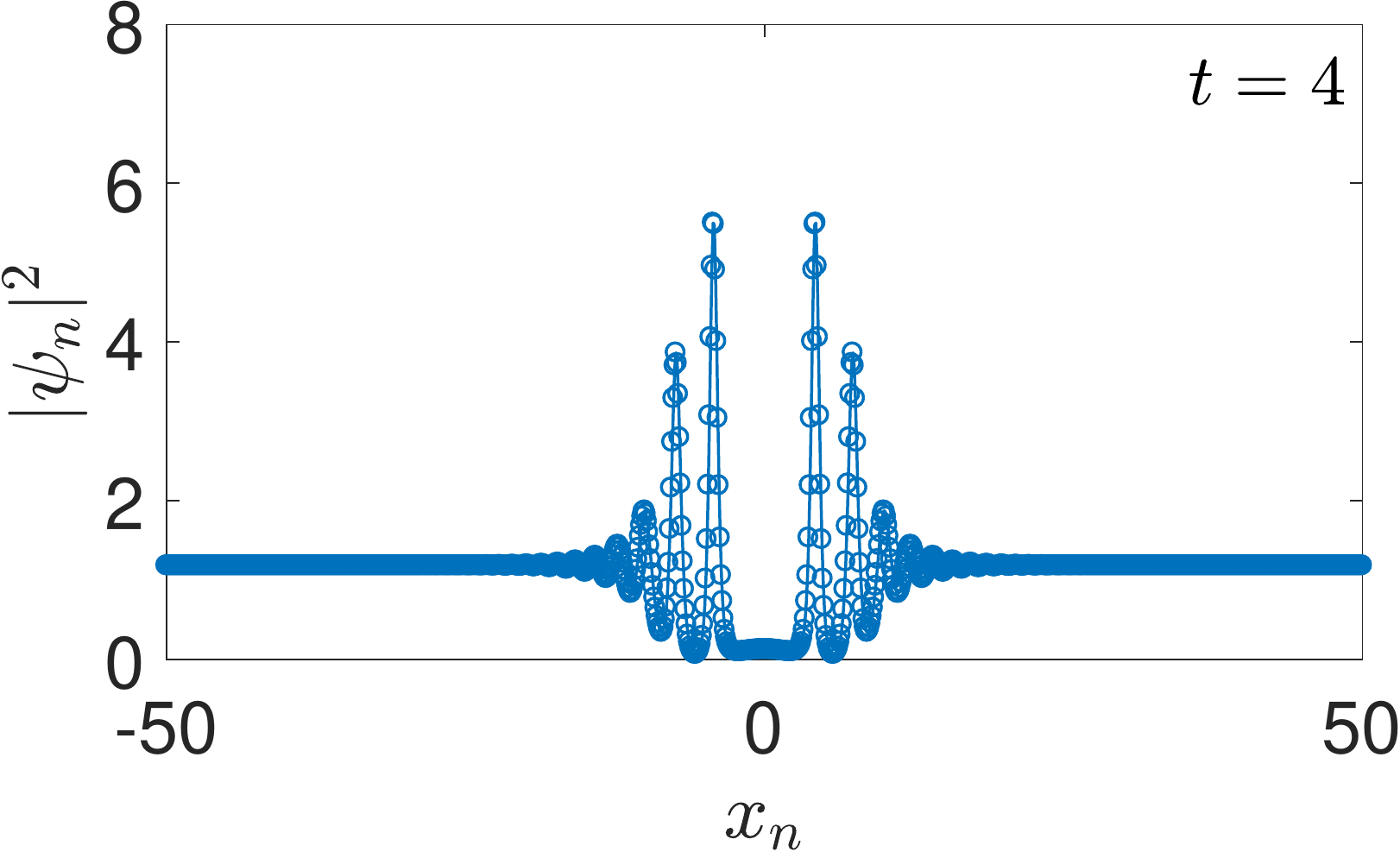}
\quad 
\includegraphics[scale=0.3]{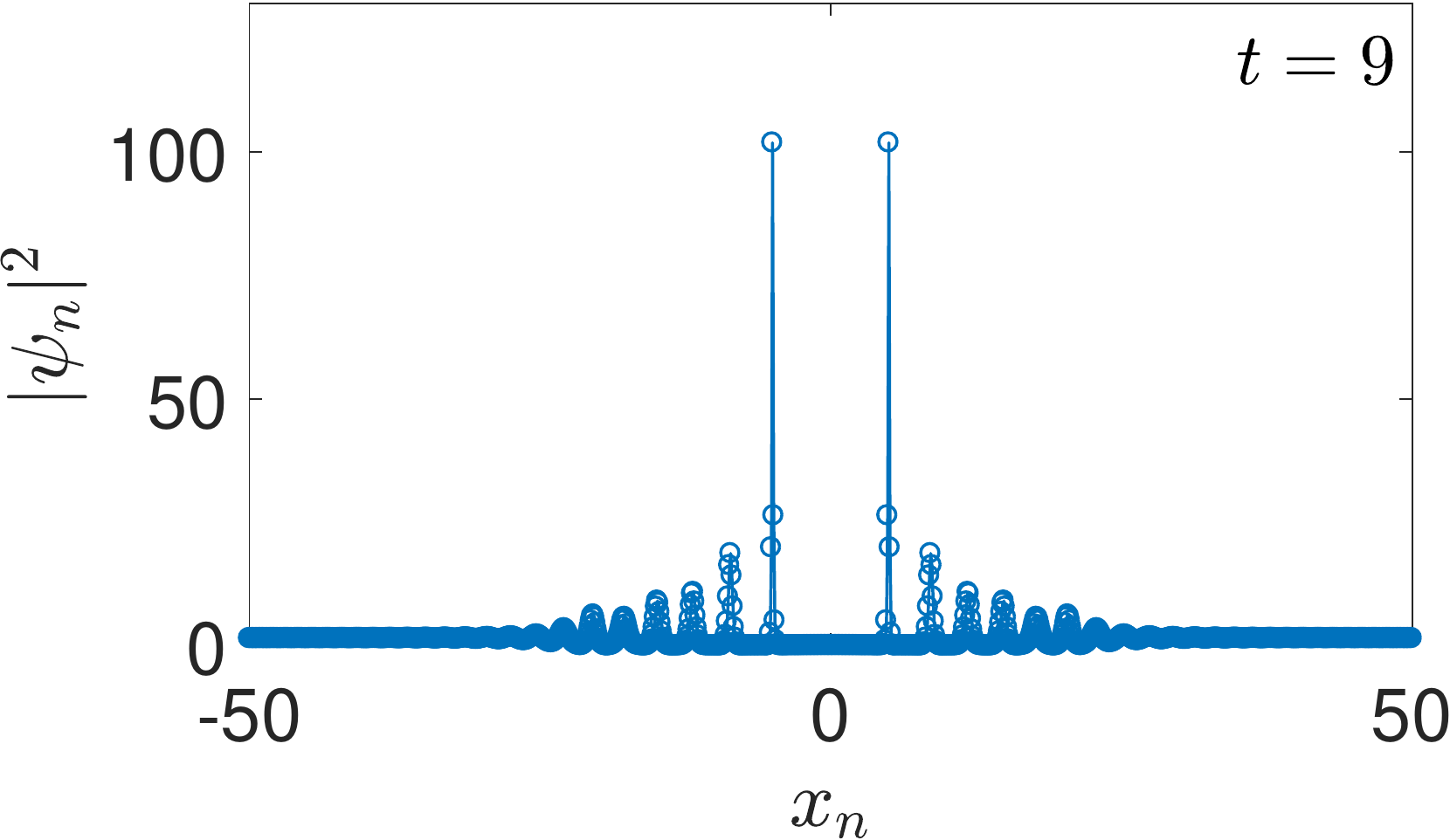}

	\caption{
Snapshots of the time-evolution of the density for the initial condition (\ref{tanh})  of amplitude $A=1$, in the focusing case $s=-1$. Other parameters:  $\gamma=0.01$,  $\delta=0.01, L=50,~\text{and}~N=1000$ ($h=0.1$).}
	\label{Fig10A}
\end{figure}

\begin{figure}[tbh!]
	\centering 
\includegraphics[scale=0.3]{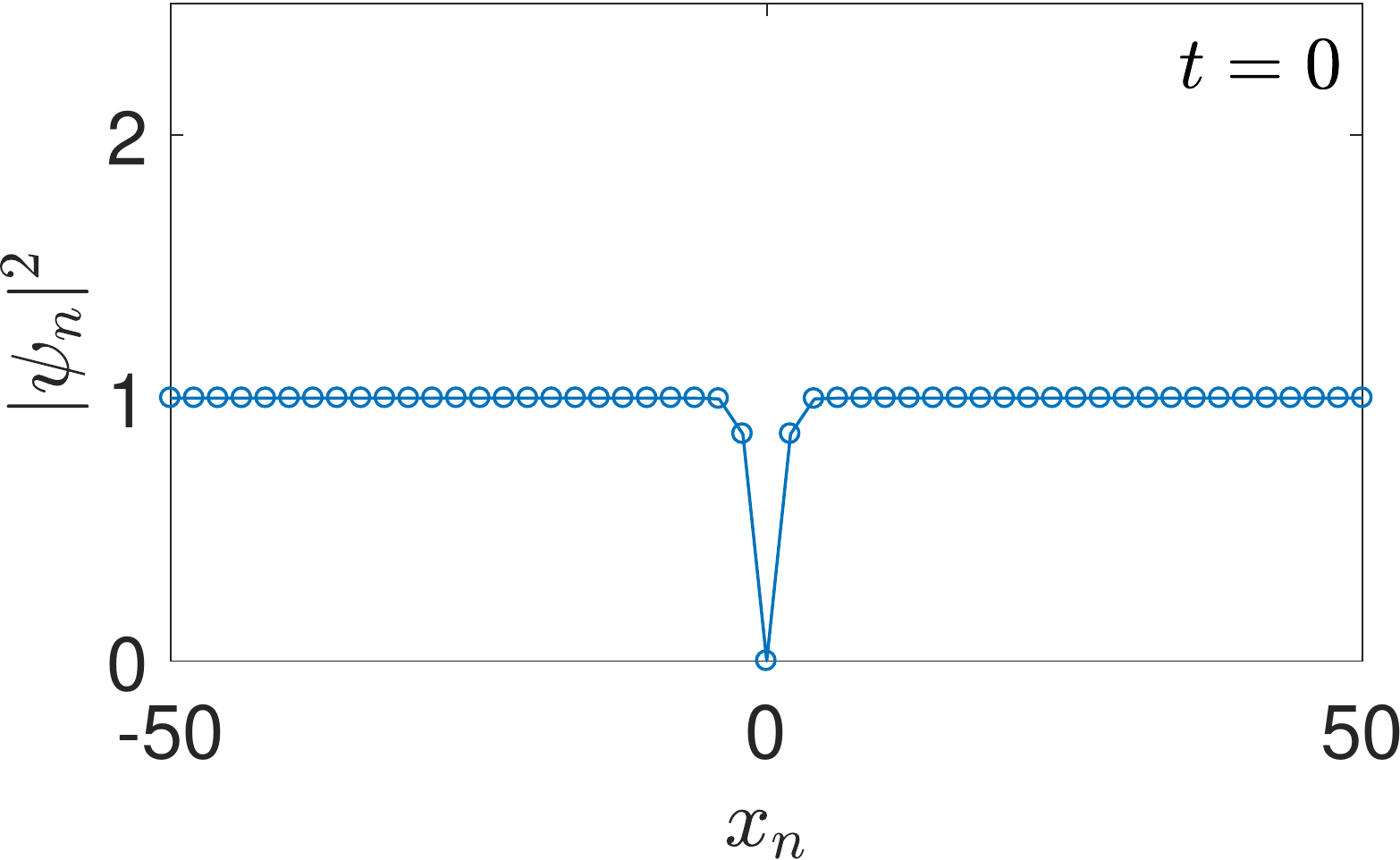}
\quad
\includegraphics[scale=0.3]{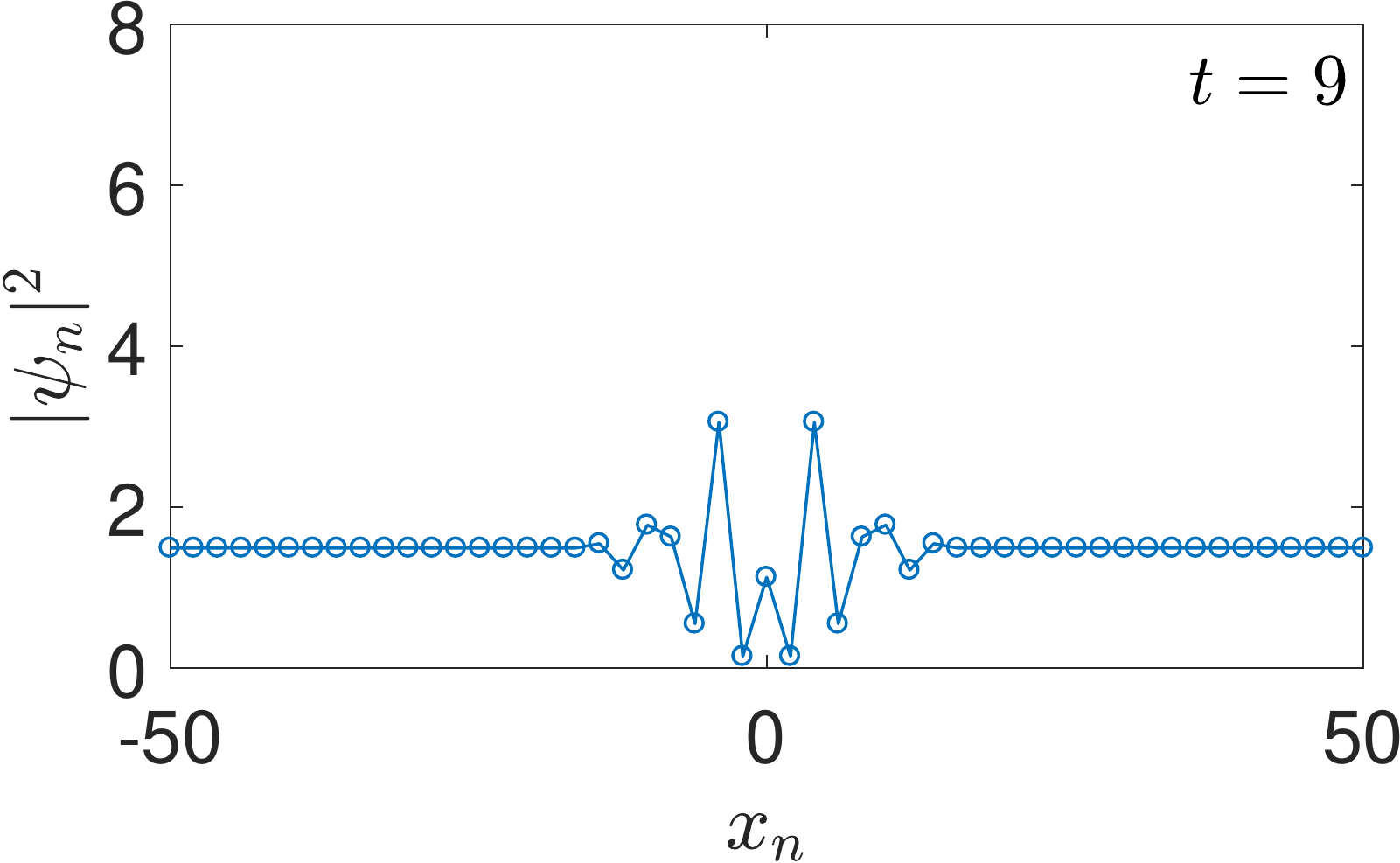}
\quad 
\includegraphics[scale=0.3]{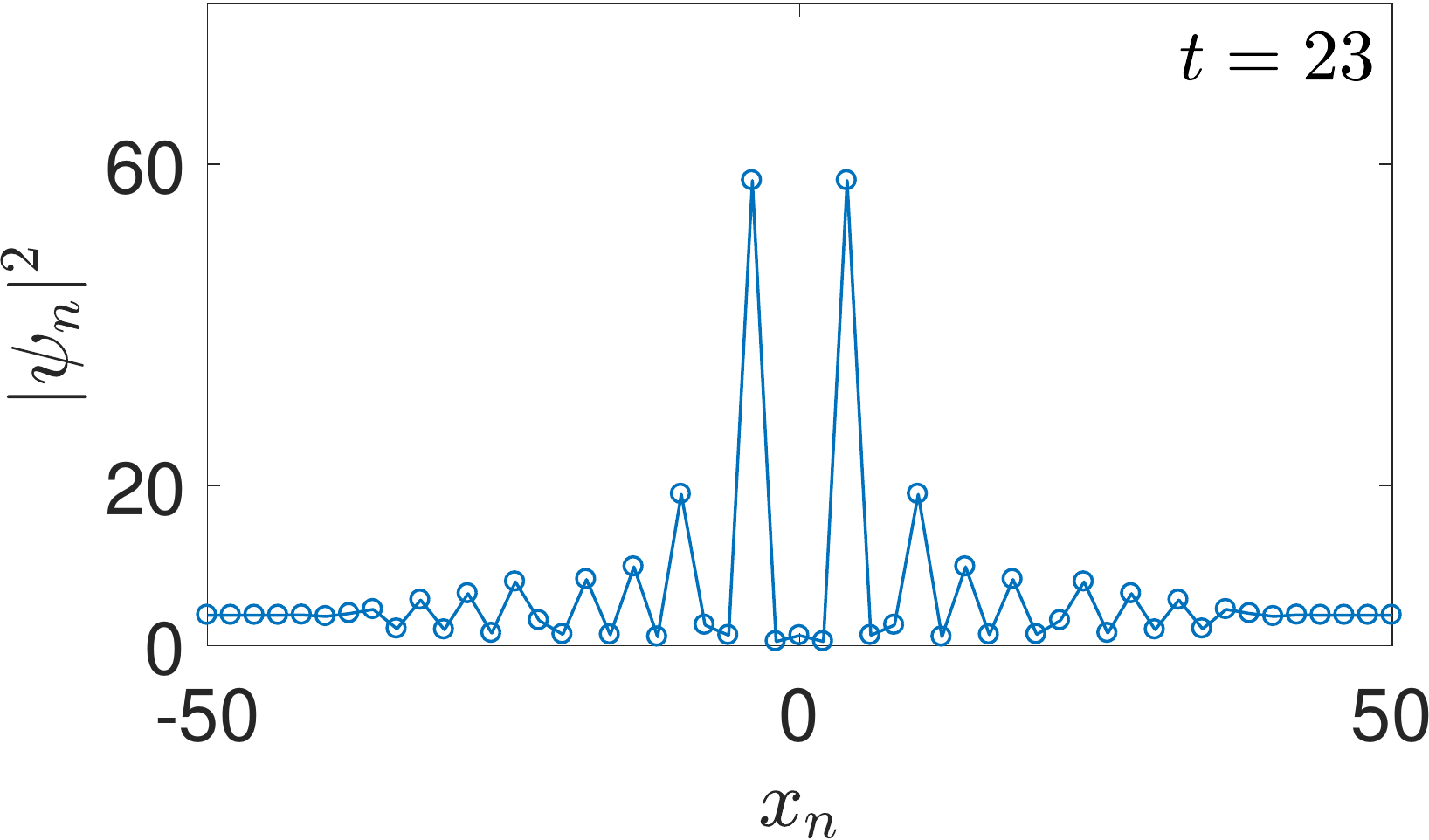}

	\caption{Snapshots of the time-evolution of the density for the initial condition (\ref{tanh})  of amplitude $A=1$, in the focusing case $s=-1$. Other parameters: $\gamma=0.01$,  $\delta=0.01, L=50,~\text{and}~N=50$ ($h=2$).}
	\label{Fig10B}
\end{figure}

This behavior contradicts in a sense, the one observed in the case of the vanishing initial conditions discussed in Section~\ref{sec3B}.  Some light on the causes of this behavior can be shed from Figures~\ref{Fig10A} and \ref{Fig10B},  which are portraying snapshots of the time-evolution of the density when $h=0.1$ and $h=2$, respectively.

%%%%%%%%%%%%%%%%%%%%%%%%%%%%%%%%%

In Figure~\ref{Fig10A}, we observe that the lattice with $h=0.1$ being focusing, tends to transform the initial ``dark discrete-soliton''-like profile,  to a  discrete ``bright'' localized wave-train, as shown in the snapshot for $t=4$, where the emergence of two peaks adjacent to the central dip is observed. At time $t=9$, these units are attaining a critically high amplitude prior to collapse against their neighbors  due to energy gain, and eventually, collapse occurs through these critical ``bright'' humps.   This highly nontrivial transient dynamics crucially differs from the one of the defocusing case (where the increase of the height of the wave background dominates), and is more compatible to the localized blow-up scenario, explaining the relevance of the lower bound $\overline{T}_\text{max}$ to the numerical blow-up times.

For $h=2$, we observe in Figure~\ref{Fig10B}, similarities of the transient dynamics to that of the case $h=0.1$, but also, a significant difference.  Although collapse is again dominated by the emergence of the critical ``bright'' solitonic structures, rather unexpectedly, their formation is accompanied by an unusual, for this discretization regime, energy dispersion along the lattice, and comparable amplitude oscillations of the adjacent units, to that of the critical  ``bright'' humps.  The whole evolution is closer to the scenario described by the extended blow-up, explaining in this case,  the proximity of the numerical blow-up times to the upper bound  $\widehat{T}_\text{max}$.

\subsection{Box-profiled initial conditions}
\label{sec3D}
\begin{figure}[!h]
	\centering 
	\includegraphics[scale=0.35]{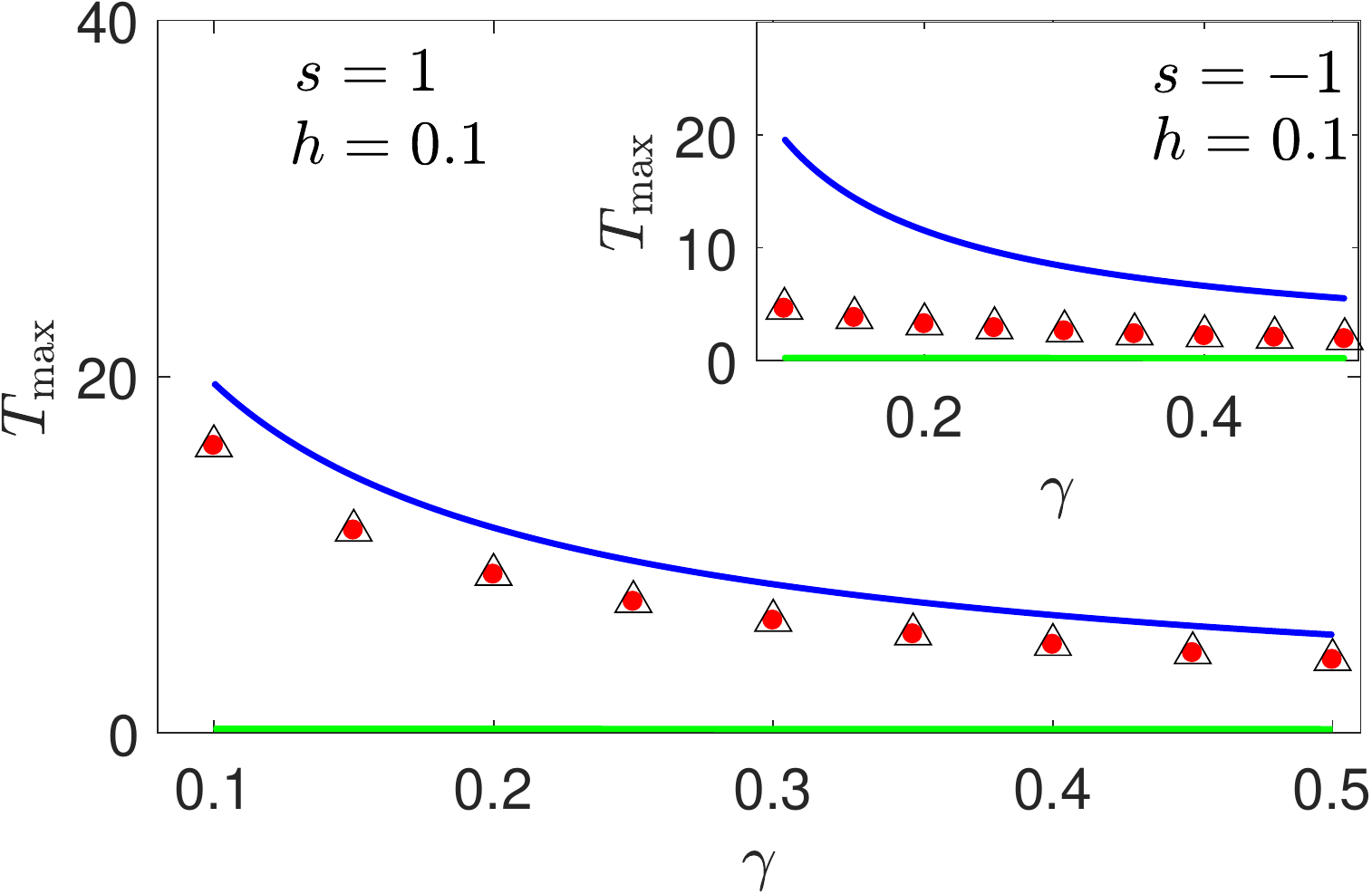}
	\hspace{0.2cm}
	\includegraphics[scale=0.35]{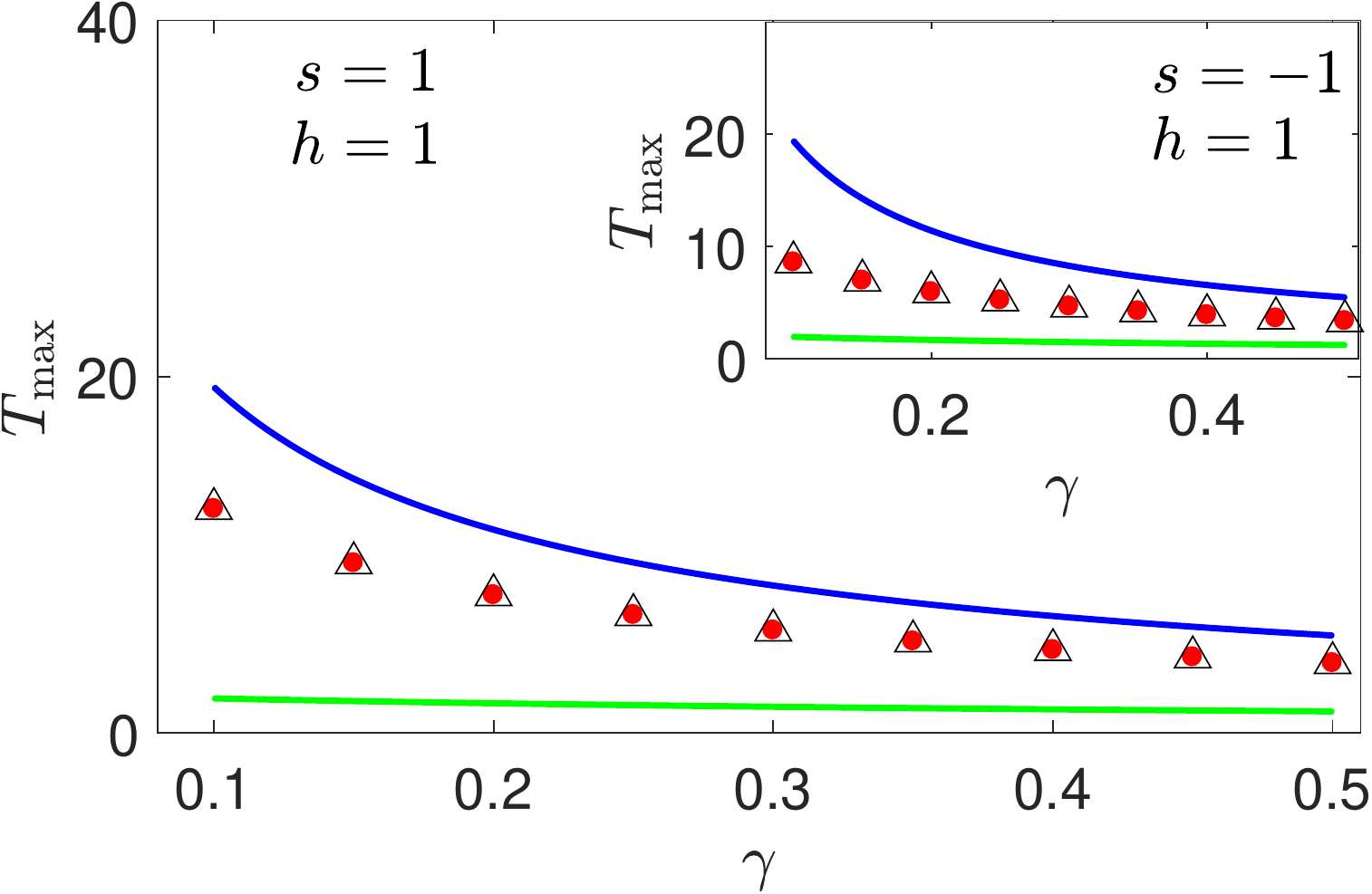}
	\hspace{0.2cm}
	\includegraphics[scale=0.35]{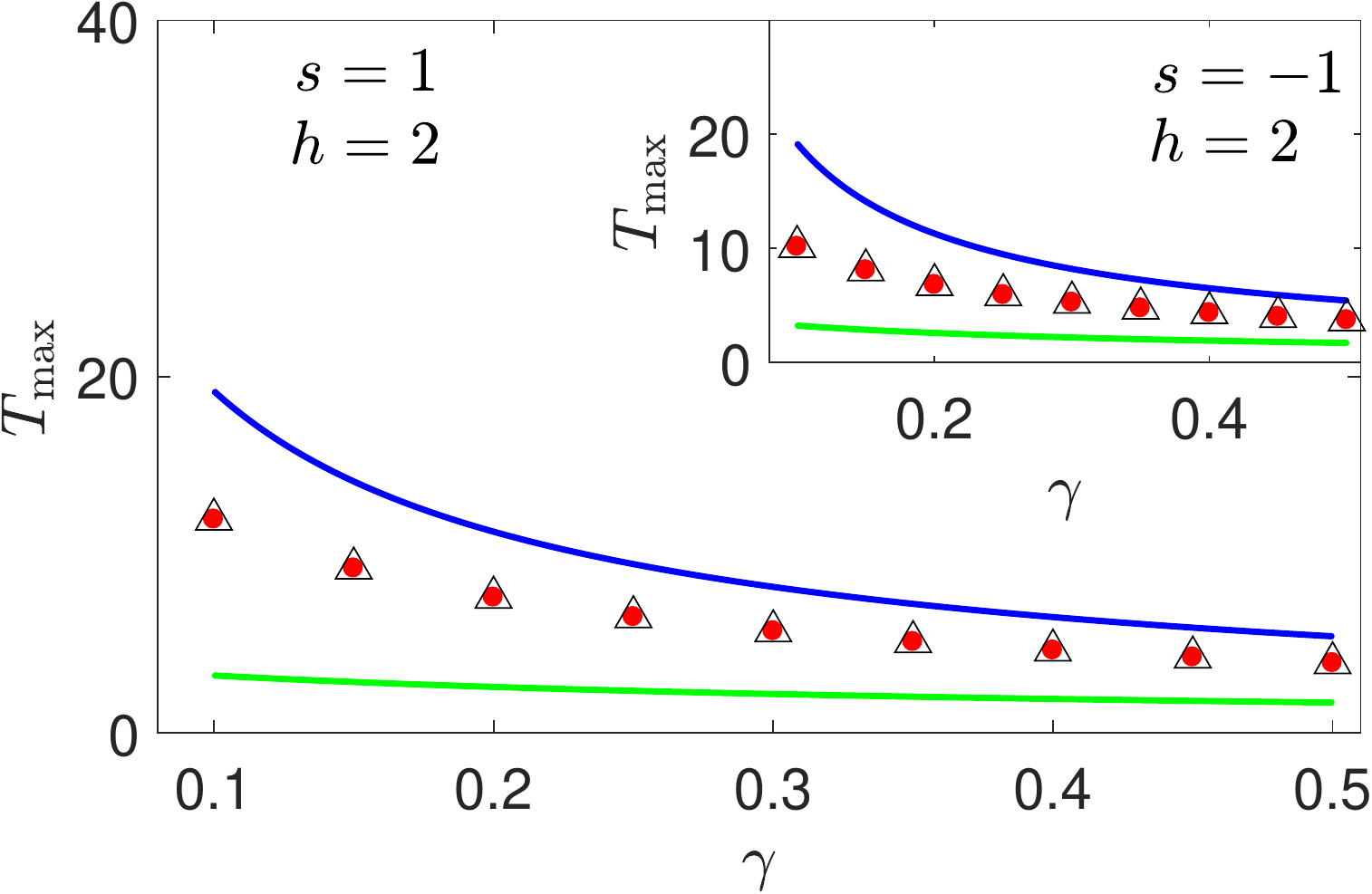}
	\caption{Numerical blow-up times [(red) dots for the problem ($\mathcal{P}$)-triangles for the problem ($\mathcal{D})$], for the rectangular box-profiled initial conditions with width $w=20$ and amplitude $A=1$,  against the analytical upper  bound [upper (blue) curve] and lower bound [lower (green) curve], as functions of $\gamma\in[0.1,0.5]$.  Main panels for the defocusing case $s=1$, and insets to the focusing one $s=-1$. In all panels and insets, $L=50$, $\delta=0.01$ and $A = 1$.   Left panel: $N=1000$ and $h=0.1$. Middle panel: $N=100$ and $h=1$. Right panel: $N=50$ and $h=2$.} 
	\label{figure11}
\end{figure}

We conclude the numerical studies by considering a final example of initial conditions possessing the form of a rectangular box.  We consider a box of width $w=20$ and amplitude $A=1$. The  numerical blow-up times are depicted in Figure~\ref{figure11}, against the analytical bounds for both the defocusing (main panels) and the focusing case (insets), as functions of the parameter $\gamma$.

%%%%%%%%%%%%%%%%%%%%%%%%%
We observe that the numerical blow-up times lie between the analytical bounds,  yet suggesting an intermediate type of collapse between the extended and localized one.  We  conclude, that a main characteristic, which determines the type of collapse is the localization width of the initial condition, despite their specific profile.   In addition, we observe (results are not shown for brevity), that as the width becomes smaller, the transient behavior prior collapse becomes reminiscent of the $\mathrm{sech}$-initiated type dynamics, while for increasing localization widths becomes reminiscent of the plane wave or $\tanh^2$-initiated type dynamics. 

\section{Conclusions}\label{sec4}
In this work, analytical studies corroborated by numerical simulations, considered the collapse dynamics of the discrete nonlinear Schr\"{o}dinger (DNLS) equation (\ref{eq01}), incorporating linear and nonlinear gain and loss. 
%The model has important applications such as in nonlinear optics, describing the dynamics of coupled waveguides where linear and nonlinear gain/losses are of importance. 
We proved by means of analytical energy arguments, upper and lower bounds for the collapse time in the relevant parametric regime. Interestingly, this collapse regime is defined by the existence of a critical value $\gamma^*$ when $\gamma<0$, $\delta>0$,  which separates the finite-time collapse from the decay of solutions. Considering both the focusing and the defocusing cases of the model, the upper bound was proved for the periodic lattice, while the lower bound was proved for the lattice supplemented with vanishing at infinity, or Dirichlet boundary conditions. In the latter case, we took advantage  of the reverse order inclusions between the spaces of summable sequences, being valid only for the inherently discrete model and not for the continuous counterpart.
  
The analytical results were tested by extended numerical simulations, for a variety of initial conditions: spatially extended, vanishing localized (resembling the profile of a ``discrete bright soliton''),  decaying to a finite background (resembling the profile of a ``discrete dark soliton'') and box-shaped ones. The numerical findings revealed that the analytical estimates can be used in order to classify distinct types of collapse.

For instance, the agreement of the analytical upper bound with the numerical blow-up times proved to be excellent for the spatially extended initial data. Their finite-time blow-up was analyzed through a reduction to relevant ODE dynamics, and defined the extreme case of the {\em extended (or weak) type of collapse}.

In the case of the localized initial conditions, the behavior of the system was proved to be more complex though well-understood. We identified two distinct extreme types of collapse dynamics. The first type is an extended type of collapse.  In this case, a large portion of the lattice or even the whole  lattice, collapses in finite time. The dynamics in this case,  is reminiscent of the evolution of spatially extended initial data towards collapse and is favored by the defocusing variant of the DNLS and the continuous/low amplitude regime. When extended collapse occurs, the numerical blow-up times approach, and in many cases, are in excellent agreement with the analytical upper bound. The second type identified is the {\em localized (or strong)} collapse. In this case, collapse occurs through a single site without transient energy dispersion. This behavior is favored by the focusing nature of the DNLS and the discrete/high amplitude regime. When localized collapse occurs, the numerical blow-up times approach, and in many cases, are in excellent agreement with the analytical lower bound. In the intermediate discreteness/amplitude regime, we identified that collapse  shares a mixing of the weak and of the strong type of collapse dynamics. In this case, the numerical blow-up times lie between the analytical bounds. In all cases, the actual blow-up times follow qualitatively the parametric functional dependencies predicted by the analytical estimates.

In the case of initial conditions bearing a density dip, we found significant differences between the defocusing and the focusing case. In the defocusing case, the exhibited dynamics resembles that of the extended type of collapse. Consequently, the numerical blow-up times are in excellent agreement with the analytical upper bound.  In the focusing case an intriguing effect was revealed. In the continuous regime the numerical blow-up times are in a proximity with the analytical lower bound, while they approach the analytical upper bound in the discrete regime. This effect drastically differs from the findings for the localized initial data, and is rather unexpected, as enhanced discreteness is associated with limited energy dispersion, which is more compatible to the localized type of collapse.  

Finally, we performed a numerical study by using rectangular box-shaped initial conditions of various localization widths. We observed that as the width of the box is decreasing, the  dynamical behavior of the system becomes reminiscent of the $\sech$-initiated collapse dynamics, while for increasing localization widths, the behavior of the system becomes similar to that of the plane wave or $\tanh^2$-initiated collapse dynamics. Hence, we concluded that the transient behavior prior collapse is chiefly determined by the localization width of the initial condition and by the focusing/ defocusing nature of the lattice. 

Our results pave the way for a better understanding of the dynamics of the considered generalized 
DNLS equation. 
An interesting direction would be to discuss the collapse dynamics for the case of higher dimensional lattices, where higher dimensionality may drastically impact the evolution corroborated with focusing/defocusing and energy gain/loss effects. On the other hand, one could consider extended models with saturable nonlinearities \cite{KevreDNLS}, or nonlinear terms describing nonlinear hopping which may allow a fast energy propagation along the lattice \cite{joha2,joha3,joha4}. These investigations are currently in progress and results will be reported in future works.

%%%%%%%%%%%%%%%%%%%%%%%%%%%%%%%%%%%%%%%%%%%%%%%%
\section*{Acknowledgments}
The author K.V. gratefully acknowledges the support of the YPATIA Doctoral Fellowship Program of the Research Unit of the University of the Aegean.
The authors N.I.K., G.F. and V.K., acknowledge that this work was made possible by the NPRP grants \# [8-764-160] and \# [9-329-1-067] from the Qatar National Research Fund (a member of Qatar Foundation). The findings achieved herein are solely the responsibility of the authors.
%% % % % % % % % % % % % % % % % % % % % % % % % % % % % % % % % % % % %	
\appendix
\numberwithin{equation}{section}

\section{Proof of Analytical Results}
\label{proofs}
In this appendix we present the proofs of the main Lemmas and Theorems of the manuscript.
\paragraph{Lemma \ref{DL}}
\begin{proof}
The summation by parts formula for two  discrete functions $\left\{\psi_n\right\}_{n=0}^{n=N}$ and $\left\{\phi_n\right\}_{n=0}^{n=N}$ reads as  
\begin{equation}\label{Apen1}
\sum_{n=0}^{N-1}(\psi_{n+1}-\psi_n)\phi_n = (\psi_{N}\phi_{N}-\psi_0 \phi_0)-\sum_{n=0}^{N-1}(\phi_{n+1}-\phi_n)\psi_{n+1}.
\end{equation}
By applying the above formula and the periodic boundary conditions (\ref{eq02}), we observe that
\begin{align}
\label{SumA}
\mathrm{Re}\sum_{n=0}^{N-1} \Delta_{d} \psi_n \overline{\psi}_n = & \mathrm{Re}\sum_{n=0}^{N-1} (\psi_{n+1} - 2\psi_n + \psi_{n-1})\overline{\psi}_n =  
\mathrm{Re}\sum_{n=0}^{N-1} (\psi_{n+1} - \psi_n)\overline{\psi}_n + \mathrm{Re}\sum_{n=0}^{N-1}(\psi_{n-1} - \psi_n)\overline{\psi}_n\nonumber\\ = & 
\mathrm{Re}\sum_{n=0}^{N-1} (\psi_{n+1} - \psi_n)\overline{\psi}_n - \mathrm{Re}\sum_{m=-1}^{N-2}(\psi_{m+1} - \psi_{m})\overline{\psi}_{m+1}\nonumber\\
= & 
\mathrm{Re}\left[\overline{\psi}_N\psi_N - \overline{\psi}_0\psi_0\right] - \mathrm{Re}\sum_{n=0}^{N-1} (\overline{\psi}_{n+1}-\overline{\psi}_n)\psi_{n+1}\nonumber \\ - &
\mathrm{Re}\left[\psi_{N-1}\overline{\psi}_{N-1}  - \psi_{-1} \overline{\psi}_{-1}\right] + \mathrm{Re}\sum_{m=-1}^{N-2}(\overline{\psi}_{m+1} - \overline\psi_{m})\psi_m .
\end{align} 
Applying the formula (\ref{Apen1}) once more, on the last term of the right-hand side of (\ref{SumA}), we have:
\begin{align}
\label{eqJ1} 
\mathrm{Re}\sum_{n=0}^{N-1} \Delta_{d} \psi_n \overline{\psi}_n = &
-\mathrm{Re}\sum_{n=0}^{N-1} (\overline{\psi}_{n+1}-\overline{\psi}_n)\psi_{n+1}  + \mathrm{Re}\left[\psi_{-1}(\overline{\psi}_0 - \overline{\psi}_{-1})\right] +  \mathrm{Re}\sum_{m=0}^{N-2}(\overline{\psi}_{m+1} -\overline{\psi}_m)\psi_{m}\nonumber \\ = & 
-\mathrm{Re}\sum_{n=0}^{N-1} (\overline{\psi}_{n+1}-\overline{\psi}_n)\psi_{n+1} + \mathrm{Re}\left[\psi_{N-1}(\overline{\psi}_N - \overline{\psi}_{N-1})\right] + \mathrm{Re}\sum_{m=0}^{N-2}(\overline{\psi}_{m+1} -\overline{\psi}_m)\psi_{m}\nonumber  \\ = & 
-\mathrm{Re}\sum_{n=0}^{N-1} (\overline{\psi}_{n+1}-\overline{\psi}_n)(\psi_{n+1}-\psi_{n})\nonumber \\ =& 
-\sum_{n=0}^{N-1}|\psi_{n+1}-\psi_{n}|^2
\end{align}
Eq.~(\ref{lp6}) readily follows from (\ref{eqJ1}), if the latter is multiplied by $h$. For Eq.~(\ref{lp7}), we repeat the same calculation on the quantity $\mathrm{Re}\sum_{n=0}^{N-1} \Delta_{d} \phi_n \overline{\psi}_n$.  
\end{proof}
%%%%%%%%%%%%%%%%%%%%%%%%%%%%%%%%%%%%%%%%%%%%%
\paragraph{Lemma \ref{DifEq}}
\begin{proof}
Since $\psi\in\mathrm{C}^1([0,T_{\mathrm{max}}(\psi^0)),\ell^2_{\mathrm{per}})$, we may differentiate Eq.~\eqref{eq06} with respect to time, and get the equation
\begin{equation}\label{eq07}
\frac{d M(t)}{dt} = -2\gamma \frac{e^{-2\gamma t}}{N} \sum_{n=0}^{N-1}|\psi_n|^2 + \frac{2e^{-2\gamma t}}{N} \re\sum_{n=0}^{N-1}\dot{\psi}_n\overline{\psi}_n.
\end{equation}
By replacing $\dot{\psi}_n$ from Eq.~\eqref{eq01} to the second term of the right-hand side of Eq.~\eqref{eq07},  we have 
$$%\begin{equation}\label{eq08}
\frac{2e^{-2\gamma t}}{N} \re\sum_{n=0}^{N-1}\dot{\psi}_n\overline{\psi}_n = \frac{2e^{-2\gamma t}}{N} \sum_{\xi=1}^4 J_{\xi},
$$%\end{equation}
where 
$$J_1 = \re \left(\rmi sk\sum_{n=0}^{N-1} \Delta_{d} \psi_n \overline{\psi}_n\right), \quad 
J_2 = \re \left(\rmi \sum_{n=0}^{N-1} |\psi_n|^2 \psi_n \overline{\psi}_n\right),$$
$$J_3 = \re \left(\gamma \sum_{n=0}^{N-1}\psi_n\overline{\psi}_n\right)=\gamma\sum_{n=0}^{N-1}|\psi_n|^2\quad\text{and }\quad  J_4 = \re \left(\delta \sum_{n=0}^{N-1}|\psi_n|^2 \psi_n \overline{\psi}_n\right)=\delta\sum_{n=0}^{N-1}|\psi_n|^4.$$  
Clearly,
$$
J_2 = \re \left(\rmi \sum_{n=0}^{N-1} |\psi_n|^4 \right) = 0.
$$
Also, due to Lemma \ref{DL}, we have that
$$
J_1 = \re \left(\rmi s k\sum_{n=0}^{N-1} \Delta_{d} \psi_n \overline{\psi}_n\right)=0.
$$
Hence, Eq.~\eqref{eq07} can be rewritten as
\begin{equation*}
\frac{d M(t)}{dt} = -2\gamma \frac{e^{-2\gamma t}}{N} \sum_{n=0}^{N-1}|\psi_n|^2+2\frac{e^{-2\gamma t}}{N}J_3 +\frac{2e^{-2\gamma t}}{N}J_4,
\end{equation*}
which is actually the claimed Eq.~(\ref{diff.form.M}).
\end{proof}
%%%%%%%%%%%%%%%%%%%%%%%%%%%%%%%%%%%%%%%%
\paragraph{Theorem \ref{The2a}}
\begin{proof}
%\begin{itemize}
%\item[1.]
By applying the Cauchy--Schwarz inequality, we  may see that
\begin{eqnarray}
\label{DCSI}
\sum_{n=0}^{N-1}|\psi_n|^2\leq \sqrt{N}\left(\sum_{n=0}^{N-1}|\psi_n|^4\right)^{\frac{1}{2}}.
\end{eqnarray}
Hence,  $M(t)$ can be estimated from above as 
$$
M(t) = \frac{e^{-2\gamma t}}{N} \sum_{n=0}^{N-1}|\psi_n|^2 \leq 
\frac{e^{-2\gamma t}}{\sqrt{N}} \left(\sum_{n=0}^{N-1}|\psi_n|^4\right)^{\frac{1}{2}},
$$
and it follows that $M(t)^2$ satisfies the inequality
\begin{equation}\label{MM}
M(t)^2 \leq \frac{e^{-4\gamma t}}{N} \sum_{n=0}^{N-1}|\psi_n|^4.
\end{equation}
By using Eq.~\eqref{diff.form.M} of Lemma \ref{DifEq}, the inequality  \eqref{MM} becomes
\begin{eqnarray}
\label{help1R}
M(t)^2 \leq\frac{e^{-2\gamma t}}{2\delta}\frac{d M(t)}{dt} \Longrightarrow
2\delta e^{2\gamma t}M(t)^2 \leq\dot{M}(t),
\end{eqnarray}
%Since,
%$$ 
%\frac{d}{dt}\left[\frac{1}{M(t)}\right] = -\frac{\frac{d M(t)}{dt}}{M(t)^2},
%$$ 
which can be also written as the differential inequality 
\begin{equation}\label{eq18}
\frac{d}{dt}\left[\frac{1}{M(t)}\right]\leq -2\delta e^{2\gamma t}.
\end{equation}
Integration of (\ref{eq18}) with respect to time, gives that
$$ \frac{1}{M(t)}\leq \frac{1}{M(0)} - 2\delta\int_0^t e^{2\gamma \tau}d\tau.$$
Since $M(t)>0,$ we see that $M(0)>0$ satisfies the inequality 
\begin{equation}\label{eq19}
2\delta\int_0^t e^{2\gamma \tau}d\tau \leq \frac{1}{M(0)}.
\end{equation}
We shall distinguish between the cases $\gamma \ne 0$ and $\gamma = 0$. 
\begin{enumerate}
\item[(i)] Assuming that the damping parameter $\gamma\neq0.$ In this case (\ref{eq19}) implies that 
$$\frac{2\delta}{2\gamma}\left(e^{2\gamma t}-1\right) \leq \frac{1}{M(0)} \Rightarrow e^{2\gamma t}\leq 1+ \frac{\gamma}{\delta M(0)}.$$ 
For $\frac{\gamma}{\delta M(0)}>-1$ we get 
\begin{equation*}\label{eq20}
t\leq \frac{1}{2\gamma} \ln \left[1+\frac{\gamma}{\delta M(0)}\right],
\end{equation*}
which proves the estimate of collapse time (\ref{eqTh1})-(\ref{eqTh2}).
\item[(ii)] For $\gamma=0$, we have from \eqref{eq19} that
\begin{equation*}\label{eq21}
2\delta t\leq\frac{1}{M(0)} 
\ \Longrightarrow
t\leq \frac{1}{2\delta M(0)},
\end{equation*}
which proves the estimate of collapse time (\ref{eqTh3}).
\end{enumerate}
\end{proof}
%%%%%%%%%%%%%%%%%%%%%%%%%%%%%%%%
\paragraph{Theorem \ref{The2}}
\begin{proof}
Lemma \ref{DifEq} is still valid in the case of the Dirichlet boundary conditions, and  the functional \begin{equation*}\label{alt06}
M(t) = \frac{e^{-2\gamma t}}{N} \sum_{n=1}^{N-1}|\psi_n|^2,
\end{equation*}
satisfies the differential equality
\begin{equation*}\label{altM}
\frac{d M(t)}{dt} = \frac{2e^{-2\gamma t}}{N}\left(\delta \sum_{n=1}^{N-1}|\psi_n|^4\right).
\end{equation*} 	
Then, by using (\ref{lp1}) for $p=4$ and $q=2$, we get that $M(t)$ satisfies
\begin{eqnarray*}
\frac{d M(t)}{dt} \leq \frac{2\delta e^{-2\gamma t}}{N}\left( \sum_{n=1}^{N-1}|\psi_n|^2\right)^2,
\end{eqnarray*} 
which can be rewritten in the form of the quadratic differential inequality 
\begin{eqnarray}\label{altM1}
	\dot{M}(t) \leq 2\delta N e^{2\gamma t}M(t)^2.
\end{eqnarray}	
Let us now consider, the initial value problem for the quadratic non-autonomous ODE
\begin{eqnarray}
\label{altode1}
\dot{x}(t)=2\delta Ne^{2\gamma t}x(t)^2,\;\;x(0)=M(0).
\end{eqnarray}
The solution of (\ref{altode1}) is
\begin{eqnarray*}
\label{altode3}
x(t)=\frac{\gamma M(0)}{\gamma+\delta N M(0)-\delta N M(0)e^{2\gamma t}}.
\end{eqnarray*}
Noticing that $NM(0)=P(0)$, its maximal interval  of existence is
\begin{eqnarray}
\label{estN}
\overline{T}_{\mathrm{max}}=\frac{1}{2\gamma} \ln \left[1+\frac{\gamma}{\delta N M(0)}\right]=\frac{1}{2\gamma} \ln \left[1+\frac{\gamma}{\delta P(0)}\right].
\end{eqnarray}
Since the function $f(t;x):=2\delta Ne^{2\gamma t}x(t)^2$ is strictly monotone increasing in the variable $x$, we may apply the comparison principle \cite[pp. 937--943]{zei85}, for the ODE (\ref{altode1}) and the differential inequality (\ref{altM1}) and get that $M(t)\leq x(t)$ for all $t\in [0,T]$-an interval where both $x(t)$ and $M(t)$ are finite. Since, by repeating the arguments of Theorem \ref{The2a}, $M(t)$ blows-up in finite time and $M(t)\leq x(t)$,  then $x(t)$ will blow earlier than $M(t)$. Therefore the new estimate $\overline{T}_{\mathrm{max}}$ of \eqref{estN} serves as a lower bound for the collapse time. In addition, as it can be verified by a simple comparison between formulas \eqref{estN} and \eqref{eqTh1}, it holds that $\overline{T}_{\max}[\gamma,\delta,P(0)]<\widehat{T}_{\max}[\gamma,\delta,M(0)]$.
\end{proof}

%%%%%%%%%%%%%%%%%%%%%%%%%%%%%%%%%%%%%%%%%%

%%%%%%%%%%%%%%%%%%%%%%%%%%%%%%%%%%%%%%%%%%%%%%

%% 
%%
\end{document}